%% file: Nearby-PM-v5r1-arXiv.tex
\documentclass{aa}

\usepackage{graphics}
\usepackage{graphicx}
\usepackage{multicol}
\usepackage{txfonts}
\usepackage{amsmath}
\usepackage{verbatim}
\usepackage{pdflscape}
\usepackage{afterpage}

\usepackage{natbib,twoopt}
\usepackage{amsmath} % For multiple line equations
\usepackage{hyperref} %% to avoid \citeads line fills
\bibpunct{(}{)}{;}{a}{}{,} %% natbib format for A&A and ApJ
\makeatletter
\newcommandtwoopt{\citeads}[3][][]{\href{http://adsabs.harvard.edu/abs/#3}%
{\def\hyper@linkstart##1##2{}%
\let\hyper@linkend\@empty\citealp[#1][#2]{#3}}}
\newcommandtwoopt{\citepads}[3][][]{\href{http://adsabs.harvard.edu/abs/#3}%
{\def\hyper@linkstart##1##2{}%
\let\hyper@linkend\@empty\citep[#1][#2]{#3}}}
\newcommandtwoopt{\citetads}[3][][]{\href{http://adsabs.harvard.edu/abs/#3}%
{\def\hyper@linkstart##1##2{}%
\let\hyper@linkend\@empty\citet[#1][#2]{#3}}}
\newcommandtwoopt{\citeyearads}[3][][]%
{\href{http://adsabs.harvard.edu/abs/#3}
{\def\hyper@linkstart##1##2{}%
\let\hyper@linkend\@empty\citeyear[#1][#2]{#3}}}
\makeatother

\usepackage{color}
\definecolor{mygreen}{RGB}{0,128,0}

\hypersetup{colorlinks=true,linkcolor=blue,citecolor=blue,urlcolor=blue}

\begin{document}

\title{Stellar and substellar companions of nearby stars from Gaia DR2}
\subtitle{Binarity from proper motion anomaly}
\titlerunning{Companions of nearby stars from Gaia DR2}
\authorrunning{P. Kervella et al.}
\author{
Pierre~Kervella\inst{1}
\and
Fr\'ed\'eric Arenou\inst{2}
\and
Fran\c{c}ois~Mignard\inst{3}
\and
Fr\'ed\'eric~Th\'evenin\inst{3}
}
\institute{
LESIA, Observatoire de Paris, Universit\'e PSL, CNRS, Sorbonne Universit\'e, Univ. Paris Diderot, Sorbonne Paris Cit\'e, 5 place Jules Janssen, 92195 Meudon, France, \email{pierre.kervella@obspm.fr}.
\and
GEPI, Observatoire de Paris, Universit\'e PSL, CNRS, 5 Place Jules Janssen, 92190 Meudon, France.
\and
Universit\'e C\^ote d'Azur, Observatoire de la C\^ote d'Azur, CNRS, Lagrange UMR 7293, CS 34229, 06304, Nice Cedex 4, France.
}
\date{Received ; Accepted}
\abstract
   {The census of stellar and substellar companions of nearby stars is largely incomplete, in particular toward the low-mass brown dwarf and long-period exoplanets. It is, however, fundamentally important in the understanding of the stellar and planetary formation and evolution mechanisms. Nearby stars are particularly favorable targets for high precision astrometry.}
   {We aim to characterize the presence of physical companions of stellar and substellar mass in orbit around nearby stars.}
   {Orbiting secondary bodies influence the proper motion of their parent star through their gravitational reflex motion.
   Using the Hipparcos and Gaia's second data release (GDR2) catalogs, we determined the long-term proper motion of the stars common to these two catalogs. We then searched for a proper motion anomaly (PMa) between the long-term proper motion vector and the GDR2 (or Hipparcos) measurements, indicative of the presence of a perturbing secondary object. We focussed our analysis on the 6741 nearby stars located within 50\,pc, and we also present a catalog of the PMa for $\gtrsim 99$\% of the Hipparcos catalog ($\approx 117\,000$ stars).
   }
   {30\% of the stars studied present a PMa greater than $3\sigma$.
The PMa allows us to detect orbiting companions, or set stringent limits on their presence.
We present a few illustrations of the PMa analysis to interesting targets.
We set upper limits of $0.1-0.3\,M_J$ to potential planets orbiting Proxima between 1 and 10\,au ($P_\mathrm{orb}=3$ to 100\,years).
We confirm that Proxima is gravitationally bound to $\alpha$ Cen.
We recover the masses of the known companions of $\epsilon$\,Eri, $\epsilon$ Ind, Ross 614 and $\beta$~Pic.
We also detect the signature of a possible planet of a few Jovian masses orbiting $\tau$\,Ceti.}
   {Based on only 22\,months of data, the GDR2 has limitations. But its combination with the Hipparcos catalog results in very high accuracy PMa vectors, that already enable us to set valuable constraints on the binarity of nearby objects. The detection of tangential velocity anomalies at a median accuracy of $\sigma(\Delta v_\mathrm{T})=1.0$\,m\,s$^{-1}$ per parsec of distance is already possible with the GDR2. This type of analysis opens the possibility to identify long period orbital companions otherwise inaccessible. For long orbital periods, Gaia's complementarity to radial velocity and transit techniques (that are more sensitive to short orbital periods) already appears to be remarkably powerful.
}
  

\maketitle

%__________________________________ Introduction
\section{Introduction}

In the present work, we examine the proper motion (hereafter PM) of the nearby stars located within 50\,pc from the Sun using the Hipparcos catalog (\citeads{2007ASSL..350.....V}, hereafter `Hip2', see also \citeads{1997A&A...323L..49P}) and Gaia's second data release (GDR2; \citeads{2016A&A...595A...1G,2018A&A...616A...1G}).
Developing the approach presented by \citetads{Kervella18} for Galactic Cepheids and RR Lyrae stars, we took advantage of the long time baseline of 24.25\,years between the Hipparcos and GDR2 position measurements to determine the mean long-term PM vectors of nearby stars with a high accuracy.
We then compared this vector to the individual measurements obtained by Gaia and Hipparcos to test for the presence of anomalies that indicate the presence of an orbiting secondary body. This is the principle employed by \citetads{1844MNRAS...6R.136B} to discover the white dwarf companion of Sirius, and more recently applied to various types of stars by \citetads{1999A&A...346..675W}, \citetads{2004ASPC..318..141J}, \citetads{2005AJ....129.2420M}, \citetads{2007A&A...464..377F}, \citetads{2008ApJ...687..566M}, \citetads{2018ApJS..239...31B} and \citetads{2018arXiv181107285B}.
For a single star, the position of the photocenter coincides with that of the center of mass.
The presence of faint secondary object results in a shift of the barycenter of the system away from its photocenter (located close to the star's position), whose orbital motion induces a "virtual" orbital displacement of the photocenter around the center of mass.
We measured this offset as a "proper motion anomaly" (PMa), that is, a difference between the long-term PM vector and the "instantaneous" PM vector from the Hip2 or GDR2 catalogs.
As the PMa depends simultaneously on the mass and orbital radius of the secondary object, it offers the possibility of setting limits on the possible combinations of these two parameters.

In Sect.~\ref{data}  we present the different sources of observational data and models that we used for our analysis (e.g., astrometry, radial velocity). Section~\ref{PMa-section} is dedicated to a description of the computation of the PMa quantity for the stars of our sample, and the sensitivity of this indicator to the presence of close-in orbiting companions. We also discuss the case of resolved binaries with stellar mass companions.
Section~\ref{overview} is dedicated to an overview of the statistical properties of the examined stars and of their companions, and we discuss in Sect.~\ref{individual_notes} a selected sample of interesting individual objects (\object{Proxima}, \object{$\epsilon$ Eri}, \object{$\tau$ Cet}, and \object{$\beta$~Pic}). Additional individual objects are discussed in Appendix~\ref{additional_individual_notes}.

%__________________________________ 
\section{Observational data\label{data}}

For the collection of most of the data used in the present work, we made extensive use of the \texttt{astroquery} set of tools \citepads{2017ascl.soft08004G} distributed as part of the \texttt{Astropy} library \citepads{2018AJ....156..123A} to access the ViZieR online database \citepads{2000A&AS..143...23O} at the CDS.
For simplicity, we note $\mu_{\mathrm{\alpha}}$ the PM along the right ascension axis $\mu_{\mathrm{\alpha}} \cos(\delta)$.

\subsection{Selected sample}

We extracted from the GDR2 catalog \citepads{2018A&A...616A...1G} the objects that are located within 50\,pc from the Sun ($\varpi > 20$\,mas).
As our goal is to compare the measurements from the Hipparcos and Gaia missions, we set a maximum $G$ band magnitude of 13, in order to cover the full range of the Hipparcos catalog, resulting in a list of 17518 GDR2 sources.
For these nearby targets, we preferred to start from the Gaia DR2 catalog rather than from the Hipparcos catalog, to take advantage of its exhaustivity and evaluate the completeness of our sample.
The sample that we analyzed is incomplete, due both to the saturation limit of Gaia that excludes the brightest stars and to the limiting magnitude of the Hipparcos catalog.
Most of the bright binaries with separations between 0.1 and $1\arcsec$ are also excluded, as their astrometric parameters are generally not provided in the GDR2 catalog.
For instance, \object{$\alpha$ Centauri AB}, \object{Sirius A}, \object{Procyon A,} and most of the brightest stars are absent from our selection, although they are present in the Hipparcos catalog.
A discussion of the completeness of our nearby star sample is presented in Sect.~\ref{starsample}.
A total of 6741 stars within 50\,pc are present in both the Hipparcos and GDR2 catalogs, and the analysis of their PMa properties is the focus of the present paper.

For completeness, we also determined the PMa of the stars located beyond 50\,pc of the Hipparcos catalog by \citetads{2007ASSL..350.....V} (Hip2).
For the cross identification of these more distant stars, we started from the catalog of 115\,562 matches by \citetads{2018ApJS..239...31B}, that we completed with a small number of cross identifications based on the position and brightness.
Our final catalog comprises 117\,206 records, out of the 117\,955 objects listed in the Hip2 catalog (99.4\%).
We provide the Hipparcos-GDR2 mean PM vectors $\vec{\mu}_\mathrm{HG}$ for 117\,189 objects, as well as the Hipparcos and GDR2 PMa for 117\,068 (99.9\%) and 115\,959 (98.9\%) objects, respectively.
We note that the identification is uncertain in some cases, particularly for close visual binaries (within $\approx 0.4\arcsec$). The stars present in the double and multiple star annex (DMSA) of the original Hipparcos catalog \citepads{1997ESASP1200.....E} are identified as such in our catalog.

\subsection{Astrometry}

\subsubsection{Hipparcos astrometry}

The position, PM, and parallax data of our sample are taken from Hip2 \citepads{2007ASSL..350.....V} and the GDR2 \citepads{2018A&A...616A...1G} catalogs. We checked that the results of our computations do not depend significantly on the chosen version of the Hipparcos reduction. We computed the PMa of a representative set of objects using the original Hipparcos catalog positions \citepads{1997A&A...323L..49P, 1997ESASP1200.....E} (Hip1) and we did not notice any significant difference in the results.

To determine the proper motion accelerations of the Hipparcos catalog stars, \citetads{2018ApJS..239...31B} recently adopted a linear combination of the Hip1 and Hip2 reductions of the Hipparcos data to estimate the star positions.
This original approach is intended to mitigate the systematics associated with each of the catalogs considered individually.
However, to ensure an easier traceability of our processing results, we preferred to rely only on the Hip2 catalog for the present analysis.

\subsubsection{Basic Gaia DR2 corrections}

The zero point of the parallax $\varpi$ of the GDR2 has been corrected by adding a constant $+29\,\mu$as to the catalog value.
This global offset correction was derived by \citetads{Lindegren18} and \citetads{2018A&A...616A..17A} (see also \citeads{2018arXiv180905112S}) from QSOs fainter than $G=17$.
The amplitude of this systematic correction may be different for bright objects, but the present analysis does not rely on a high accuracy of the parallax.

We corrected the GDR2 PM vectors for the rotation of the Gaia reference frame \citepads{Mignard18} using the following expressions from \citetads{Lindegren18}:
\begin{gather}
\mu_{\alpha, \mathrm{corr}} = \mu_{\alpha} + w_x \sin(\delta) \cos(\alpha) + w_y \sin(\delta) \sin(\alpha) - w_z \cos(\delta) \\
\mu_{\delta, \mathrm{corr}} = \mu_{\delta} - w_x \sin(\alpha) + w_y \cos(\alpha),
\end{gather}
where the rotation parameters are
$w_x = -0.086 \pm 0.025\,\mathrm{mas\,a}^{-1}$,
$w_y = -0.114 \pm 0.025\,\mathrm{mas\,a}^{-1}$ and
$w_z = -0.037 \pm 0.025\,\mathrm{mas\,a}^{-1}$.
As determined by \citetads{Lindegren18}, the systematic uncertainty on the GDR2 PM vectors is $\sigma_\mathrm{sys} = 32\,\mu$as\,a$^{-1}$ for bright stars ($G<13$) and $66\,\mu$as\,a$^{-1}$ for fainter stars.
We quadratically added this uncertainty to the GDR2 catalog error bars on both the $\alpha$ and $\delta$ axes.

As a complement, we retrieved the record of the stars present in the Gaia DR1 catalog \citepads{2016A&A...595A...1G,2016A&A...595A...2G, 2017A&A...599A..50A}, that are discussed for specific cases in Sect.~\ref{individual_notes}.
For Proxima and Barnard's star, we also considered the PM and parallax measurements from the Hubble Space Telescope's fine guidance sensor (FGS) reported by \citetads{1999AJ....118.1086B}, whose mean measurement epochs were derived using the observing log listed by \citetads{1998AJ....116..429B}.
They are discussed respectively in Sects.~\ref{proxima} and \ref{barnard}.

\subsubsection{Renormalized unit weight noise\label{ruwe}}

To check for the peculiar behavior of a given object of the GDR2 catalog compared to the objects of similar brightness and color, we adopted the formalism of the renormalized unit weight error (RUWE, hereafter noted $\varrho$) introduced by Lindegren.
It is based on the combination of the astrometric $\chi^2$, the number of good observations $N$, the $G$ magnitude and the color index $C = G_\mathrm{BP} - G_\mathrm{RP}$ through:
\begin{equation}
\varrho = \frac{\sqrt{\chi^2/(N-5)}}{u_0(G,C)}.
\end{equation}
The values of the empirical function $u_0(G,C)$ are available on ESA's Gaia website\footnote{\url{https://www.cosmos.esa.int/web/gaia/dr2-known-issues}}. In absence of color index $C$, a single dimensional $u_0(G)$ is also available.
Lindegren proposes a limit of $\varrho < 1.4$ below which a record can be considered "well behaved" in the GDR2 catalog.
Most of the stars in our sample satisfy this criterion, however, due to the general difficulty of the observation of bright objects with Gaia, the derived companion masses using the astrometric excess noise $\epsilon_i$ (see Sect.~\ref{excess-noise}) are unreliable for the stars near the Gaia saturation limit ($G\approx3$) as, for example, $\beta$~Pic (Sect.~\ref{betapic}).
In some rare cases, for example, \object{Ross 614} (see Sect.~\ref{ross614}), the RUWE is high ($\varrho = 11.9$) due to the presence of a massive stellar companion (\object{Ross 614 B}) located very close to the target and contributing to its astrometric position measurement error.
For such an object, the measured astrometric excess noise $\epsilon_i$ is due to the unresolved stellar companion contribution in flux and color.

\subsection{Radial velocity \label{radialvel}}

Although the determination of the PMa relies on the comparison of the tangential PM vectors at two epochs, the radial velocity (RV) must be taken into account for nearby targets to extrapolate the GDR2 parallax to the Hipparcos epoch (Sect.~\ref{lighttime}), and also to account for the changing geometrical projection of the space velocity vector with time.
For the RV of the stars of our sample, we considered the following, in order of decreasing priority:
\begin{enumerate}
\item \citetads{2002ApJS..141..503N} (RV for 889 late-type stars),
\item \citetads{2018A&A...616A...7S} (GDR2 catalog of RV standards),
\item \citetads{2007A&A...475..519H} (Geneva-Copenhagen survey),
\item \citetads{2018A&A...616A...1G} (GDR2 RVS measurements; see also \citeads{2018A&A...616A...5C,2018arXiv180409372K}),
\item \citetads{2012AstL...38..331A} (XHIP catalog).
\end{enumerate}
These catalogs provide respectively 9\%, 28\%, 24\%, 27\%, and 9\% of the RVs of the 6741 stars located within 50\,pc. For our full Hipparcos sample of 117\,206 records, the corresponding fractions are 0.7\%, 3.4\%, 8.7\%, 54.5\% and 12.7\%. There is no RV for 20.2\% of the full sample, as well as for 3\% of the nearby star sample (174 stars). In this case, the value of the RV is set to zero in the computations with an uncertainty of $\pm 50$\,km\,s$^{-1}$.
The radial velocities of white dwarfs (hereafter WD) are notoriously difficult to measure and they are unavailable for a significant part of our small WD sample.
The impact of RV projection effects on the determined PMa is usually negligible for targets located at distances $\gtrsim 5$\,pc, but for closer targets with no RV measurement, the determined PMa should be considered provisional.
If not already corrected for in the searched catalogs, we corrected the RVs for the differential gravitational redshift of each star with respect to the solar value ($+633$\,m\,s$^{-1}$) using the following stellar mass and radius estimates:
\begin{itemize}
\item The $(K,V-K)$ surface brightness-color relations by \citetads{2004A&A...426..297K} for the radius, and the isochrones by \citetads{2000A&AS..141..371G} for the masses of stars with $M_K \leqslant 4.5$.
\item The mass and radius relations calibrated by \citetads{0004-637X-804-1-64} based on the absolute $K$ band magnitude for K and M dwarfs ($4.5 < M_K \leqslant 10$).
\item The WD physical parameters were taken from the catalog by \citetads{2016MNRAS.462.2295H}. When not available for a given star, the $(K,V-K)$ surface brightness-color relation by \citetads{2004A&A...426..297K} was employed to estimate the radius, and a fixed mass of $0.6 \pm 0.2\,M_\odot$ was assumed (from the mass distribution by \citeads{1997ApJ...488..375F}; see also \citeads{2012ApJS..199...29G}).
\end{itemize}
The gravitational redshift correction usually has a negligible effect on the determined PMa, except for the nearest WDs.
The convective blueshift (see e.g., \citeads{2018A&A...611A..11C}) has not been corrected. It adds an uncertainty on the order of 300\,m\,s$^{-1}$, that was taken into account in the computation but that has a fully negligible effect on the resulting PMa uncertainty.
The transverse relativistic redshift is negligible in all cases.

For Proxima (\object{GJ 551}), we adopt the absolute barycentric RV of $v_\mathrm{r} = -22\,204 \pm 32$\,m\,s$^{-1}$ determined by \citetads{2017A&A...598L...7K} at a mean measurement epoch of J2012.554.
For the WD \object{Wolf 28}, we adopt the RV of $v_\mathrm{r} = 15 \pm 20$\,km\,s$^{-1}$ determined by \citetads{1993AJ....105.1033A}.
For Proxima, Barnard's star (\object{GJ 699}) and Kapteyn's star (\object{GJ 191}), we took into account  secular accelerations of +0.45\,m\,s$^{-1}$\,a$^{-1}$ \citepads{2008A&A...488.1149E}, +4.50\,m\,s$^{-1}$\,a$^{-1}$ \citepads{2003A&A...403.1077K} and -0.18\,m\,s$^{-1}$\,a$^{-1}$ \citepads{2014MNRAS.443L..89A}, respectively.
At the current accuracy of the Gaia measurements, these corrections have no effective impact on the PMa of any star of our sample.

\subsection{Additional data}

We completed the information on each target with the $m_K$ magnitude from the 2MASS catalog \citepads{2006AJ....131.1163S}, the visible $m_V$
 magnitude from the NOMAD catalog \citepads{2004AAS...205.4815Z} as well as the information on known binary and multiple systems present in the WDS catalog \citepads{2001AJ....122.3466M}. The interstellar reddening was neglected for all stars within 50\,pc. For the more distant objects in our extended catalog, we adopted the color excess $E(B-V)$ predicted by the \emph{Stilism}\footnote{\url{https://stilism.obspm.fr}} three-dimensional (3D) model of the local interstellar medium \citepads{2014A&A...561A..91L, 2017A&A...606A..65C}.

%__________________________________ 
\section{Proper motion anomaly \label{PMa-section}}

\subsection{Definition}

For an isolated single star with no intrinsic morphological change (from, e.g., spots, or convection, see \citeads{2011A&A...528A.120C, 2018MNRAS.476.5408M,2018A&A...617L...1C}), the motion of its photocenter is rectilinear and uniform, and the PM vector is therefore constant in direction and norm.
For a binary system, the presence of the secondary mass will shift the barycenter (center of mass) away from the primary star.
Due to the photometric contribution from the secondary, the photocenter of the system will also be displaced.
For an unresolved, ideal system of two perfectly identical stars (in masses and luminosities), the positions of the barycenter and of the photocenter remain identical, and no time-dependent variation of the $\vec{\mu}$ vectors are detectable.
However, in the general case, the ratio $p = L_2/L_1$ of the luminosity of a low mass companion to that of its parent star is considerably smaller than the ratio of their masses $q = m_2/m_1$.
This results in a shift of the center of mass relatively to the photocenter.
As both stars revolve around their center of mass, the photocenter (that is close to the geometrical center of the primary when $q\ll1$) follows a "virtual" orbit around the barycenter.

%______________ Figure
\begin{figure}
\includegraphics[width=\hsize]{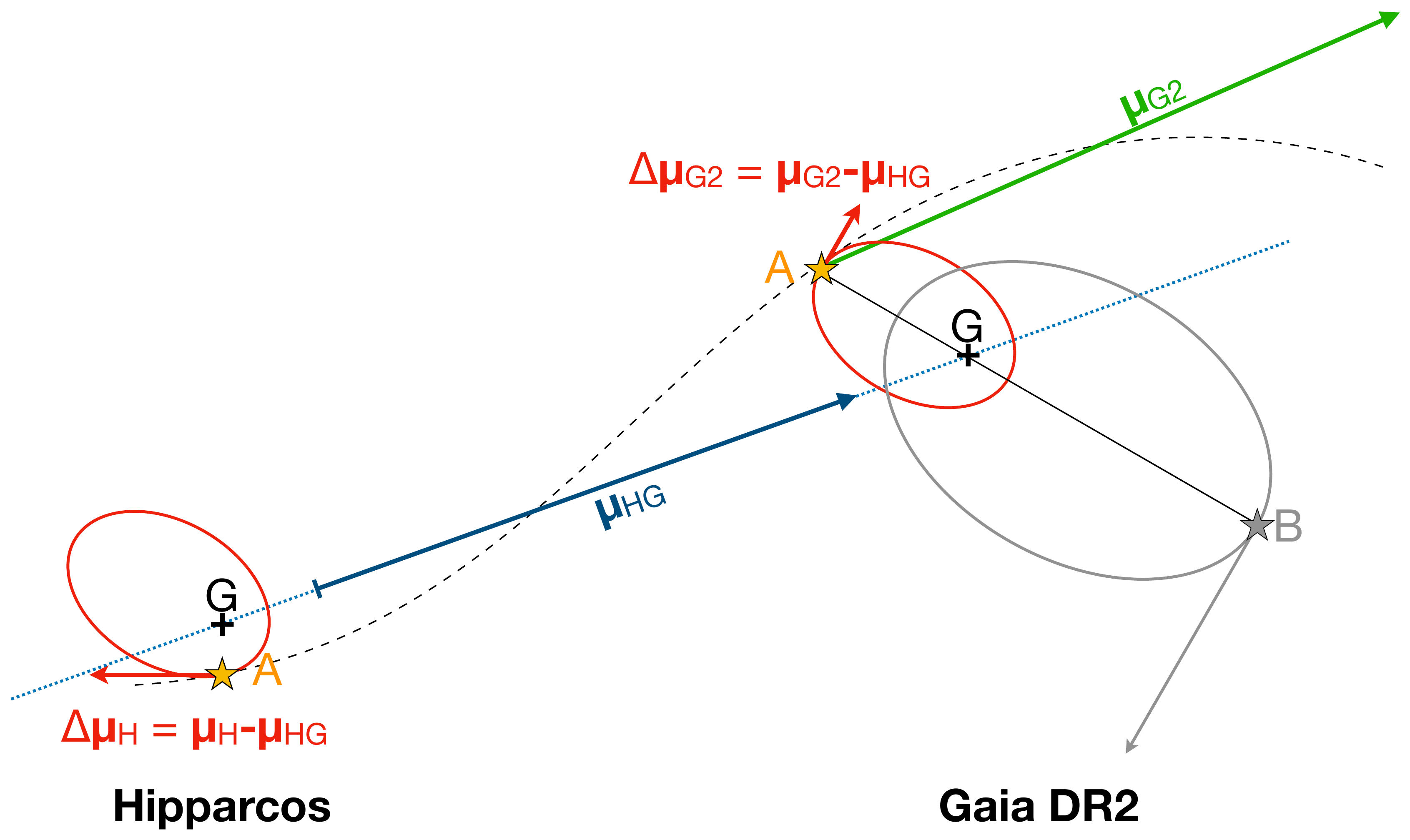}
\caption{Principle of the proper motion anomaly $\Delta \vec{\mu}_\mathrm{H/G2}$ determination. We assume in the figure that the secondary object B has a negligible photometric contribution, and that the photocenter of the system is at the position of star A.\label{PMa-principle}}
\end{figure}

As shown in Fig.~\ref{PMa-principle}, the PM vector of the photocenter of a binary system will vary with time, and thus differ from the PM vector of the barycenter that remains constant for an isolated system.
In the following analysis, we define the PMa vector $\Delta \vec{\mu}_\mathrm{H/G2}$ as the result of the subtraction of the long-term PM vector $\vec{\mu}_\mathrm{HG}$ from the PM vectors of the two catalogs:
\begin{equation}
\Delta \vec{\mu}_\mathrm{H/G2} = \vec{\mu}_\mathrm{H/G2} - \vec{\mu}_\mathrm{HG},
\end{equation}
with $\vec{\mu}_\mathrm{H/G2}$ the PM vectors from the Hip2 or GDR2 catalogs and $\vec{\mu}_\mathrm{HG}$ the mean PM vector determined from the difference in astrometric position $(\alpha,\delta,\varpi)$ between the two catalogs, expressed at the corresponding comparison epoch.
The "instantaneous" $\vec{\mu}_\mathrm{Hip/G2}$ PM vectors contain the sum of the barycenter velocity and the "virtual orbital" velocity of the photocenter.
The differential PMa vector $\Delta \vec{\mu}$ thus corresponds to the projected velocity vector of the photocenter around the barycenter at the Hipparcos or GDR2 epochs.

\subsection{Light travel time \label{lighttime}}

Due to the radial velocity of the targets, the difference in distance between the Hip2 and GDR2 epochs results in a different light propagation time.
The amplitude of this effect (equivalent in its principle to the classical aberration of light) can be considerable. For example,~for the very fast moving \object{Kapteyn's star} ($v_\mathrm{r} \approx +245$\,km\,s$^{-1}$) whose geometric distance to the Sun increased by 1200\,au between the Hipparcos and GDR2 epochs. This corresponds to an additional light propagation time of 7.2\,days for the GDR2 epoch, and a considerable tangential shift on the sky of 171\,mas.
To compare the absolute PM vectors $\vec{\mu}_\mathrm{HG}$ and $\vec{\mu}_\mathrm{G2}$ of the star, we must either correct both vectors for the light travel time effect, or correct neither of them and work in the "aberrated" referential.
As the GDR2 pipeline does not correct for the light travel time effect in the computation of the astrometric solution, we choose not to correct the mean $\vec{\mu}_\mathrm{HG}$ vector to be in the same "aberrated" referential as the GDR2 PMs.

As the radial velocity is unaffected by the light travel time effect, we took into account the differential light travel time correction to extrapolate the stellar parallaxes from the GDR2 epoch to the Hipparcos epoch (Sect.~\ref{longterm}).
The magnitude of this correction is negligible for all but the nearest stars.

\subsection{Long-term proper motion \label{longterm}}

The projection of the 3D space velocity vector of a star on the tangential plane (orthogonal to the line of sight) varies with time due to the change in geometrical perspective induced by its space motion.
As a consequence, for the nearest targets (Proxima, Barnard's star...) a three-dimensional (3D) computation is mandatory to properly determine the long term PM vector $\vec{\mu}_\mathrm{HG}$ in spherical $(\mu_\alpha, \mu_\delta)$ coordinates at a given epoch, that is the quantity reported in the Hipparcos and GDR2 catalogs.
The overall principle we adopted is to determine the 3D space velocity of the star in the cartesian coordinate system. In this frame, this velocity is constant for an isolated star.
We subsequently projected this vector at the spherical International Celestial Reference System (ICRS) coordinates of the star at the relevant epochs to be compared with the catalog values.
In practice, we first transformed the ICRS equatorial six-dimensional (6D) position-velocity vector of the target $(\alpha, \delta, \varpi, \mu_\alpha, \mu_\delta, v_\mathrm{rad})$ into Cartesian coordinates $(x,y,z,v_x,v_y,v_z)$.
This allowed us to take into account the correlations between the RV and the transverse PM components induced by the perspective effects.
For the stars located beyond 50\,pc, we used a simplified two-dimensional (2D) computation of the PM vector $\vec{\mu}_\mathrm{HG}$ in the spherical ICRS frame, as the geometrical RV projection effect is negligible.
For the astrometric transformations between coordinate systems, we used extensively the tools available in the \texttt{Astropy}\footnote{\url{http://www.astropy.org}} library version 3.0 \citepads{2018AJ....156..123A}, and the uncertainties were propagated using a Monte Carlo (MC) approach.

The GDR2 parallaxes typically have one order of magnitude higher accuracy than that of Hip2 (except for some very bright stars).
To achieve better accuracy on the parallax at the Hipparcos epoch, we affected to the Hipparcos 6D vector the GDR2 $\varpi_\mathrm{G2}$ parallax, extrapolated to the J1991.25 epoch using the GDR2 space velocity vector.
The long-term PM vector $\vec{\mu}_\mathrm{HG}$ is a differential quantity in the tangential plane, this extrapolation along the distance axis has thus a negligible influence on its value, even for the nearest stars.
This operation is however useful to obtain more accurate estimates of the uncertainties on the $\vec{\mu}_\mathrm{HG}$ vector through a better accounting of the geometric correlations.
This extrapolation in time is also justified as we do not aim to detect a change of velocity along the radial axis from the astrometric data, but only in the tangential plane.
In this computation, we took into account the light travel time effect for the extrapolation of the GDR2 parallax to the Hipparcos epoch (Sect.~\ref{lighttime}).
From this procedure, we obtain the long term Hipparcos-Gaia PM vector expressed in cartesian coordinates $(v_x,v_y,v_z)_\mathrm{HG}$.

\subsection{Proper motion anomaly computation\label{PMacomputation}}

From the determined long-term PM vector in cartesian coordinates and the individual velocity vectors $(v_x,v_y,v_z)_\mathrm{Hip/G2}$, we computed the tangential velocity anomaly vectors at the Hipparcos and GDR2 epochs $\Delta (v_x,v_y,v_z)_\mathrm{Hip/G2}$ through:
\begin{equation}
\Delta (v_x,v_y,v_z)_\mathrm{Hip/G2} = (v_x,v_y,v_z)_\mathrm{Hip/G2} - (v_x,v_y,v_z)_\mathrm{HG}.
\end{equation}
This differential vector was finally transformed back into spherical ICRS angular coordinates to obtain the tangential $\Delta \mu_\mathrm{Hip/G2}$ PMa vector at the Hipparcos and GDR2 epochs (and optionally at the Gaia DR1 or HST-FGS epochs, when available).
This quantity is the final product of the processing chain, together with the corresponding linear tangential velocity.

We uniformly applied this process to our full sample. The PMa vectors of the nearest stars of our sample are listed in Table~\ref{pm_binaries1}, and the results for the stars up to 50\,pc are available electronically from the CDS.
We provide in Table~\ref{pm_binaries1} the long term proper motion vector $\vec{\mu}_{HG}$ for the Hipparcos and Gaia DR2 epochs separately. Although the 3D space velocity vector is constant in space, its tangential projection changes for such very nearby stars as a function of time, due to the changing geometrical perspective. This effect is, in practice, negligible for the stars located farther than a few tens of parsecs.

% Table of the PM anomalies of the nearest stars
% ========================================
\input{Tables/PM-table}

\subsection{Companion mass\label{companionspma}}

The velocity $v_1$ of a star orbiting on a circular orbit of radius $r$ around the center of mass of a binary system is:
\begin{equation}
v_\mathrm{1} = \sqrt{\frac{G\,m_2^2}{(m_1+m_2)\,r}}
\end{equation}
with $m_1$ and $m_2$ the masses of the two components.
When $m_2 \ll m_1$, and for an orbital plane perpendicular to the line of sight, we have the simpler expression:
\begin{equation}\label{m2mass}
\frac{m_2}{\sqrt{r}} = \sqrt{\frac{m_1}{G}}\,v_\mathrm{1} = \sqrt{\frac{m_1}{G}}\,\left( \frac{\Delta \mu [\mathrm{mas\,a}^{-1}] }{\varpi [\mathrm{mas}]} \times 4740.470 \right)
\end{equation}
where $\Delta \vec{\mu}$ is the PMa, identified to the tangential orbital velocity of the primary star and $\varpi$ its parallax.
The constant multiplicative term in brackets in Eq.~\ref{m2mass} is intended to transform the ratio of the $\Delta \mu$ and $\varpi$ quantities in their usual units (indicated in brackets) into a velocity expressed in m\,s$^{-1}$.
From a single $\Delta \vec{\mu}$ measurement and an estimation of the mass $m_1$ of the primary, it is therefore possible to derive an estimate of the mass $m_2$ of a companion normalized to the square root of the orbital radius $\sqrt{r}$.
We adopted the notation $m_2^\dag = m_2/\sqrt{r}$ in the rest of the present paper to refer to the normalized companion mass, with a physical unit of $M_J\,\mathrm{au}^{-1/2}$ except otherwise noted.
We estimate the mass $m_1$ of our target stars using the approach described in Sect.~\ref{radialvel}.

\subsection{Statistical and systematic corrections}

\subsubsection{Effect of orbital inclination\label{inclination}}

The measured PMa concerns only the two tangential components of the orbital velocity of the primary.
We thus obtain from Eq.~\ref{m2mass} a lower limit for the true normalized mass $m_2^\dag$ of the companion.
The inclination of a randomly oriented orbit in space is $i = 60^{+21}_{-27}\,\deg$.
From statistical and geometrical considerations, the norm of the measured 2D PM vector corresponds to $\eta = 87^{+12}_{-32}\%$ of the norm of the 3D orbital PM vector. We determined this factor from a MC simulation of the projection of the orbital velocity vector, including the effect of the orbital phase, eccentricity and inclination.
The distributions of these parameters were considered uniform in the simulation.
\citetads{2016MNRAS.456.2070T} show that the eccentricity distribution of long period binaries peaks around $e \approx 0.5$, and this maximum increases to $e \approx 0.6$ for extremely long periods. We checked that adopting these distributions has no significant impact on the $\eta$ factor.
We multiplied the observed $\Delta \vec{\mu}$ by $\eta^{-1}$ to estimate the deprojected distribution of the companion mass $m_2^\dag$ (propagating the associated uncertainties).

\subsubsection{Observing window smearing}

The PM measurements listed in the Hip2 and GDR2 catalogs are not instantaneously measured quantities.
They result from the adjustment of a set of transit observations obtained over a period of $\delta t_\mathrm{H} = 1227$\,d \citepads{1997A&A...323L..49P} and $\delta t_\mathrm{G2} = 668$\,d \citepads{2018A&A...616A...1G}, respectively.
The PMa $\Delta \vec{\mu}$ is therefore a time average of the intrinsic velocity vector of the star over the observing period $\delta t$.
As a consequence, if the orbital period of the system is significantly shorter than $\delta t$, then the PMa will be decreased by the temporal smearing of the signal. If the orbital period is exactly $\delta t$, no $\Delta \vec{\mu}$ will be detectable (for a uniform distribution of the transits over the observing period).
The differential astrometric signal ($\Delta \vec{\mu}$) will be decreased by the following factor $\gamma$:
\begin{equation}
\gamma = \frac{1}{\delta t \left| \Delta \vec{ \mu}(0)\right|} \left| \int_{0}^{\delta t} \Delta \vec{\mu}(t) \,dt  \right|
\end{equation}
with $\mu_{\alpha,\delta}$ the RA and Dec components of the differential PMa.
For a circular face-on orbit, we can represent the PM vector as
\begin{equation}
\Delta \mu_{\alpha,\delta}(t) = K \, \left[ \sin \left(\frac{2 \pi t}{P}\right) ,  \cos \left (\frac{2 \pi t}{P}\right) \right]\\
\end{equation}
where $P$ is the orbital period and $K$ is the amplitude of the astrometric wobble.
We can thus write the $\gamma$ components in $\alpha$ and $\delta$:
\begin{equation}
\gamma_{\alpha,\delta} = \frac{1}{\delta t} \left[ \int_{0}^{\delta t} \sin \left(\frac{2 \pi t}{P}\right) \,dt, \int_{0}^{\delta t} \cos \left(\frac{2 \pi t}{P}\right) \,dt \right]
\end{equation}
\begin{equation}
\gamma_{\alpha,\delta} = \frac{P}{2 \pi \delta t} \left[ \left(1 - \cos \frac{2 \pi \delta t}{P}\right), \left( \sin \frac{2 \pi \delta t}{P}\right) \right]
\end{equation}
The global $\gamma$ factor on the norm of the PMa is thus:
\begin{equation}
\gamma(P,\delta t)  = \frac{ P}{\sqrt{2} \pi \delta t} \sqrt{1 - \cos \frac{2 \pi \delta t}{P} },
\end{equation}
that we can rewrite using the normalized period $\bar{P} = P/\delta t$:
\begin{equation}\label{eqgamma}
 \gamma(\bar{P})  = \frac{\bar{P}}{\sqrt{2} \pi} \sqrt{1 - \cos \frac{2 \pi}{\bar{P}} }.
\end{equation}

%______________ Figure
\begin{figure}
\includegraphics[width=\hsize]{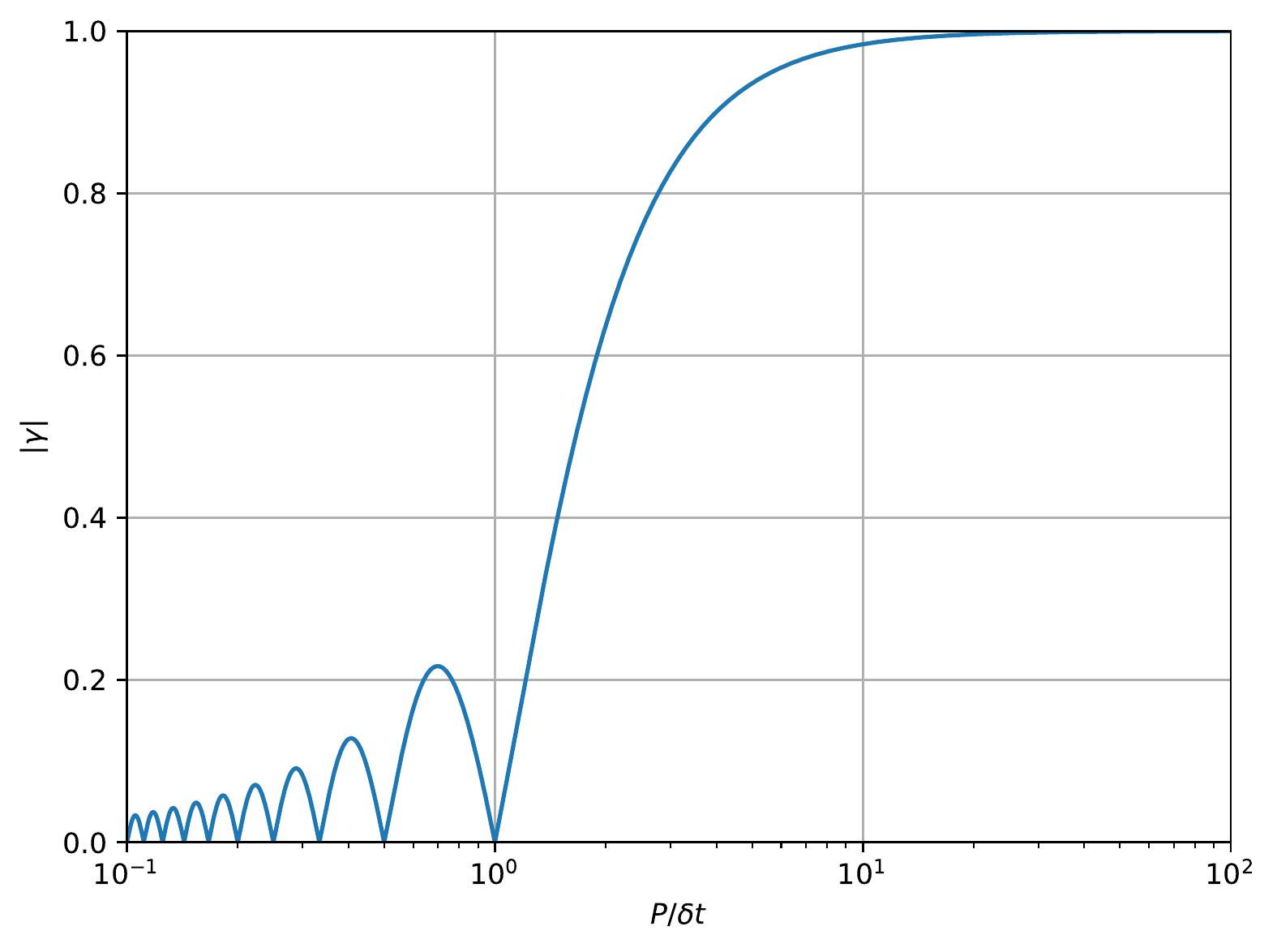}
\caption{Relative sensitivity variation $\gamma$ of the PMa to orbiting companions, due to the observing window smearing effect. The $\gamma$ parameter is represented as a function of the orbital period $P$ normalized to the observing window $\delta t$ (e.g., $\delta t_\mathrm{G2} = 668$\,days for GDR2).\label{gamma}}
\end{figure}

For $P \gg \delta t$, we verify that $\gamma =1$, and for $P = \delta t$, $\gamma =0$.
The shape of the $\gamma$ function is presented in Fig.~\ref{gamma}.
The orbital radius $r_\mathrm{G2}$ corresponding to the observing period $\delta t_\mathrm{G2}$ of GDR2 is listed in Table~\ref{phys-properties}, together with the companion mass sensitivity $m_2$ normalized at 1\,au. 
To take into account the change in sensitivity induced by the observing window smearing, we normalize in the following analysis the achieved companion sensitivity from the PMa by multiplying it by $\gamma^{-1}$.

\subsubsection{Efficiency as a function of orbital period \label{efficiency-period}}

\begin{figure}
\includegraphics[width=\hsize]{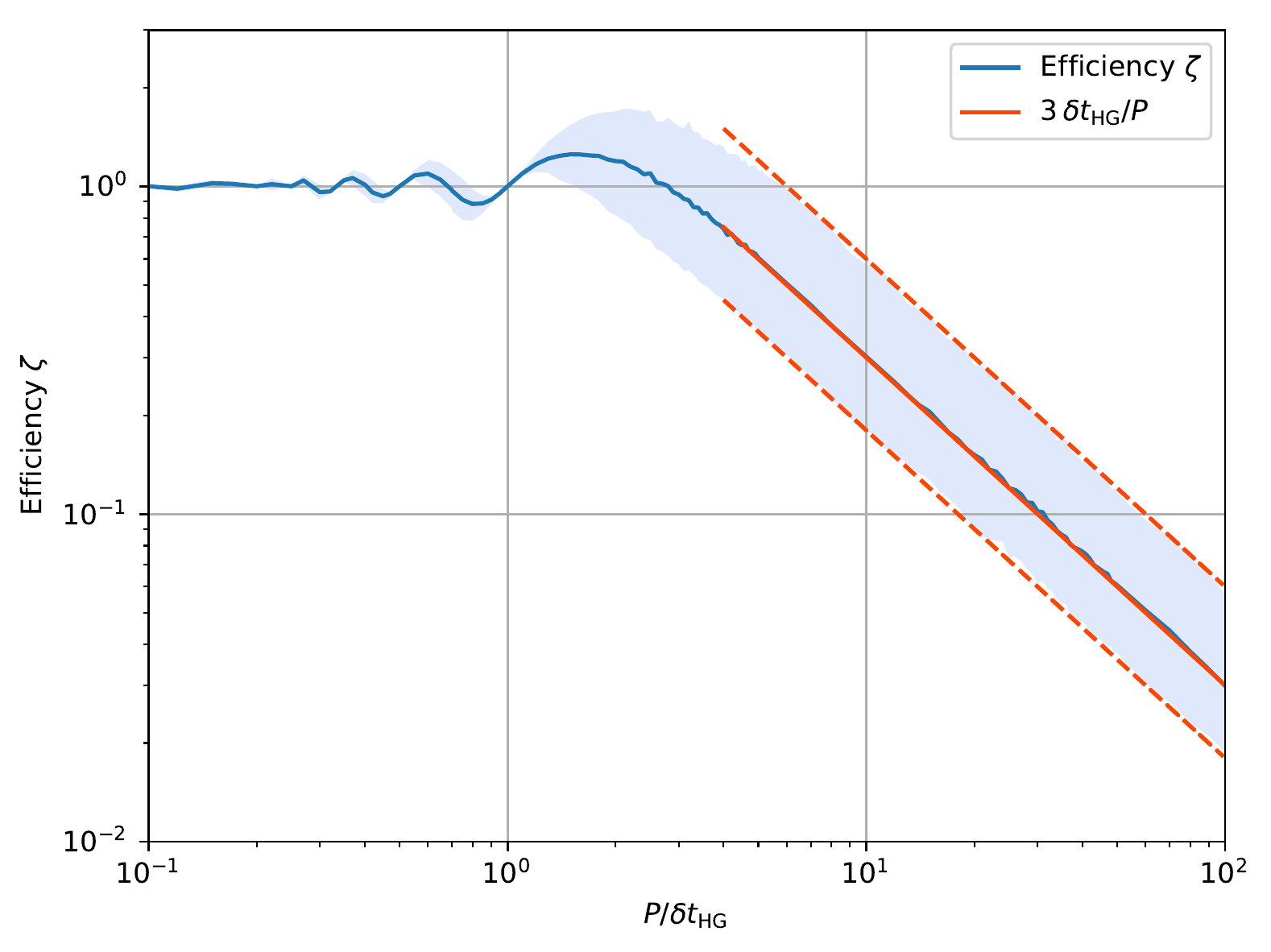}
\caption{Multiplicative efficiency $\zeta$ of the PMa estimation. It is represented as a function of the orbital period $P$ normalized to the observing time base between the Hip and GDR2 measurements $\delta t_\mathrm{HG} = 24.25$\,years).\label{zeta}}
\end{figure}

The computation process of the PMa involves the subtraction of the long-term $\mu_\mathrm{HG}$ PM vector from the short-term PM vector measured at the Hip2 ($\mu_\mathrm{Hip}$) or GDR2 ($\mu_\mathrm{G2}$) epochs (Sect.~\ref{PMacomputation}). These two epochs are separated by 24.25\,years. If the orbital period of a binary system is significantly longer than this time base, this subtraction biases the PMa vector, as part of the orbital velocity is subtracted with the long-term PM vector $\mu_\mathrm{HG}$.
This leads to a bias in the estimation of the companion mass-radius domain.
This bias is difficult to determine analytically, due to the intermixing of the geometrical (e.g., orientation in space, or eccentricity) and temporal (orbital period) parameters. We therefore adopted a numerical MC approach to determine the multiplicative bias that affects the PMa estimation.
We considered a large number of orbits, with uniformly distributed geometrical parameters, for a broad range of orbital periods between 0.1 and 500\,times the Hip-GDR2 time base (that is, between 2.5 and 12\,000\,years).
We then simulated the observational determination of the PMa, and estimated the ratio $\zeta$ of the measured PMa vector to the true orbital value:
\begin{equation}
\zeta \left(\frac{P}{\delta t_\mathrm{HG}} \right) = \left| \frac{\vec{\mu}_\mathrm{G2} - \vec{\mu}_\mathrm{HG}}{\vec{\mu_\mathrm{orb}}} \right|
.\end{equation}
The $\zeta$ function is represented in Fig.~\ref{zeta}. Its behavior for $P/\delta t_\mathrm{HG}>4$ is very close to $\zeta = 3 \delta t_\mathrm{HG}/P$ (solid red line in Fig.~\ref{zeta}).
The limits of its 68\% confidence interval in this limit case (dashed red lines in Fig.~\ref{zeta}) are 0.6$\times$ and 2.0$\times$ its median value.
We note that $\zeta$ is slightly larger than unity for orbital periods of approximately one to three times the $\delta t_\mathrm{HG}$ time base, that is, orbital periods of 25 to 75\,years. The efficiency $\zeta$ is precisely unity for $P=\delta t_\mathrm{HG}$.
In the following, we multiplied the companion mass-radius sensitivity curve using the numerical $\zeta^{-1}$ function.

%______________ Figure

\subsection{Normalized companion mass sensitivity curve\label{companionspma}}

The secondary body mass $m_2(r)$ is a function of the detected linear tangential PMa $| \Delta \vec{\mu}_\mathrm{G2} |$ and of the linear orbital radius $r$ through:
\begin{equation}\label{eqm2}
m_2(r) = \frac{\sqrt{r}}{ \gamma \left[ P(r)/\delta t \right] } \sqrt{\frac{m_1}{G}}\,\frac{\Delta v_\mathrm{T,G2} }{\eta\  \zeta}
\end{equation}
where $P(r)$ is the period corresponding to the orbital radius $r$ (for $m_2 \ll m_1$):
\begin{equation}
P(r)= \sqrt{\frac{4 \pi^2 r^3}{G\,m_1}}
\end{equation}
and $\Delta v_\mathrm{T,G2}$ is the norm of the tangential PMa vector converted to linear velocity using the GDR2 parallax.
The parameters $m_1$ and $\Delta v_\mathrm{T,G2}$ are listed for each star in Table~\ref{phys-properties}.
This allows us to compute the possible $(m_2,r)$ combinations corresponding to a given observed tangential velocity anomaly.

The normalized mass $m_2^\dag$ is proportional to the square root of the primary star mass $\sqrt{m_1}$ and to the tangential velocity anomaly $\Delta v_\mathrm{T}$. The sensitivity of the detectable normalized mass thus decreases as the primary mass increases (as a function of $\sqrt{m_1}$) and decreases linearly with the increasing distance.
The accuracy of $\Delta v_\mathrm{T}$ is in principle set jointly by the PM accuracy and the parallax. But as this is a differential quantity, the parallax uncertainty has, in practice, a negligible influence on its error bar for nearby stars.

The median accuracy of the norm of the PMa vector $\Delta \mu_\mathrm{G2} = \vec{\mu}_\mathrm{G2}-\vec{\mu}_\mathrm{HG}$ of the $\approx 6700$ tested stars within 50\,pc is $\sigma(\mu) = 234\,\mu$as\,a$^{-1}$ ($209\,\mu$as\,a$^{-1}$ for the full Hip2 sample) and the corresponding median $\Delta v_\mathrm{T}$ accuracy is therefore $\sigma(\Delta v_\mathrm{T}) = 1.11$\,m\,s$^{-1}$ per parsec of distance (0.99\,m\,s$^{-1}$\,pc$^{-1}$ for all stars).
The median accuracy of the $\mu_\mathrm{HG}$ long-term proper motion is $\sigma(\mu_\mathrm{HG}) = 38\,\mu$as\,a$^{-1}$ ($48\,\mu$as\,a$^{-1}$ for all stars), and its contribution to the uncertainty on the PMa is therefore negligible compared to the GDR2 uncertainty.
For a nearby solar mass star, the achieved PMa accuracy of $234\,\mu$as\,a$^{-1}$ corresponds to a theoretical sensitivity on the detection of companions of
\begin{equation}
\sigma \left(m_2^\dag \right) = 0.039\,M_J\,\mathrm{au}^{-1/2}\,\mathrm{pc}^{-1} = 13\,M_\oplus\,\mathrm{au}^{-1/2}\,\mathrm{pc}^{-1}
.\end{equation}
It is however not the practical sensitivity that can be achieved with the PMa technique for all orbital radii, due to the smearing by the observing window $\delta t$ (Fig.~\ref{gamma}) and the degraded efficiency for very long orbital periods (Fig.~\ref{zeta}).
Considering the $\delta t_\mathrm{G2}=668$\,d observing window of Gaia, the effective GDR2 sensitivity is limited for a solar twin to approximately $0.050\,M_J\,\mathrm{pc}^{-1}=16\,M_\oplus\,\mathrm{pc}^{-1}$ at an orbital radius of 3\,au ($P_\mathrm{orb}=5.2$\,years).

% Table of the PM anomalies of the nearest stars
% ========================================
\input{Tables/Phys-table}

\subsection{Companions from astrometric excess noise \label{excess-noise}}

For orbital periods shorter than $\delta t$, the astrometric wobble of the star around the center of mass will appear as a noise on the astrometric solution of the star. 
The "excess noise" quantity provided in the GDR2 catalog (\texttt{epsi}, here noted `$\epsilon_i$') corresponds to the extra noise that must be added to the Gaia individual observations to reach a reduced $\chi^2$ of 1 in the astrometric fit \citepads{Lindegren18}.
Bright objects are subject to a number of possible biases induced in particular by saturation, which may make the astrometric excess noise $\epsilon_i$ parameter unreliable.
We considered  the RUWE (Sect.~\ref{ruwe}) as an indicator of the reliability of a given GDR2 record (among others provided in the catalog).
As \texttt{epsi} is sensitive to any deviation from the "pure" parallactic and linear PM trajectory of a source, it is related to the value of $a_1$, and thus $m_2\,r / m_1$.
Assuming that no specific instrumental bias is present, this quantity is thus an indicator for the mass of the companion $m_2$ multiplied by its orbital radius $r$. It is thus a tracer for the presence of short period orbiting companions (as noticed, e.g., by \citeads{2018A&A...619A...8G}).
For a face-on circular orbit, we have the simple relation:
\begin{equation}
m_2(r) =   \frac{\sqrt{2} \, \epsilon_i}{\varpi \, r}\,m_1.
\end{equation}
The factor $\sqrt{2}$ in this expression is due to the fact that $\epsilon_i$ is a standard deviation (expressed in mas) while the companion mass is linked to half of the peak-to-peak amplitude.
Contrary to the PM anomaly, the sensitivity of the $\epsilon_i$ parameter to companion mass decreases for long periods ($P>\delta t$), as only a fraction of the orbit is covered during the observations, thus reducing the astrometric signature.
In principle, it is therefore a complementary indicator to the PMa.
The sensitivity also decreases for the very short periods due to the minimal astrometric wobble.

\subsection{Resolved binary stars}

Resolved binary stars exhibit orbital velocity vectors with precisely opposite directions ($\theta_B = \theta_A \pm 180^\circ$).
For well characterized resolved binary stars, it is possible to compute the PM of their center of mass, using an a priori estimate of the masses of the two stars (see, e.g., \citeads{2016kervella} for $\alpha$ Cen AB). 
We repeated this computation for both the Hipparcos and GDR2 catalog positions, and usef the two derived positions to determine the barycenter PM vector $\mu_\mathrm{HG,AB}$.
In this approach, we adopted model values of the masses of the two stars to determine the position of their center of mass.
This is however not needed if radial velocities are available for the two stars at two different epochs. In this case, it is possible to determine the mass ratio from the ratio of the orbital velocities (see, e.g., \citeads{2016A&A...593A.127K}).
The uncertainties on the masses of the two stars must be taken into account, and degrade the accuracy of the estimated barycenter position.
The subtraction of $\mu_\mathrm{HG,AB}$ from the catalog PM vectors of each star then provides their individual tangential orbital velocity vectors.
Sample applications of the PMa analysis to the nearby binaries \object{61 Cyg}, \object{GJ 725} and \object{GJ 338} are presented in Sect.~\ref{61Cyg_binary} and \ref{binary-individuals}.

%__________________________________ Overview
\section{Overview of results \label{overview}}

\subsection{Star sample and binarity fraction\label{starsample}}

%______________ Figure
\begin{figure}
\includegraphics[width=\hsize]{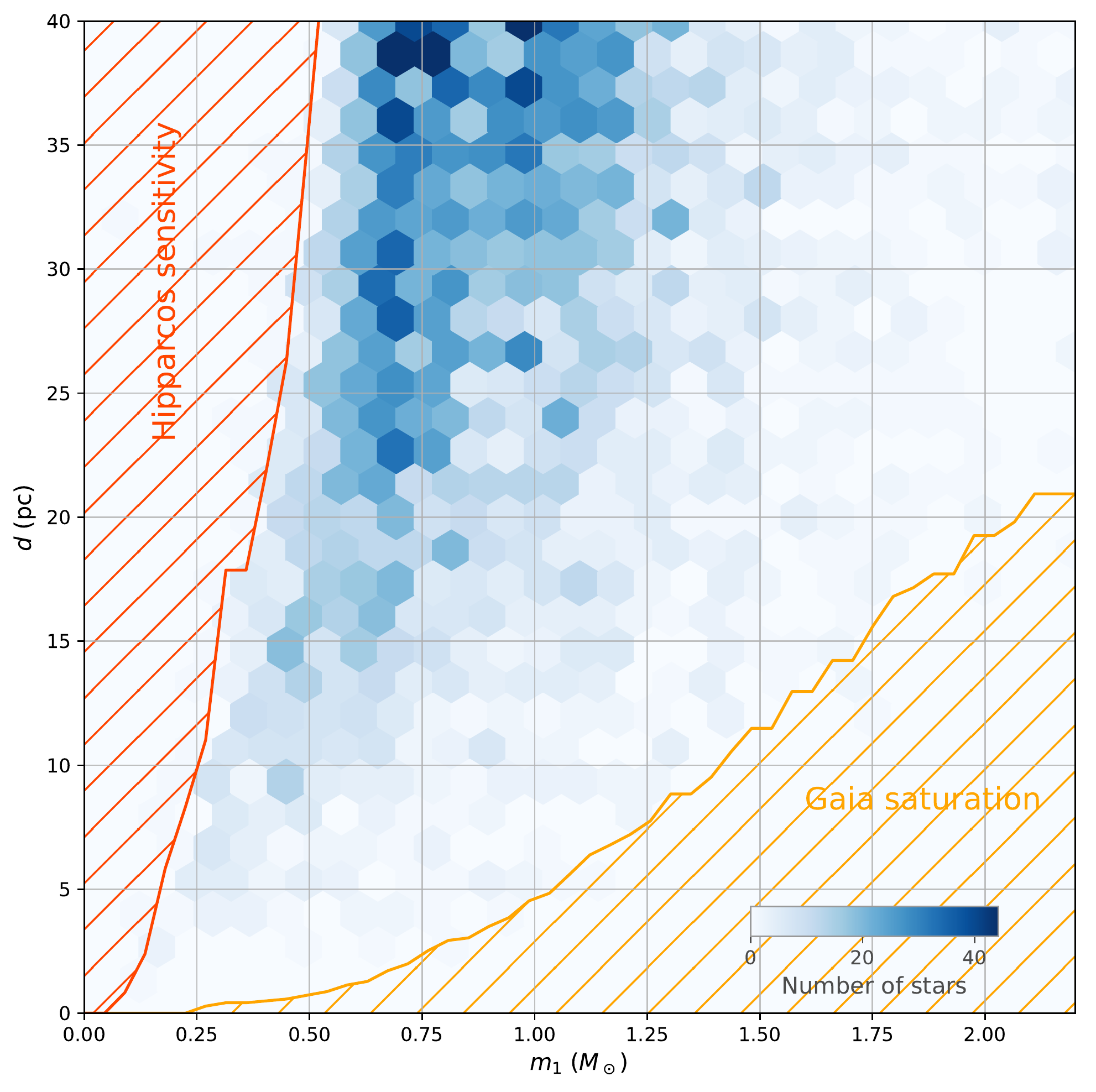}
\caption{2D histogram of the distances $d$ as a function of the primary mass $m_1$.
The excluded domains from the sensitivity limit of Hipparcos and the saturation limit of Gaia ($G\approx3$) are shown as hatched areas, considering the mass-luminosity relation of main sequence stars from \citetads{2013ApJS..208....9P}.
\label{GaiaDR2_50pc-m1dist}}
\end{figure}

The median mass of the 6741 nearby stars of our 50\,pc PMa sample is $\bar{m_1} = 0.90^{+0.40}_{-0.25}\,M_\odot$, and their median radius is $\bar{R_1} = 0.91^{+0.65}_{-0.26}\,R_\odot$. The coincidence of these two numerical values expressed in solar units is expected from the linearity of the mass-radius relation for low mass main sequence stars (see e.g., \citeads{2009A&A...505..205D}).
The completeness of our nearby star sample is limited both by the photometric sensitivity limit of Hipparcos ($V = 12.4$), that sets a maximum distance to the observed stars of a given spectral type, and the saturation limit of Gaia ($G = 3$) that prevents the observation of the nearest bright stars. In Fig.~\ref{GaiaDR2_50pc-m1dist} we show the 2D histogram of the $(m_1,d)$ combinations of our sample.

%______________ Figure
\begin{figure}
\includegraphics[width=\hsize]{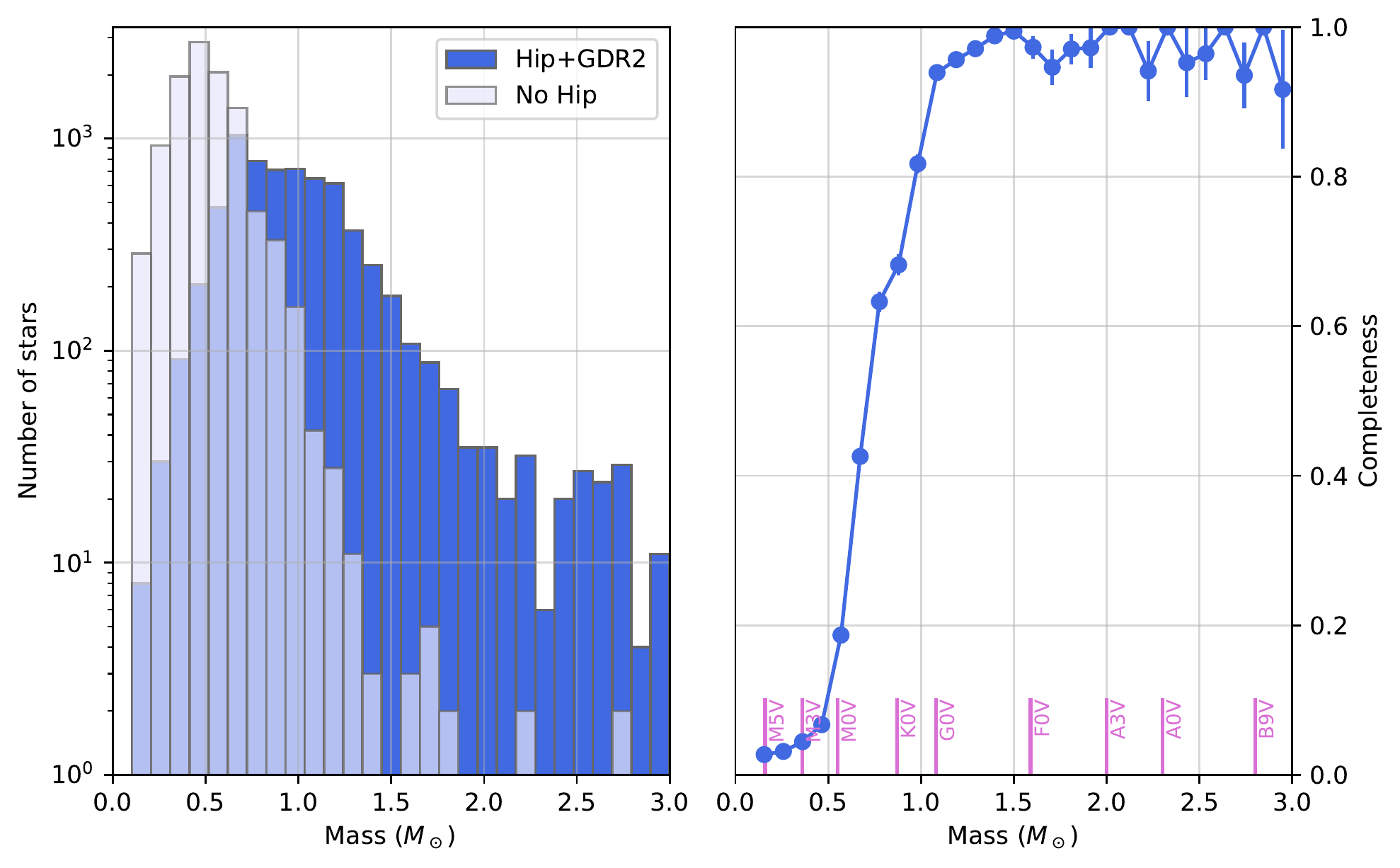}
\caption{\textit{Left panel:} Histogram of the masses of the stars of our 50\,pc sample that are present in the Hipparcos and GDR2 catalogs (dark blue histogram) and that are present in the GDR2 (with $G<13$) but absent from Hip2 (light blue histogram). \textit{Right panel:} Completeness level of the sample of stars tested for the presence of a PMa as a function of their mass. The corresponding spectral types on the main sequence are shown in pink. The error bars represent the binomial proportion 68\% confidence interval.
\label{GaiaDR2_50pc-completeness}}
\end{figure}

Figure~\ref{GaiaDR2_50pc-completeness} shows the mass distribution of the stars that are present in Hip2+GDR2 or only in the GDR2 (with a limit in brightness of $G<13$), and the resulting completeness level of our sample as a function of the stellar mass.
As expected, a severe limitation in terms of completeness comes from the limiting magnitude of Hipparcos, that cuts the majority of the mid-K and M spectral types below $m_1 \approx 0.6\,M_\odot$.
The overall completeness level of the sample of stars tested for PMa (Hip2+GDR2) is above 95\% for the stars with $m_1\geqslant1\,M_\odot$ and of 25\% only for the stars less massive than the Sun.
The very low mass stars are particularly interesting as their PMa vectors give access to low mass planets, as shown for example, by the analysis of Proxima (Sect.~\ref{proxima}). The limiting magnitude of Gaia ($G=21$ in the GDR2 catalog) corresponds to a complete survey of the objects more massive than $m_1 \approx 0.075\,M_\odot$ (the hydrogen burning limit) within 10\,pc of the Sun, and down to $m_1 \approx 0.097\,M_\odot$ (M6.5V spectral type) at 50\,pc.
To give a qualitative idea of the overall completeness of our sample with respect to the total population of nearby objects within 50\,pc, a query of the GDR2 catalog with the only constraint that $\varpi_\mathrm{G2}>20$\,mas (without magnitude limit) returns $\approx 73\,246$ sources. As we examined 6741 sources located within the same distance, this indicates that the measurement of the PMa of $\approx 66\,000$ additional objects (essentially of low and very low mass) will be possible with future Gaia data releases. This may reveal a large number of planetary mass companions.

%______________ Figure
\begin{figure}
\includegraphics[width=\hsize]{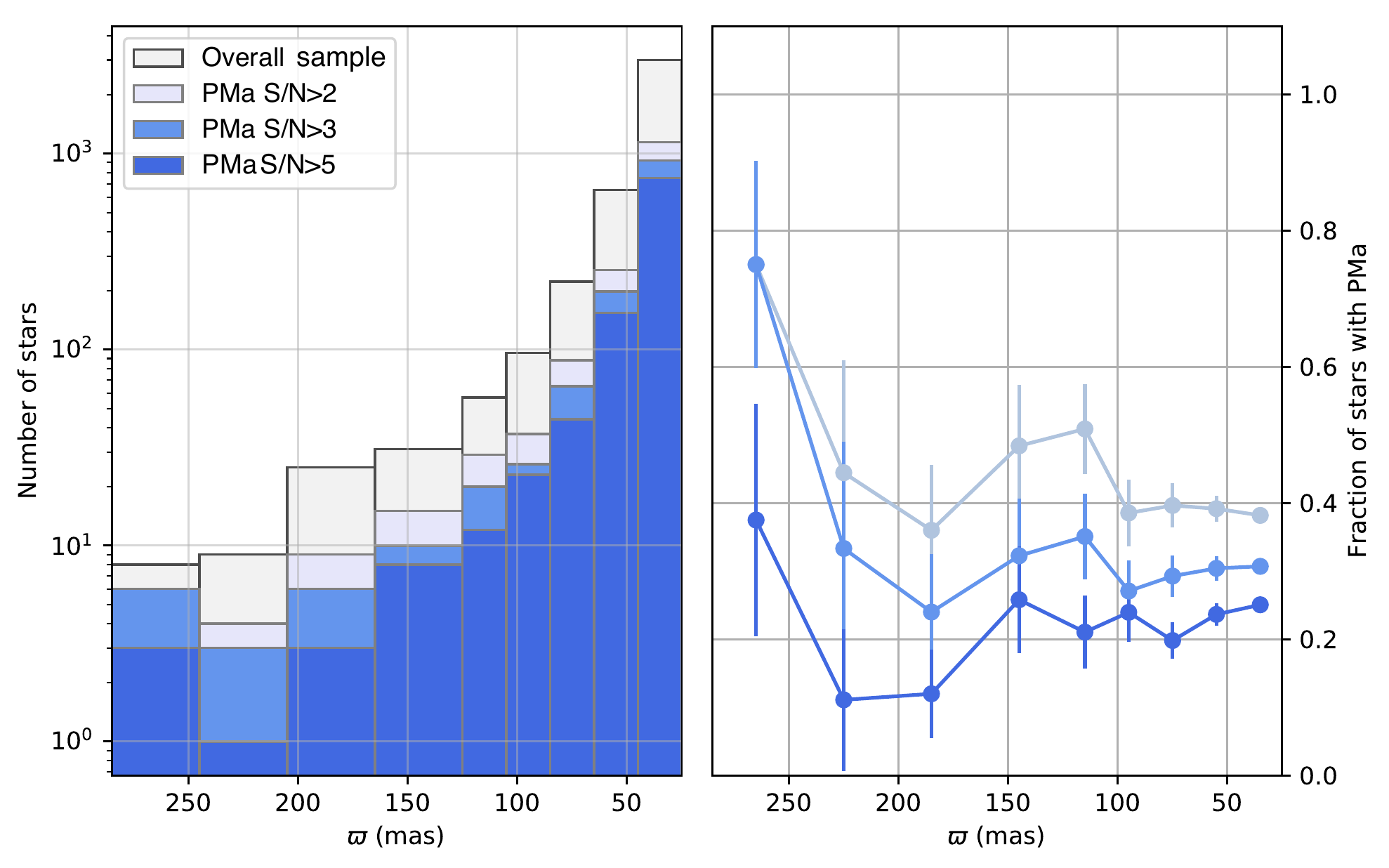}
\caption{\textit{Left:} Number of stars showing a PMa S/N $\Delta_\mathrm{G2}$ with thresholds of 2, 3, and 5, as a function of parallax for stars with parallaxes $\varpi_\mathrm{G2}$ between 25 and 285\,mas.
\textit{Right:} Binarity fraction as a function of parallax for $\Delta_\mathrm{G2}$ thresholds of 2, 3, and 5 (light blue, medium blue and dark blue lines, respectively).\label{GaiaDR2_50pc-PMa-histo}}
\end{figure}

In the present work, we consider that a GDR2 PMa $\Delta_\mathrm{G2} > 3$ indicates a 
bona fide binary detection, while $2 < \Delta_\mathrm{G2} < 3$ points at a suspected binary.
The histogram of the GDR2 signal-to-noise ratio $\Delta_\mathrm{G2}$ of the measured PMa for our 50\,pc sample is presented in Fig.~\ref{GaiaDR2_50pc-PMa-histo} as a function of the parallax.
The overall fraction of stars within 50\,pc exhibiting a PMa is 38.5\% for $\Delta_\mathrm{G2} > 2$, 30.6\% for $\Delta_\mathrm{G2} > 3$ and 24.6\% for $\Delta_\mathrm{G2} > 5$.

\begin{figure}
\includegraphics[width=\hsize]{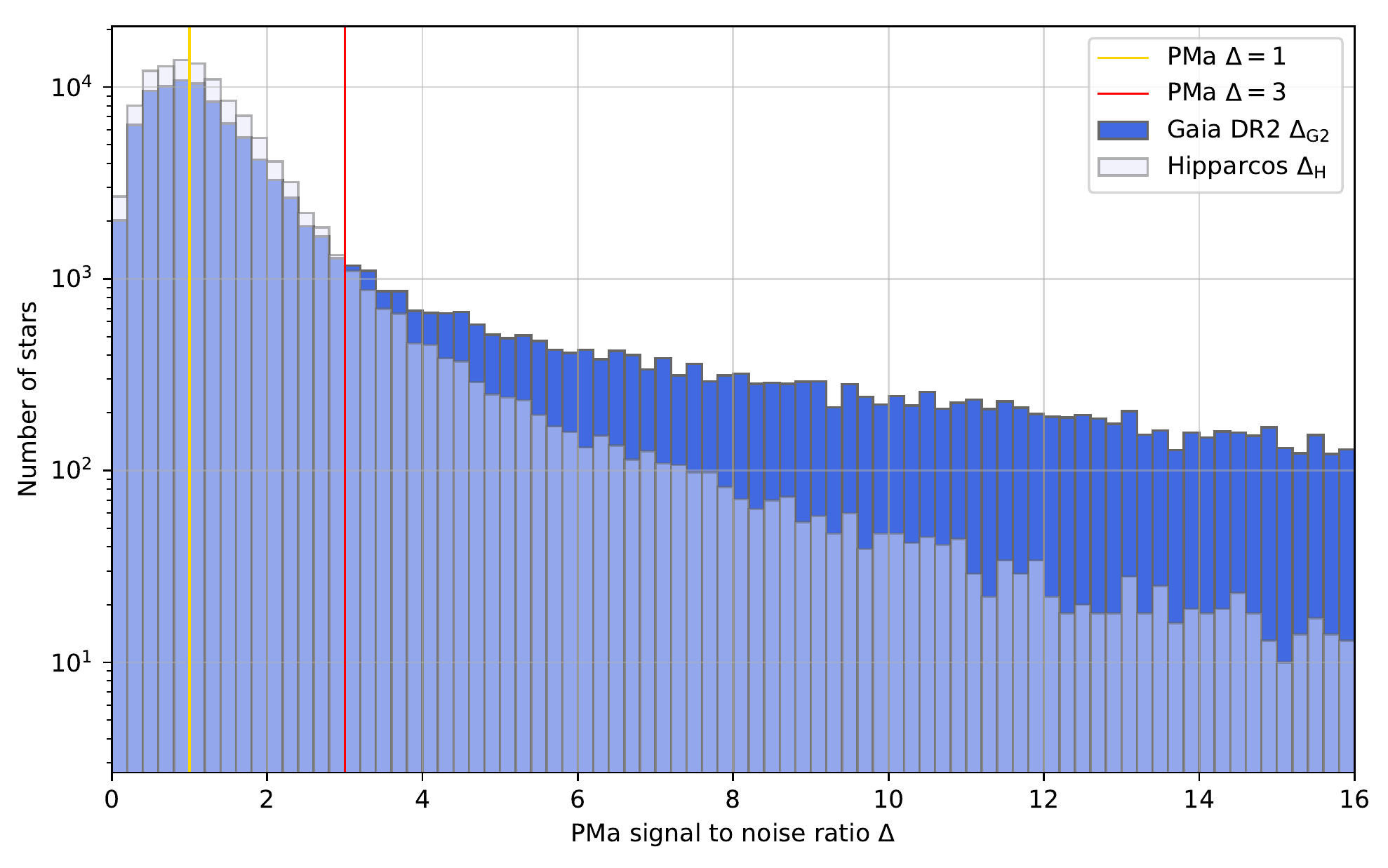}
\caption{Histogram of the observed GDR2 and Hip2 PMa signal-to-noise ratios $\Delta$ for the full Hipparcos sample of 115\,959 stars. Our binary detection limit of $\Delta = 3$ is marked with a red line, and the orange line marks the signal-to-noise ratio $\Delta=1$. \label{PMa-All-histo}}
\end{figure}

The fraction of the 115\,959 stars in the full Hip2 sample that shows a GDR2 PMa is comparable to that of our 50\,pc sample at 36.1\% for $\Delta_\mathrm{G2} > 2$, 26.9\% for $\Delta_\mathrm{G2} > 3$ and 20.2\% for $\Delta_\mathrm{G2} > 5$.
The histogram of the observed PMa signal-to-noise ratio for the Hip2 and GDR2 PM vectors is presented in Fig.~\ref{PMa-All-histo}. In both cases, the PMa distribution peaks slightly below $\Delta = 1$, indicating a satisfactory estimate of the uncertainties. The considerable improvement in accuracy of the GDR2 PM vectors compared to Hipparcos results in a larger number of detected binaries.

%______________ Figure

%______________ Figure
\begin{figure}
\centering
\includegraphics[width=\hsize]{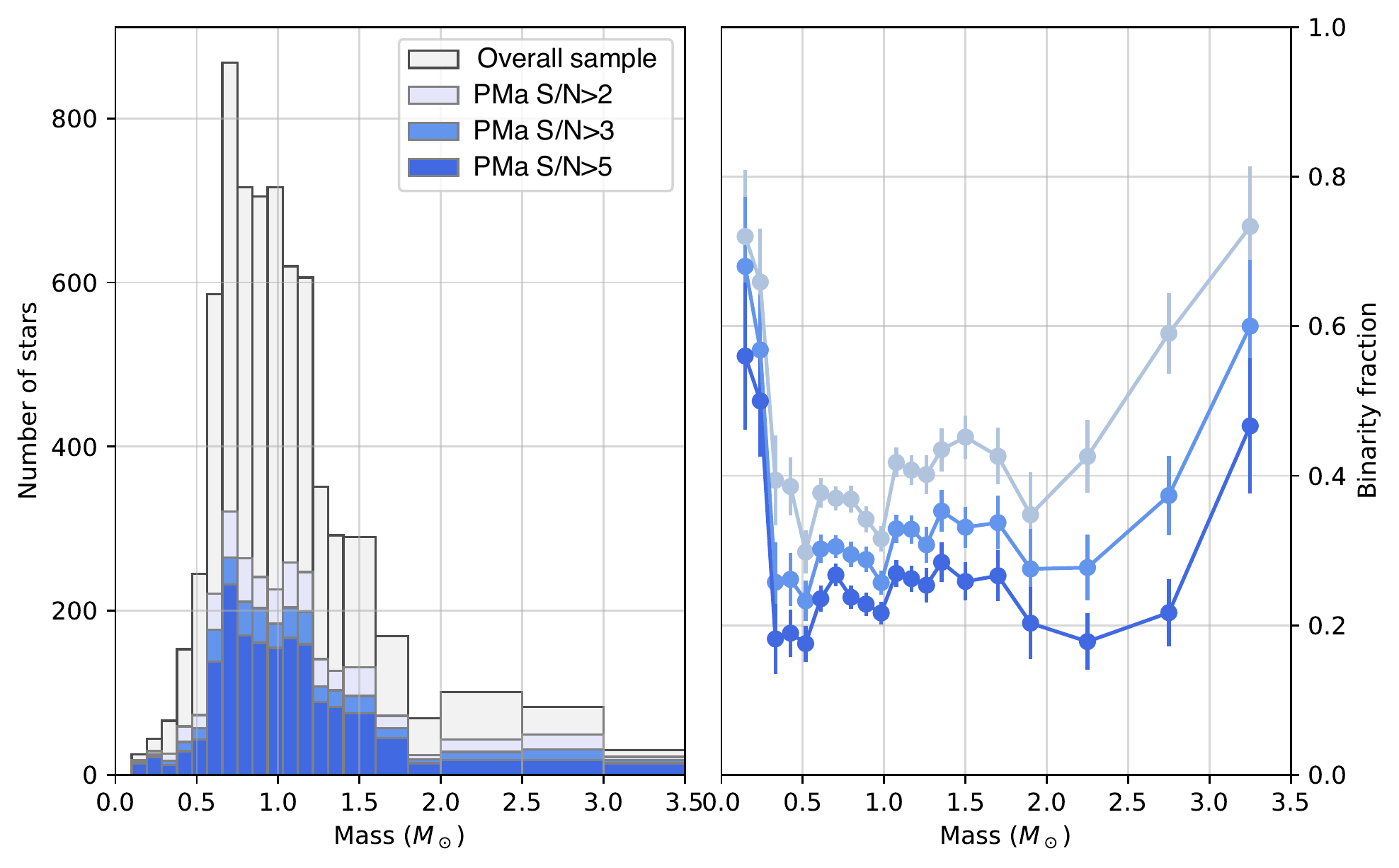}
\caption{\textit{Left:} Number of stars within 50\,pc showing a PMa S/N $\Delta_\mathrm{G2}$, with thresholds of 2, 3, and 5 (light blue, medium blue and dark blue histograms, respectively) as a function of the primary star mass $m_1$.
The histogram of the overall sample is represented in light gray.
\textit{Right:} Binarity fraction as a function of the primary star mass $m_1$. The binaries detected considering thresholds of $\Delta_\mathrm{G2}=2$, 3, and 5 are represented with light blue, medium blue, and dark blue lines, respectively.
\label{GaiaDR2_50pc-PMa-binarity}}
\end{figure}

In our nearby stars sample, we observe only a limited variation of the binarity fraction with the parallax for $\varpi < 200$\,mas. This indicates a weak dependence of the sensitivity of our survey with respect to the distance for these nearby stars, and therefore a satisfactory completeness level of the companions detections for companion masses $m_2^\dag \gtrsim 2\,M_J\,\mathrm{au}^{-1/2}$  up to our distance limit of 50\,pc (Sect.~\ref{companionspma}).
As shown in Fig.~\ref{GaiaDR2_50pc-PMa-binarity}, the binarity fraction (including stellar and substellar mass companions) in our nearby star sample exhibits a relatively mild evolution with the primary star mass between $m_1=0.35\,M_\odot$ and $1.95\,M_\odot$, as the fraction of stars with $\Delta_\mathrm{G2} > 2$ increases from $\approx 35\%$ to $\approx 45\%$.
A very different statistics is observed for the binarity fraction of low mass red dwarfs of $m_1<0.35\,M_\odot$. It shows a very steep decreasing gradient, with a binarity fraction of more than 60\% in the $m_1=0.15 \pm 0.05\,M_\odot$ mass bin that goes down to 35\% at $0.35\,M_\odot$.
This behavior is due to the relatively bright limiting magnitude of Hipparcos, which restricts our sample of very low mass stars to the very nearest objects (Fig.~\ref{GaiaDR2_50pc-completeness}). For the red dwarfs located within a few parsec of the Sun, our sensitivity to low mass orbiting planets is considerably better than for more distant and more massive stars. The limit for these low mass stars is usually in the Saturn mass regime or below, depending on the orbital radius (see, e.g., Sections.~\ref{proxima}, \ref{barnard}, and \ref{kapteyn}).
Assuming a continuous binarity fraction as a function of primary mass, we can tentatively interpret the higher binarity fraction below $0.35\,M_\odot$ as an indication that at least $30$\% of the stars of our PMa sample host planets with masses between $m_2^\dag \approx 0.1\,M_J\,\mathrm{au}^{-1/2}$ and $1\,M_J\,\mathrm{au}^{-1/2}$.
This fraction comes in addition to the $\approx 40\%$ of the stars that show evidence of the presence of companions with $m_2^\dag \gtrsim 1\,M_J\,\mathrm{au}^{-1/2}$.
Although the number of very low mass stars is limited by the Hipparcos sensitivity in the present analysis, with the future Gaia data releases it will be possible to probe the evolution of the binarity fraction as a function of stellar mass in the very low mass regime.

\begin{figure*}
\centering
\includegraphics[width=\hsize]{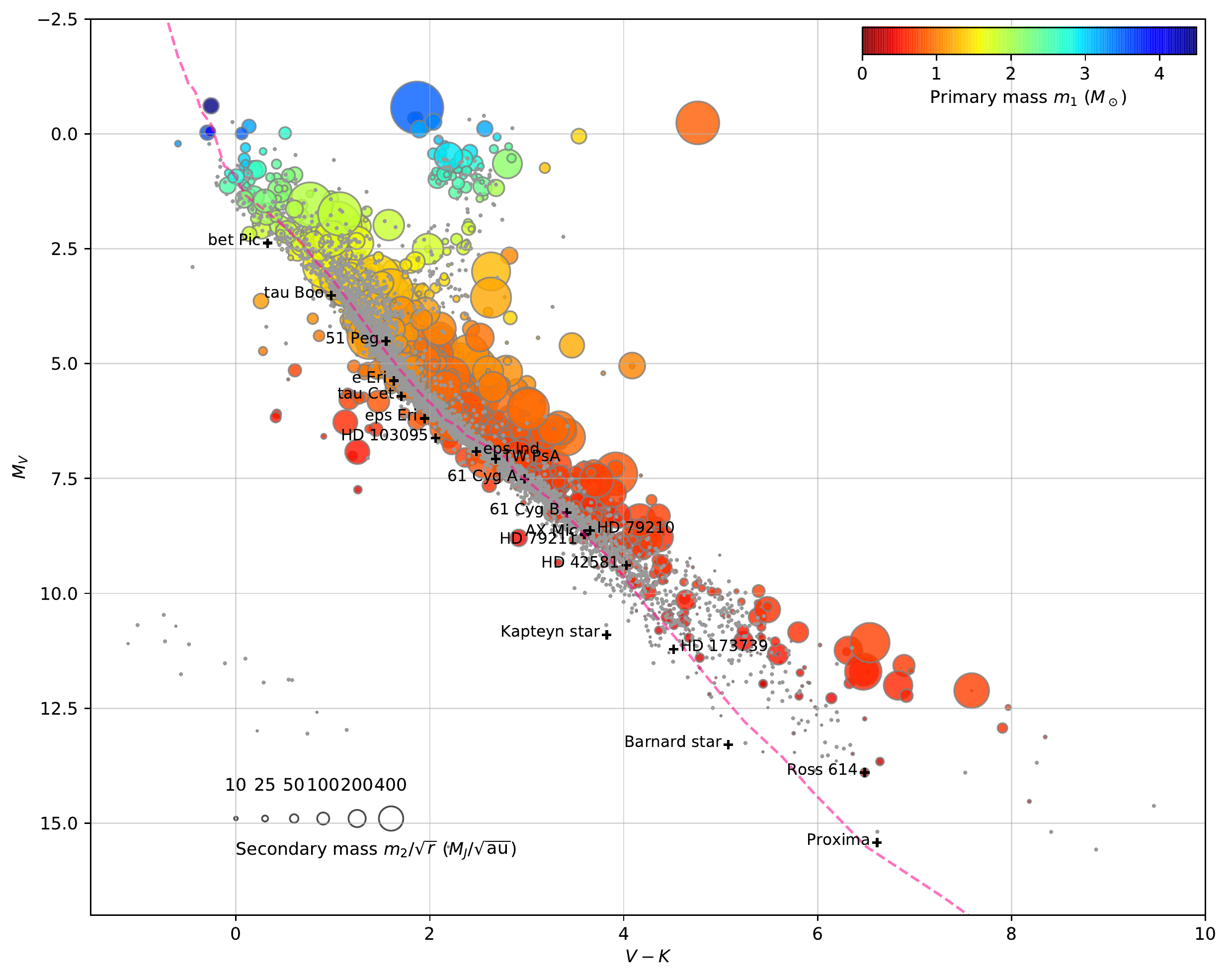}
\caption{Positions of the stars of our sample within 50\,pc in the $[M_V,V-K]$ color-magnitude diagram.
The stars exhibiting a PMa signal-to-noise ratio $\Delta_\mathrm{G2}>2$ are represented in colors.
The symbol size correspond to the normalized companion masse $m_2^\dag$ following the scale displayed in the lower left corner), and the primary mass $m_1$ is represented in color using the scale represented in the upper right corner.
The stars with a PMa signal-to-noise ratio $\Delta_\mathrm{G2}\leqslant2$ are represented using light gray dots, and the objects discussed individually in Sect.~\ref{individual_notes} are indicated with "+" symbols.
To guide the eye, the main sequence from \citetads{2013ApJS..208....9P} is represented by a dashed pink curve.
\label{GaiaDR2_50pc-HR-diagram}}
\end{figure*}

The positions of the stars of our 50\,pc sample in the $[M_V,V-K]$ color-magnitude diagram are shown in Fig.~\ref{GaiaDR2_50pc-HR-diagram}. Due to the relatively small number of massive stars close to the Sun, the statistics are limited in the top left region of the diagram.
The presence of companions of main sequence stars with normalized masses in the stellar or substellar regimes appears ubiquitous.

%______________ Figure

\subsection{Companion mass distribution}

%______________ Figure
\begin{figure}
\centering
\includegraphics[width=\hsize]{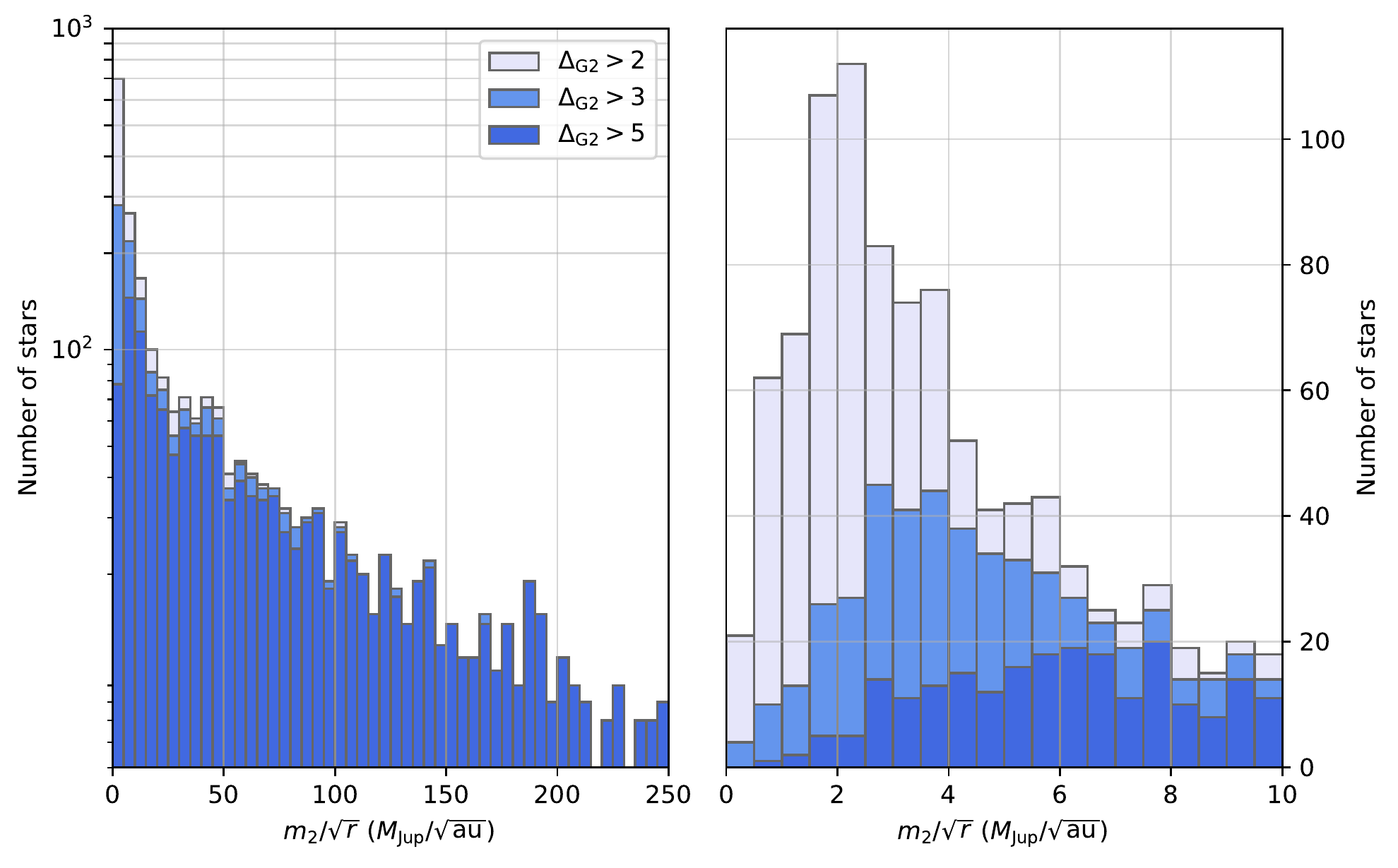}
\caption{\textit{Left:} Histogram of the normalized companion masses $m_2^\dag$ for different levels of PMa signal-to-noise ratio $\Delta_\mathrm{G2}$.
\textit{Right:} Enlargement of the histogram of normalized companion masses between 0 and $10\,M_J\,\mathrm{au}^{-1/2}$. \label{mBstathisto}}
\end{figure}

As detailed in Sect.~\ref{companionspma}, the PMa provides an estimate of the mass of the companions normalized to the square root of their orbital radius. We present in Fig.~\ref{mBstathisto} the distribution of the derived $m_2^\dag$ values up to the stellar mass regime.
In the right panel of this figure, we notice that a significant population of companions with low normalized masses below $2\,M_J\,\mathrm{au}^{-1/2}$ is suspected with a signal-to-noise ratio of between two and three.
The improvement in accuracy of the PM vector expected in the Gaia DR3 will likely confirm the existence of a large fraction of these very low mass companions, and estimate their orbital radii and masses.

%______________ Figure
\begin{figure}
\centering
\includegraphics[width=\hsize]{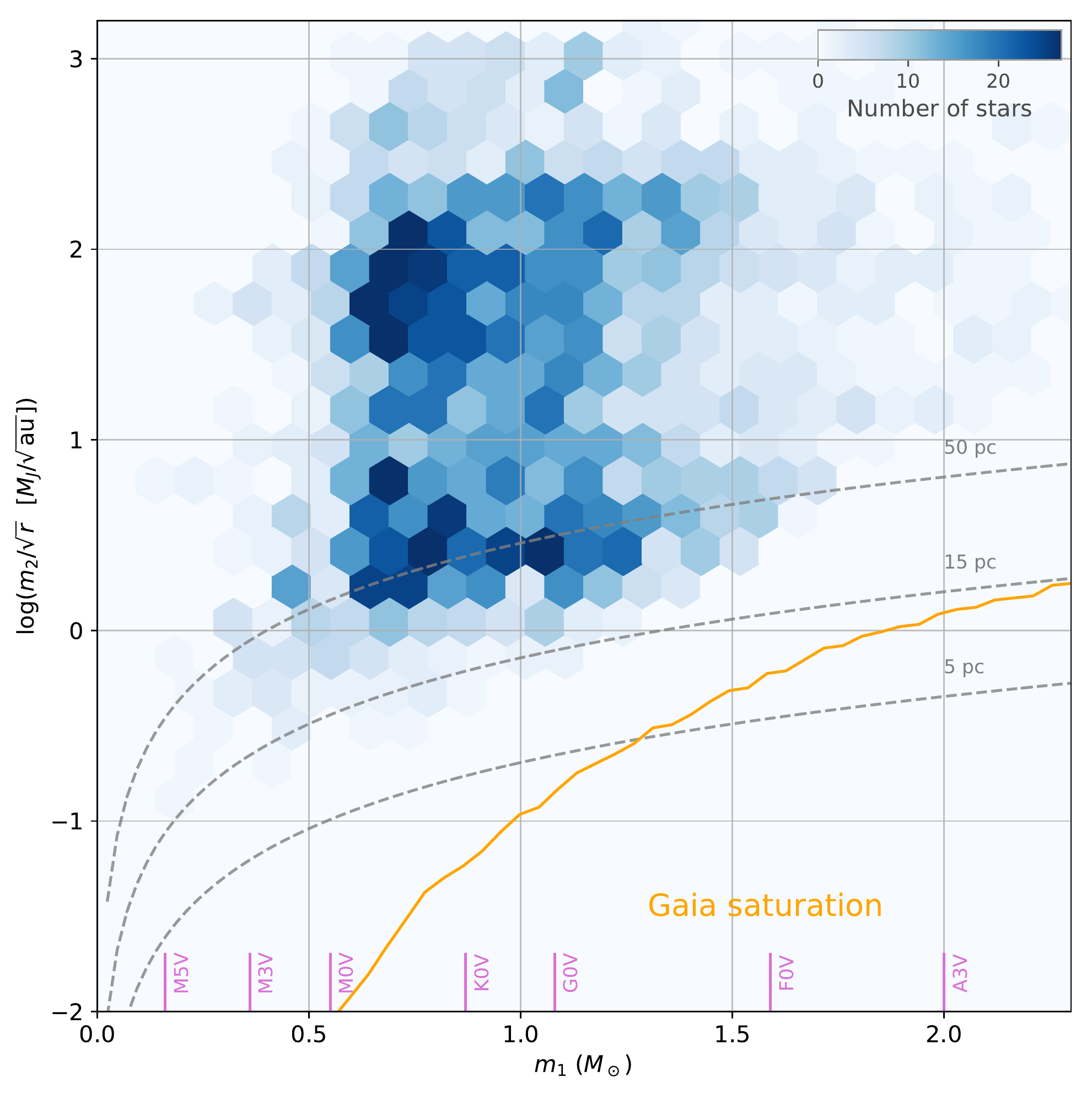}
\caption{2D histogram of the estimated normalized masses $m_2^\dag$ as a function of the primary mass $m_1$ for the stars exhibiting a PMa signal-to-noise ratio $\Delta_\mathrm{G2} > 2$.
The normalized companion mass detection limits for the GDR2 for a star located at distances of 5, 15, and 50\,pc are shown with dashed gray curves (see Sect.~\ref{companionspma} for details).
The area below the orange curve is the companion sensitivity domain excluded by the saturation limit of Gaia ($G\approx3$), considering the mass-luminosity relation of main sequence stars from \citetads{2013ApJS..208....9P}.
\label{GaiaDR2_50pc-m1m2}}
\end{figure}

The two-dimensional (2D) histogram of the $(m_1,m_2^\dag)$ combinations for stars within 50\,pc is presented in Fig.~\ref{GaiaDR2_50pc-m1m2}. The detection sensitivity limits are represented for different distances of the primary star as dashed curves. The apparent deficit of low mass planets around massive stars (lower right quadrant of the histogram) is due to the saturation limit of Gaia that excludes the massive nearby stars from our sample. This deficit is also due to a combination of the overall rarity of massive stars in the solar neighborhood, and the decreased sensitivity to very low mass companions for massive primaries. The Hipparcos limiting magnitude reduces the number of dwarfs with masses lower than $\approx 0.5\,M_J$ compared to their actual frequency (Sect.~\ref{starsample}).

\subsection{White dwarfs}

%______________ Figure
\begin{figure}
\centering
\includegraphics[width=\hsize]{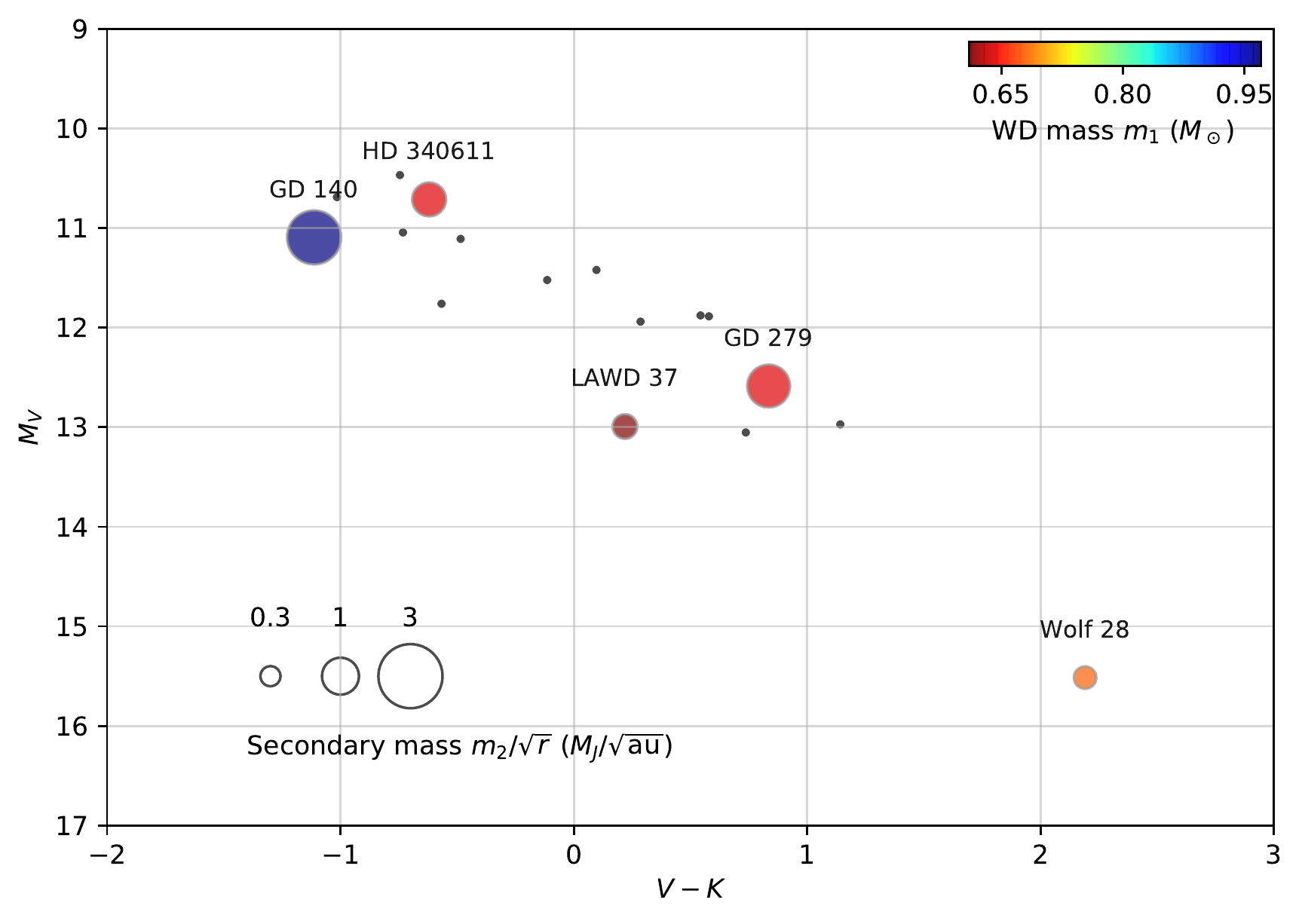}
\caption{Positions of the white dwarfs of our sample in the $[M_V,V-K]$ color-magnitude diagram.
The primary mass and the secondary normalized mass are represented with color and symbol size, respectively, only for the stars with a PMa signal-to-noise ratio $\Delta_\mathrm{G2}\geqslant2$.
The WD with $\Delta_\mathrm{G2}<2$ are represented with a gray dot.
\label{GaiaDR2_50pc-HR-WD}}
\end{figure}

Our sample comprises 17 bright WDs of the solar neighborhood in common between the Hip2 and GDR2 catalogs.
  As detailed in Table~\ref{wd-PMa}, two of these WDs show a significant PMa signal-to-noise ratio $\Delta_\mathrm{G2}$ higher than 3: \object{LAWD 37} (see Sect.~\ref{lawd37}) and \object{GD 140}. In addition, \object{Wolf 28} (Sect.~\ref{wolf28}), \object{HD 340611} and \object{GD 279} show marginal indications of binarity with $\Delta_\mathrm{G2}\geqslant2$.
The binarity fraction (including triple and quadruple stars) of the WD population has been estimated by \citetads{2009JPhCS.172a2022H} to $32 \pm 8$\,\% (see also \citeads{2017A&A...602A..16T}). Including the suspected binary systems, the WD binarity fraction that we obtain from our 17 stars with a PMa analysis is $29 \pm 11 \%$, in line with the expected frequency. The positions of the WD of our sample in the $[M_V,V-K]$ color-magnitude diagram are presented in Fig.~\ref{GaiaDR2_50pc-HR-WD}.

We are currently obliged to use the Hipparcos positions to determine the mean PM $\vec{\mu}_\mathrm{HG}$ and the PMa quantity. This limitation severely reduces the number of accessible WDs due to Hipparcos' lower photometric sensitivity.

\begin{table}
 \caption{Properties and observed tangential velocity anomaly $\Delta v_\mathrm{T,G2}$ for the 17 white dwarfs of our sample. The estimated mass of the WD is listed in the $m_1$ column, $\Delta_\mathrm{G2}$ is the signal-to-noise ratio of the PMa and $m_2^\dag$ is the normalized mass of the companion (or an upper limit).
The column $\Delta$ is set to $\bullet$ for 3<$\Delta_\mathrm{G2}$<5 and $\circ$ for 2<$\Delta_\mathrm{G2}$<3.
 \label{wd-PMa}}
 \centering
 \renewcommand{\arraystretch}{1.2}
 \small
  \begin{tabular}{lcrrcr}
  \hline
  \hline
         Name    & $m_1$ & $\Delta v_\mathrm{T,G2}$ & $\Delta_\mathrm{G2}$ & $\Delta$ & $m_2^\dag$  \\ 
             & ($M_\odot)$ & (m\,s$^{-1}$) & & & ($M_J$\,au$^{-1/2}$)  \\ 
  \hline  \noalign{\smallskip}
\object{Wolf 28} & $0.68_{0.02}$ & $11.0_{5.2}$ & 2.1 & $\circ$ & $0.37^{+0.18}_{-0.16}$ \\
\object{LAWD 37} & $0.61_{0.01}$ & $14.1_{2.9}$ & 4.9 & $\bullet$ & $0.44^{+0.15}_{-0.09}$ \\
\object{CPD-69 177} & $0.68_{0.02}$ & $5.7_{8.6}$ & 0.7 &  & $0.19^{+0.26}_{-0.25}$ \\
\object{LAWD 74} & $0.62_{0.02}$ & $7.1_{11.6}$ & 0.6 &  & $0.22^{+0.33}_{-0.32}$ \\
\object{DN Dra} & $0.75_{0.03}$ & $13.4_{9.7}$ & 1.4 &  & $0.47^{+0.32}_{-0.30}$ \\
\object{CD-38 10980} & $0.68_{0.02}$ & $15.0_{14.1}$ & 1.1 &  & $0.50^{+0.43}_{-0.41}$ \\
\object{HD 340611} & $0.64_{0.03}$ & $26.1_{11.7}$ & 2.2 & $\circ$ & $0.85^{+0.40}_{-0.34}$ \\
\object{GD 140} & $0.97_{0.03}$ & $53.7_{16.1}$ & 3.3 & $\bullet$ & $2.14^{+0.82}_{-0.60}$ \\
\object{BD-07 3632} & $0.53_{0.08}$ & $31.0_{19.3}$ & 1.6 &  & $0.91^{+0.56}_{-0.51}$ \\
\object{CD-30 17706} & $0.61_{0.02}$ & $7.4_{14.6}$ & 0.5 &  & $0.23^{+0.41}_{-0.40}$ \\
\object{GD 279} & $0.64_{0.03}$ & $41.8_{15.1}$ & 2.8 & $\circ$ & $1.35^{+0.57}_{-0.45}$ \\
\object{LAWD 23} & $0.69_{0.03}$ & $37.6_{47.1}$ & 0.8 &  & $1.26^{+1.42}_{-1.38}$ \\
\object{HIP 117059} & $0.56_{0.05}$ & $29.6_{74.4}$ & 0.4 &  & $0.90^{+1.97}_{-1.96}$ \\
\object{EGGR 141} & $0.62_{0.02}$ & $21.8_{21.3}$ & 1.0 &  & $0.69^{+0.62}_{-0.59}$ \\
\object{EGGR 150} & $0.63_{0.02}$ & $27.7_{26.3}$ & 1.1 &  & $0.89^{+0.78}_{-0.74}$ \\
\object{LAWD 52} & $0.58_{0.01}$ & $28.0_{22.8}$ & 1.2 &  & $0.86^{+0.65}_{-0.62}$ \\
\object{Feige 22} & $0.59_{0.02}$ & $37.7_{35.1}$ & 1.1 &  & $1.17^{+1.00}_{-0.96}$ \\
  \hline
\end{tabular}
\end{table}

\subsection{Brown dwarf desert \label{bddesert}}

The concept of "brown dwarf desert" designates the rarity of brown dwarf companions (with a mass of 5 to $80\,M_J$) orbiting solar mass stars at distances shorter than $r=5$\,au ($P \lesssim 11$\,years).
\citetads{2016A&A...588A.144W} determined the masses of the companions of a sample of 15 solar type stars from a combination of radial velocity and Hipparcos astrometry.
They find regularly distributed companion masses, including over the brown dwarf desert \citepads{2000PASP..112..137M}.
The mass distribution of our sample of stars peaks at a lower mass than that of the Sun, but \citetads{2012AJ....144...64D} determined that the brown dwarf desert extends to a broad range of masses of the primary star.
The 2D histogram in Fig.~\ref{GaiaDR2_50pc-m1m2} shows a double peaked distribution of companion normalized masses for primaries of masses $m_1$ between approximately 0.5 and $1.5\,M_\odot$.
The main peak is located at a normalized mass around $m_2^\dag \approx 1-2\,M_J\,\mathrm{au}^{-1/2}$, while the second peak is located at $m_2^\dag \approx 50-100\,M_J\,\mathrm{au}^{-1/2}$.
The gap between these two peaks is compatible with the expected extent of the brown dwarf desert.
Figure~\ref{GaiaDR2_50pc-m1m2} shows indications that the existence of the brown dwarf desert may be restricted to stellar masses similar to that of the Sun.
However, our sample comprises a too limited number of targets above $1.5\,M_\odot$ and below $0.5\,M_\odot$ to reliably explore the persistence of the deficit of brown dwarf companions in these mass regimes.
Moreover, the resolution of the present degeneracy between $m_2$ and $r$ is necessary to firmly conclude on the distribution of companion masses in the brown dwarf desert.

\section{Notes on individual targets\label{individual_notes}}

We present in this section the application of the PMa analysis technique to a few representative objects. Further examples are provided in Appendix~\ref{additional_individual_notes}.

\subsection{Resolved binary 61 Cyg AB (GJ 820 AB) \label{61Cyg_binary}}

\begin{figure}
\centering
\includegraphics[width=\hsize]{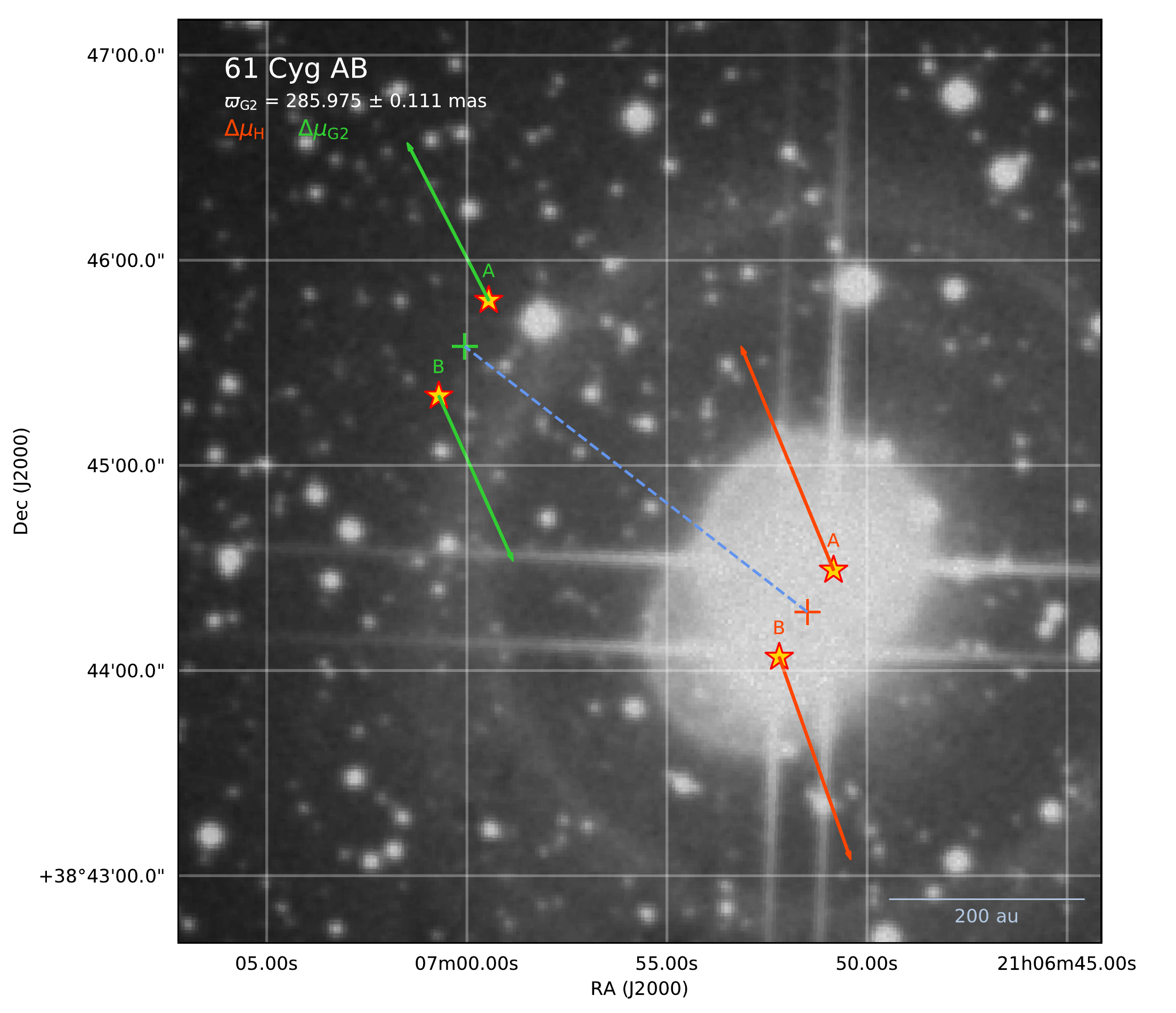}
\caption{Positions of the components of the binary star 61 Cyg AB (K5V+K7V) at the Hipparcos and GDR2 epochs, with the corresponding PMa vectors. The positions of their center of mass are indicated with a "$+$" symbol. The background image is taken from the Second Generation Digitized Sky Survey Red (DSS2-Red).\label{61CygABfig}}
\end{figure}

The nearby binary star 61 Cyg AB (\object{GJ 820} AB, \object{ADS 14636 AB}) is a pair of K5V+K7V red dwarfs.
We present in this paragraph a combined analysis of their respective PMa vectors.
The secondary component 61 Cyg B is one of the 34 Gaia benchmark stars \citepads{2014A&A...564A.133J, 2015A&A...582A..49H}.
We adopted the masses $m_A = 0.708 \pm 0.053\,M_\odot$ and $m_B= 0.657 \pm 0.057\,M_\odot$, estimated from the absolute $K$ band magnitude from 2MASS and the relations by \citetads{0004-637X-804-1-64}.
These values are close to the masses derived by \citetads{2008A&A...488..667K} from evolutionary modeling of the two stars ($m_A = 0.690\,M_\odot$, $m_B = 0.605\,M_\odot$).
We determined the following PM vector for the barycenter of the two stars:
\begin{equation}
\vec{\mu}_\mathrm{HG,AB} = [+4133.04 \pm 0.81, +3203.81 \pm 0.17]\,\mathrm{mas}\,\mathrm{a}^{-1}.
\end{equation}
The subtraction of this barycenter PM vector from the individual Hip2 PM vectors provides the following residual PM anomalies:
\begin{gather}
\vec{\Delta \mu}_\mathrm{H,A} = [+34.7_{6.6},  +65.4_{12.1}]\,\mathrm{mas}\,\mathrm{a}^{-1} \\
\vec{\Delta \mu}_\mathrm{H,B} = [-26.76_{0.87},  -59.13_{0.47}]\,\mathrm{mas}\,\mathrm{a}^{-1},
\end{gather}
and the same computation for the GDR2 PM vectors gives
\begin{gather}
\vec{\Delta \mu}_\mathrm{G2,A} = [+30.55_{0.84},  +46.03_{0.32}]\,\mathrm{mas}\,\mathrm{a}^{-1} \\
\vec{\Delta \mu}_\mathrm{G2,B} = [-27.84_{0.82},  -48.20_{0.21}]\,\mathrm{mas}\,\mathrm{a}^{-1}.
\end{gather}
We represent these PMa vectors on the sky in Fig.~\ref{61CygABfig}.
%
%______________ Figure
We can check a posteriori that the mass ratio we adopted is reasonable from the ratio of the norms of the PMa vectors:
\begin{equation}
\frac{m_B}{m_A} = \left(\frac{\Delta \mu_A}{\Delta \mu_B}\right)
.\end{equation}
For the GDR2 epoch, we obtain $m_B/m_A = 0.993 \pm 0.022$, higher than but statistically compatible with the adopted $m_B/m_A = 0.93 \pm 0.11$.
The derived mass ratio from the PMa is also noticeably higher than the ratio $m_B/m_A = 0.877$ found by \citetads{2008A&A...488..667K}.
This difference can be interpreted as the orbital PM $\Delta \mu_B$ of component 61 Cyg B being too slow for its "photometric" mass, thus indicating the possible presence of a dark mass contributor in orbit around this star.
We also observe a difference in position angle of the PM anomalies of components A and B from GDR2:
\begin{equation}
\theta_A = 33.57 \pm 0.75\,\deg\ \mathrm{and}\ \theta_B = 210.01 \pm 0.75 \deg.
\end{equation}
The difference $\Delta \theta_{AB} = (\theta_A - \theta_B) \mod {180^\circ} = 3.56 \pm 1.06 \deg$ is statistically significant, while it should be zero if the photocenters of each component coincided with their respective centers of mass.
This PMa offset between 61 Cyg A and B points at the possible presence of a third body in the system, likely orbiting around 61 Cyg B.
This hypothesis will be testable with the Gaia astrometry in the future data releases.

\subsection{Proxima\label{proxima}}

Our nearest stellar neighbor \object{Proxima Centauri} (\object{GJ551}, \object{V645 Cen}, and \object{$\alpha$ Cen C}) is a very low mass red dwarf of spectral type M5.5V.
It hosts the nearest exoplanet, \object{Proxima b} \citepads{2016Natur.536..437A}, a telluric mass planet with a very short orbital period (11.2\,d) and orbital distance (0.05\,au) that place it inside the habitable zone of its parent star.
The search for transiting exoplanet signals have been unsuccessful up to now \citepads{2018AJ....155..228B, arXiv1901.07034}.
Proxima is an active star, with frequent flares that were detected over a broad range of wavelengths \citepads{2018ApJ...855L...2M, 2018ApJ...860L..30H}.

%\subsubsection{Maximum possible orbital radius}
%
%We here consider the triple system $\alpha$\,Cen AB + Proxima as equivalent to a binary, due to the very large semi-major axis of Proxima's orbit ($a_\mathrm{Prox} = 8.7$\,kau; \citeads{2017A&A...598L...7K}) compared to that of the $\alpha$\,Cen AB system ($a_{AB}=23.5$\,au; \citeads{2016kervella}). The eccentricity of Proxima's orbit is $e=0.50$ and the mass ratio of the system is $q = m_\mathrm{Prox}/(m_A+m_B+m_\mathrm{Prox}) = 0.056$. \citetads{1999AJ....117..621H} established the range of stable orbits for planets orbiting in binary systems, and for the parameters of Proxima's orbit, they predict a maximum stable orbital radius for planets orbiting Proxima of $r_\mathrm{max} = 0.20\ a_\mathrm{Prox} = 1.7$\,kau. In the following, we adopt this limit to estimate the range of possible companions orbiting Proxima.

\subsubsection{Companion mass sensitivity}

The fast PM of Proxima coupled with the high accuracy of the position measurements by Hip2 and GDR2 results in an extraordinary accuracy of the PM vector coordinates: its norm is estimated to $\mu_\mathrm{HG} = 3859.110 \pm 0.069$\,mas\,a$^{-1}$ (at GDR2 epoch) which corresponds to a signal-to-noise ratio of $5.6 \times 10^4$.
Subtracting the long-term PM from the GDR2 vector, we measure a tangential velocity anomaly of $\Delta v_\mathrm{tan,G2} = 2.7 \pm 1.5$\,m\,s$^{-1}$ at GDR2 epoch (Table~\ref{phys-properties}) significant at a $1.8\sigma$ level.
The accuracy on the tangential velocity anomaly is limited by the precision of the GDR2 PM vector, which will improve in the future data releases.

In addition to the Hip2 and GDR2 catalogs, we also tested the PM vector obtained by \citetads{1999AJ....118.1086B} for the presence of a PMa.
We find a marginal PMa at a level of signal-to-noise ratio of $\Delta_\mathrm{FGS} = 2.6$, but the reliability of the PM vector coordinates of Proxima is uncertain as the FGS measurement is based on differential astrometry with background stars, whose PM vectors were poorly constrained at the time of the original data reduction.
We note that the parallax of Proxima $\varpi_\mathrm{FGS}=768.7 \pm 0.3$\,mas determined by \citetads{1999AJ....118.1086B} is in perfect agreement with the GDR2 value ($\varpi_\mathrm{G2}=768.53 \pm 0.22$\,mas). The ground-based parallax of $\varpi_\mathrm{L14}=768.1 \pm 1.0$\,mas measured by \citetads{2014AJ....148...91L} is also perfectly consistent with the GDR2.
\citetads{2017MNRAS.466L.118M} established mass and radius limits to companions of Proxima from adaptive optics imaging with the SPHERE instrument, setting a maximum mass of a planet orbiting beyond 2\,au from the star to $4\,M_J$.
From ground based astrometric measurements, \citetads{2014AJ....148...91L} set the maximum mass of possible companions of Proxima to $2\,M_J$ at 0.8\,au and $1\,M_J$ at 2.6\,au.
Closer to Proxima, \citetads{2008A&A...488.1149E} set very low mass limits using the radial velocity technique, of $4\,M_\oplus$ at $r=0.1$\,au and $15\,M_\oplus = 0.05\,M_J$ at $r=1$\,au.
This limit was further decreased with the detection of the $m_b\,\sin i = 1.3\,M_\oplus$ Proxima b \citepads{2016Natur.536..437A}.

\begin{figure}[]
\centering
\includegraphics[width=9cm]{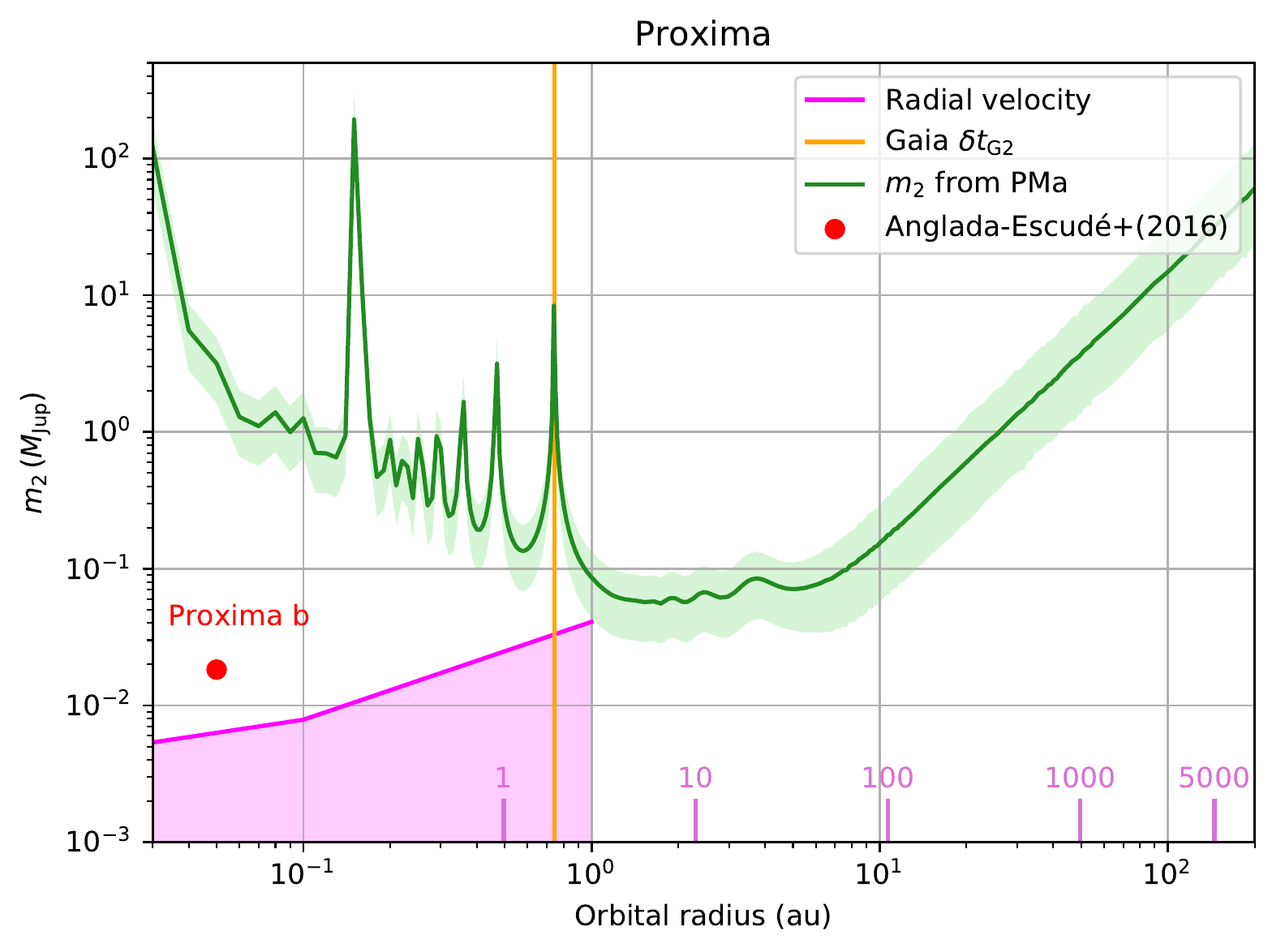}
\caption{Possible companions of Proxima. The possible companion mass and orbital radius $(m_2,r)$ combinations explaining the observed $\Delta\mu_\mathrm{G2}$ PMa from the GDR2 are shown with the green curve, taking into account the GDR2 time window.
The shaded green region corresponds to the $1\sigma$ uncertainty domain.
The orange vertical line marks the orbital radius whose period corresponds to the GDR2 time window ($\delta t_\mathrm{G2} = 668$\,d).
The permitted $(m_2,r)$ domain for short-period planets from radial velocity searches is shaded in pink.
The orbital radii corresponding to periods of $P_\mathrm{orb}=1$ to 5000\,years are indicated with pink vertical marks, with corresponding period values in pink.
 \label{proxima-m2r}}
\end{figure}

The possible $(m_2,r)$ combinations (with $m_2$ the companion mass and $r$ its orbital radius) corresponding to the detected GDR2 tangential velocity anomaly $\Delta v_\mathrm{tan,G2}$ are presented in Fig.~\ref{proxima-m2r} in green color.
The possible domain of $(m_2,r)$ combinations delineated by \citetads{2008A&A...488.1149E} is represented as the shaded pink area.
We exclude at a $1\sigma$ level the presence of a planet with a mass $m_2 > 0.1 M_J$ (two Neptune masses) at $r<3$\,au and $m_2 > 0.3\,M_J$ (one Saturn mass) at $r < 10$\,au ($P_\mathrm{orb} < 100$\,years).
Planetary companions on wider orbits between 10 and 50\,au ($P_\mathrm{orb} = 100$ to $1000$\,years) are also excluded with mass limits ranging from $m_2 = 0.3$ to $8\,M_J$.
These stringent constraints on the presence of planets around Proxima emphasize the remarkable complementarity of the Gaia astrometry and radial velocity searches for short and long period planets.

%______________ Figure

\subsubsection{Effect of Proxima b}

The amplitude of the astrometric wobble induced by the presence of \object{Proxima b} \citepads{2016Natur.536..437A} of a mass $m_b\,\sin i = 1.3\,M_\oplus$ is $\approx 2.8\,\mu$as in position, with a period of $P_\mathrm{orb} = 11.2$\,days. This amplitude is likely to remain undetected by Gaia whose final average position accuracy is foreseen around $4\,\mu$as. 
The Doppler radial velocity amplitude of the reflex motion of Proxima is 1.4\,m\,s$^{-1}$, as measured by \citetads{2016Natur.536..437A}.
This represents a lower limit for the tangential velocity amplitude.
Due to the very short orbital period of Proxima b, the PMa quantity, as a tracer of the orbital velocity, is not sensitive enough to measure the tangential reflex velocity of Proxima induced by planet b, as shown in Fig.~\ref{proxima-m2r} (at $r = 0.05$\,au).
However, knowing the precise ephemeris of the planet from the spectroscopic orbit, a differential processing of the epoch astrometry data may reveal the wobble signal in position and PM velocity. The intrinsic morphological properties of the star (spots, rotation, and flaring) may complicate the analysis.

\subsubsection{The orbit of Proxima around $\alpha$\,Cen}

\citetads{2017A&A...598L...7K} showed that Proxima is gravitationally bound to $\alpha$\,Cen, possibly captured by its bright binary neighbor \citepads{2018MNRAS.473.3185F}.
We took advantage of the improved accuracy of the $\vec{\mu}_\mathrm{HG}$ vector of Proxima provided by the combination of the Hip2 and GDR2 positions to check the orbital parameters of the orbit of Proxima around $\alpha$ Cen (\object{WDS J14396-6050AB}, \object{GJ559AB}).
The revised parameters are statistically identical, and the accuracy of the relative velocity of Proxima with respect to $\alpha$ Cen is slightly improved to $v_\mathrm{\alpha-Prox} = 280 \pm 48 \,\mathrm{m\,s}^{-1}$ giving a $-5.5\,\sigma$ difference with the unbound velocity value ($v_\mathrm{max}=546 \pm 10$\,m\,s$^{-1}$).
The error bar on this differential velocity is dominated by the uncertainty on the absolute radial velocity of Proxima ($\pm 32$\,m\,s$^{-1}$) and the correction of the convective blueshift.
The orbital motion of Proxima cannot explain the observed residual tangential velocity, as the total change on Proxima's PM vector induced by its orbital motion is only $dv_\mathrm{tan} = 0.053$\,m\,s$^{-1}$ ($d\Delta \mu = 8.6\,\mu$as\,a$^{-1}$) over the 24.25 years separating the Hip and GDR2 epochs. 
The final PM accuracy of Gaia is expected to be on the order of $\sigma(\mu) \approx 3\,\mu$as\,a$^{-1}$ (compared to $\sigma(\mu) \approx 250\,\mu$as\,a$^{-1}$ for Proxima in GDR2), so the orbital acceleration of Proxima may be detectable with Gaia's full data set. We note, however, that the presence of a long-period planet will complicate this detection.

As discussed by \citetads{2009MNRAS.399L..21B, 2011Ap&SS.333..419B} and \citetads{2018MNRAS.480.2660B}, very high precision astrometry of Proxima could provide a test of the modified Newtonian dynamics (MOND) theory \citepads{1983ApJ...270..365M, 2004PhRvD..70h3509B}.
\citetads{2018MNRAS.480.2660B} estimated that the relative positions of Proxima with respect to $\alpha$\,Cen AB computed using the MOND or Newtonian gravity theories would differ by $\approx 7\,\mu$as after ten years (their Fig.~13).
Although this seems in principle within reach of Gaia, unfortunately $\alpha$\,Cen AB is too bright and the position of the barycenter of AB will not be measurable with the required accuracy.
The change in radial velocity over the same Hip-GDR2 period of 24.25\,years is $dv_\mathrm{r} = 0.025$\,m\,s$^{-1}$, also beyond the accuracy of the current instrumentation.

\subsection{$\epsilon$ Eri}

From the GDR2 catalog, we find an insignificant PMa on the young K2V dwarf \object{$\epsilon$~Eri} (\object{GJ 144}, \object{HD 22049}) of $\Delta v_\mathrm{tan,G2} = 5.7 \pm 13.1$\,m\,s$^{-1}$, compatible with zero.
The tangential velocity anomaly measurement from Hip2 is more accurate at $\Delta v_\mathrm{tan,H} =7.6 \pm 4.4$\,m\,s$^{-1}$, and we discuss its implications hereafter.

%______________ Figure
\begin{figure}
\centering
\includegraphics[width=\hsize]{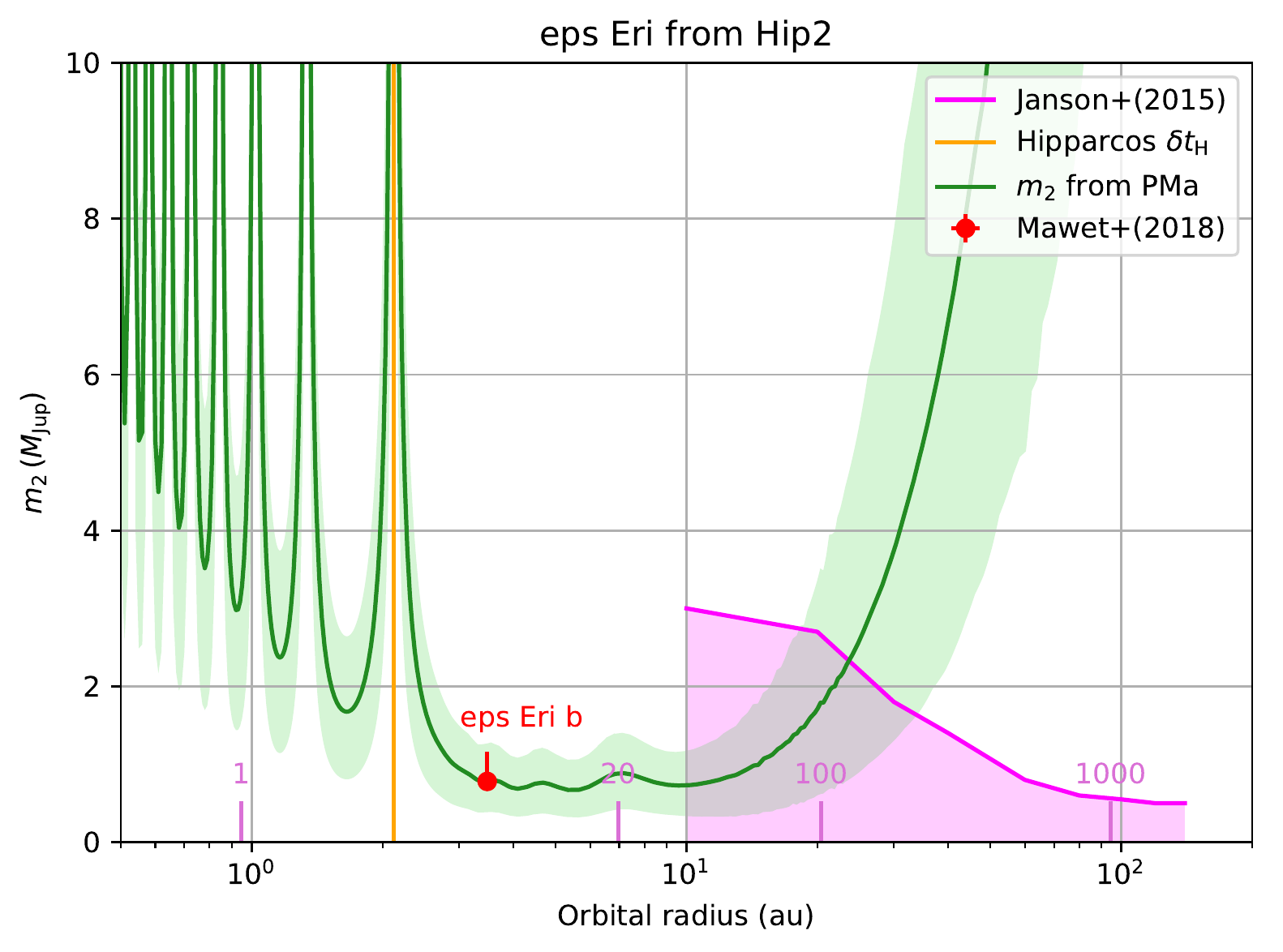}
\caption{Limits on the companion $(m_2,r)$ combinations for $\epsilon$ Eri. The parameters of the giant planet $\epsilon$ Eri b determined by  \citetads{1538-3881-157-1-33} are represented with a red dot. The shaded pink area delineates the permitted $(m_2,r)$ combinations between 10 and 140\,au determined by \citetads{2015A&A...574A.120J} from \textit{Spitzer} imaging.\label{eps-Eri-m2r}}
\end{figure}

The massive planet \object{$\epsilon$ Eri b} ($m_2 = 0.86\,M_J$) was discovered by \citetads{2000ApJ...544L.145H} on an eccentric orbit ($e=0.6$, $r=3.4$\,au). \citetads{1538-3881-157-1-33} recently confirmed its mass ($m_2 = 0.78^{+0.38}_{-0.12}\,M_J$) and orbital radius ($r=3.48 \pm 0.02$\,au), but obtained a much lower eccentricity ($0.07^{+0.06}_{-0.05}$). This mass and orbital radius of $\epsilon$ Eri b provide an excellent match to the observed $\Delta v_\mathrm{tan,H}$ (red dot in Fig.~\ref{eps-Eri-m2r}).
A detailed analysis of the epoch astrometry of Hip2 as conducted by \citetads{2018NatAs.tmp..114S} and \citetads{2019ApJ...871L...4D} on $\beta$~Pic (see also Sect.~\ref{betapic}) may provide valuable informations on $\epsilon$\,Eri~b, but it is beyond the scope of the present work.
\citetads{2015A&A...574A.120J} searched for massive companions of $\epsilon$\,Eri using \textit{Spitzer} imaging in the infrared, and established mass limits of $m_2=0.5-2.5\,M_J$ for separations of $r=140-20$\,au (for an age of 800\,Ma). The corresponding permitted range of $(m_2,r)$ combinations is represented in shaded pink in Fig.~\ref{eps-Eri-m2r}.

\subsection{$\tau$ Cet \label{tauceti}}

The nearby solar analog \object{$\tau$ Ceti} (\object{GJ 71}, \object{HD 10700}, spectral type G8V) is an old \citepads{2005A&A...436..253T}, low metallicity dwarf, that hosts a dust disk \citepads{2007A&A...475..243D}.
\citetads{2017AJ....154..135F} recently confirmed the detection of four telluric mass planets orbiting $\tau$\,Cet, that were initially announced by \citetads{2013A&A...551A..79T}. Their orbital semi-major axes range from 0.1 to 1.3\,au, and their minimum masses from 1.9 to $3.9\,M_\oplus$. These planets have a negligible contribution to the GDR2 astrometry.

\begin{figure}
\centering
\includegraphics[width=\hsize]{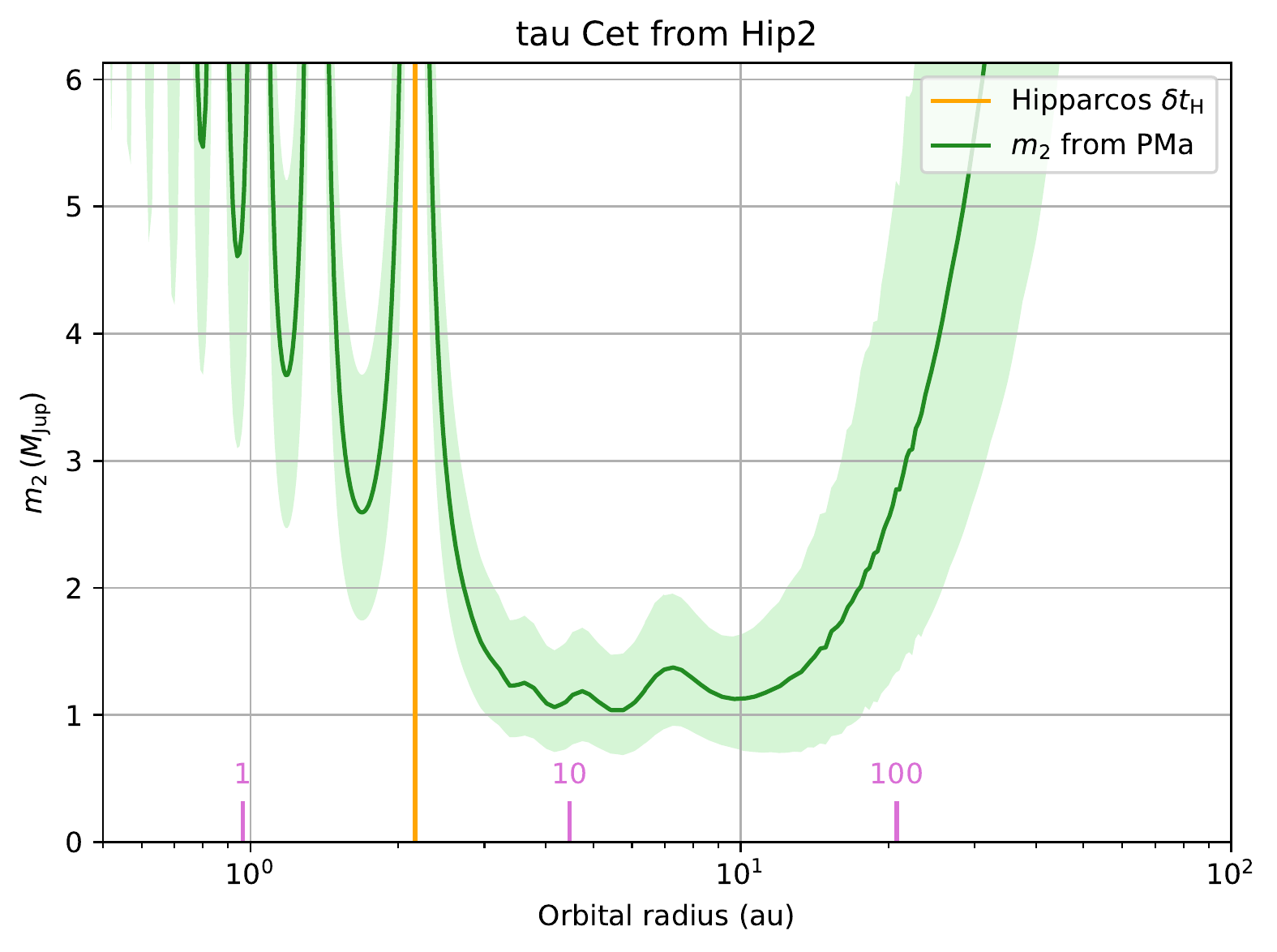}
\caption{Possible companion $(m_2,r)$ combinations for $\tau$ Ceti.\label{tau-Cet-m2r}}
\end{figure}

We detect a high tangential velocity anomaly on $\tau$\,Ceti at a level of $\Delta v_\mathrm{tan,G2} = 132 \pm 28$\,m\,s$^{-1}$ in the GDR2 data, but the excess noise is high at $\epsilon_i = 1.7$\,mas, while the RUWE is low at $\varrho = 1.1$.
The accuracy of the GDR2 PM vector is much lower than that of Hip2, probably due to the high brightness of $\tau$\,Ceti ($m_G = 3.2$), causing a saturation of the Gaia detectors.
Therefore for this work we have analysed the Hipparcos PM vector, giving a tangential velocity anomaly of $\Delta v_\mathrm{tan,H} = 11.3 \pm 4.0$\,m\,s$^{-1}$. This low but significant signal corresponds to a range of possible $(m_2,r)$ pairs shown in Fig.~\ref{tau-Cet-m2r}.
The observed signal could be explained for example, by a Jupiter analog orbiting at 5\,au.
We exclude ($1\sigma$) the presence of a planet more massive than $5\,M_J$ between 3 and 20\,au from the star.

%______________ Figure

\subsection{$\beta$~Pic \label{betapic}}

\object{$\beta$~Pic} (\object{GJ 219}, \object{HD 39060}) is a young hot dwarf (spectral type A6V) surrounded by a dust disk, that hosts a giant planet, \object{$\beta$ Pic b} \citepads{2009A&A...493L..21L,2010Sci...329...57L}.
The orbital parameters of this planet were determined for example, by \citetads{2016AJ....152...97W} (see also \citeads{2012A&A...542A..41C,2015ApJ...811...18M}).
It has a semi-major axis of $a=9.7$\,au and an orbital period of $P_\mathrm{orb} = 22.5$\,years.
We discuss separately the analysis of the GDR2 and Hipparcos measurements of $\beta$~Pic in the following sections.

\subsubsection{Gaia astrometry}

$\beta$~Pic is absent from the Gaia DR1, and its GDR2 measurement of the PM is affected by large uncertainties of 0.74\,mas\,a$^{-1}$, that is, five times larger than the PM error bars in the Hip2 catalog (that are around $\pm 0.13$\,mas\,a$^{-1}$).
They result in similarly large uncertainties in the corresponding PMa (Table~\ref{pm_binaries1}).
The high brightness of $\beta$~Pic ($m_G=3.7$) is close to the saturation limit of Gaia, and the measurements are probably affected by instrumental effects that degrade their accuracy.
The RUWE (Sect.~\ref{excess-noise}) however remains limited at $\varrho = 1.2$, showing that the quality of the astrometric fit of $\beta$~Pic does not differ significantly from the other stars of similar color and brightness, and that the error bars on the determined parameters are likely reliable.

The GDR2 record of $\beta$~Pic lists a high value of the astrometric excess noise $\epsilon_i = 2.14$\,mas (Table~\ref{pm_binaries1}), that indicates that the astrometric fit shows the presence of significant residuals.
It is unlikely that these residuals are caused by the astrometric reflex motion induced on $\beta$~Pic by planet b, as its orbital period is too long compared to the 22-month observing window of the GDR2. Its influence will be limited to a shift of the PM vector of star A and not induce a noise on the astrometric fit. This is even truer as the planet position was close to conjunction at epoch 2015.5, meaning that the reflex motion of $\beta$~Pic was very close to linear uniform.
So the presence of excess noise could be interpreted in terms of the presence of a second planet with an orbital period on the order of $\delta t_\mathrm{G2}$. However, due to the difficulty to measure saturated star images, instrumental origin of the noise cannot be excluded.

\subsubsection{Hipparcos astrometry}

Following an approach similar to ours, \citetads{2018NatAs.tmp..114S} determined from a combined analysis of the Hipparcos epoch astrometry and GDR2 position that the mass of the planet $\beta$~Pic b is $m_b = 11 \pm 2\,M_J$.
\citetads{2018NatAs.tmp..114S} determined the PMa of $\beta$~Pic by subtracting the Hip-GDR2 mean PM vector that they computed from the Hip2 epoch astrometry positions of $\beta$~Pic.
We note the excellent agreement of our $\vec{\mu}_\mathrm{HG}$ mean PM vector with their determination.
With a comparable analysis, based on a combination of the Hip1 and Hip2 catalogs, \citetads{2019ApJ...871L...4D} determined a slightly higher mass of $m_b = 13 \pm 3\,M_J$.
We estimated the shift of the GDR2 position of $\beta$~Pic relative to Hip due to the reflex orbital motion of planet b to $[d\alpha,d\delta] = [+1.3,+0.8]$\,mas (for $m_b = 11\,M_J$), in agreement with \citetads{2018NatAs.tmp..114S}. We have, however, neglected this position shift for homogeneity with our other targets.

Adopting the orbital parameters from \citetads{2016AJ....152...97W}, a mass of $m_A = 1.70\,M_\odot$ for the primary and the mass $m_b=11\,M_J$ determined by \citetads{2018NatAs.tmp..114S}, we expect a tangential reflex orbital velocity of $\Delta v_\mathrm{tan,A} = 73.5$\,m\,s$^{-1}$ at a position angle $\theta = 211.3\,\deg$ for $\beta$~Pic~A in average over the observing period of Hipparcos (1989.85 to 1993.21).
If we adopt the mass of $m_b = 12.9\,M_J$ determined by \citetads{2017AJ....153..182C} from spectroscopy of the planet, the predicted reflex velocity becomes $\Delta v_\mathrm{tan,A} = 86.1$\,m\,s$^{-1}$.
The tangential velocity anomaly that we obtain from the Hip2 catalog PM vector corresponds to $\Delta v_\mathrm{tan,A} = 82.7 \pm 17.5$\,m\,s$^{-1}$, within $0.5\sigma$ with the predicted velocity using both mass values for planet b.
The position angle of the PMa vector is $\theta = 199.1 \pm 8.1\,\deg$ is also similar to the predicted value within $1.5\sigma$.
At the known orbital radius of planet b ($r=9.7$\,au), we obtain a mass range of $m_2 = 13.7^{+6}_{-5}\,M_J$.
%, ignoring the orbital phase of the planet.

\begin{figure*}
\centering
\includegraphics[width=9cm]{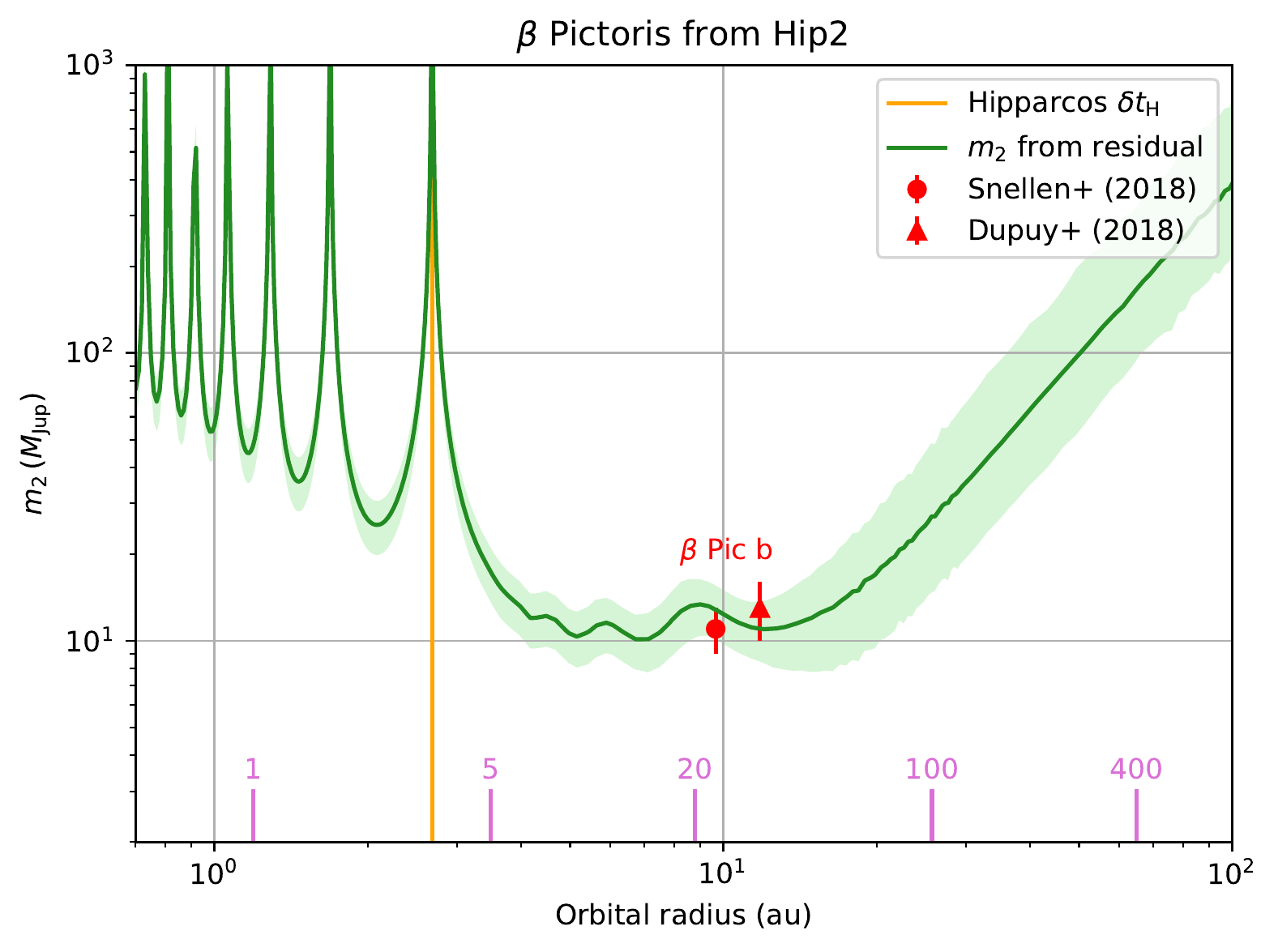}
\includegraphics[width=9cm]{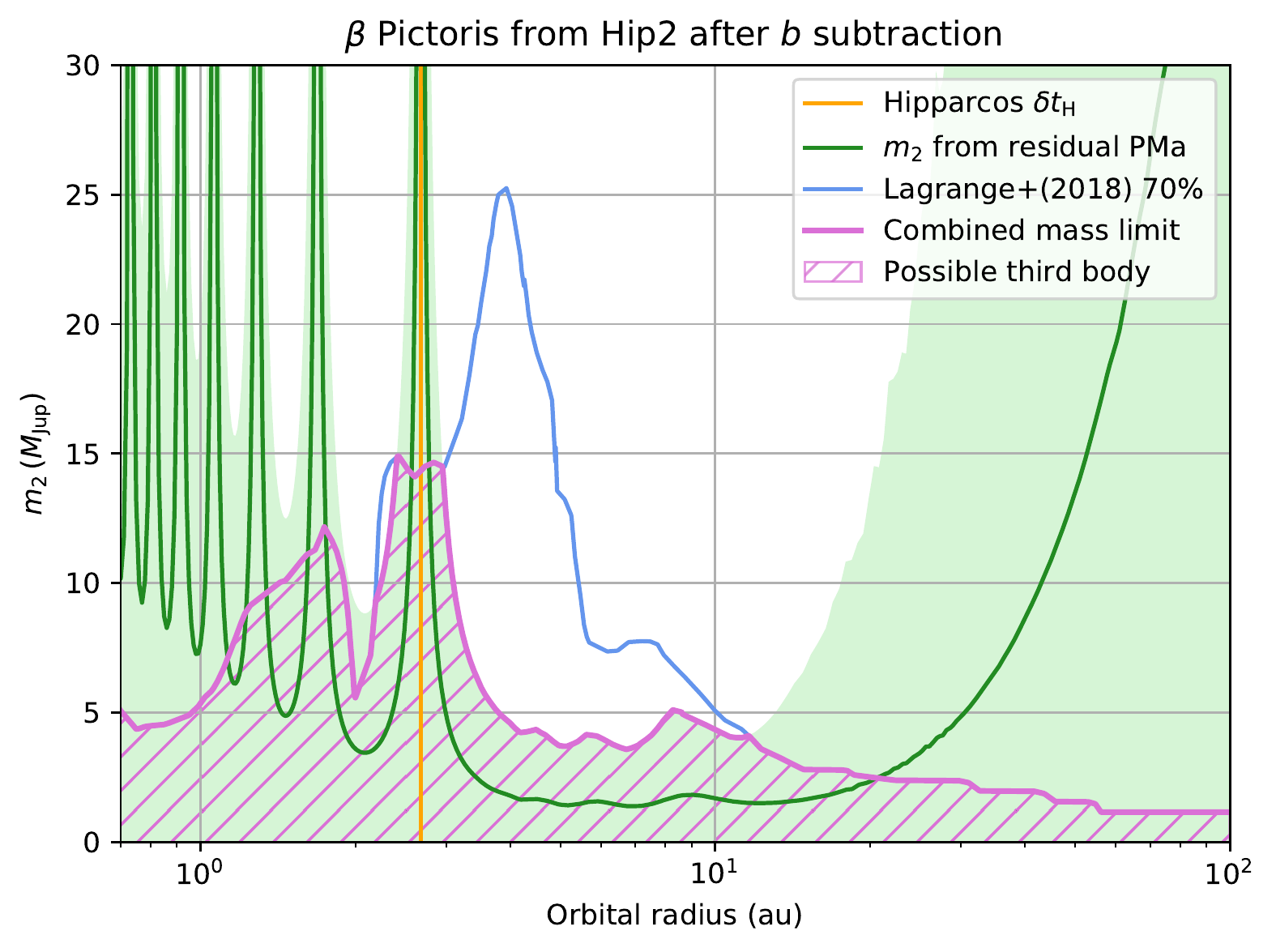}
\caption{\textit{Left:} Possible $(m_2,r)$ combinations for companions of $\beta$~Pictoris, compared to the properties of the known exoplanet $\beta$~Pic b from \citetads{2018NatAs.tmp..114S} and \citetads{2019ApJ...871L...4D} (red symbols). \textit{Right:} Mass limits on additional planetary mass companions from its PMa after subtraction of planet b (green curve and shaded green area), and the analysis by \citetads{2018A&A...612A.108L} (light blue curve). The combined mass limit is shown as a pink curve, and the permitted $(m_3,r)$ range for a second planet is represented as the hatched area. \label{bet-Pic-m2r}}
\end{figure*}

The left panel of Fig.~\ref{bet-Pic-m2r} shows the $(m_2,r)$ domain that we predict for planet b from the PMa of Hipparcos, compared to the masses of $\beta$~Pic~b derived by \citetads{2018NatAs.tmp..114S} and \citetads{2019ApJ...871L...4D} (red symbols).
Although the agreement is good between the observed Hip2 PMa and the expected value from $\beta$~Pic b, we can test for the presence of a residual after subtraction of its contribution (assuming a mass of $11\,M_J$).
The range of possible planet-radius combinations $(m_2,r)$ corresponding to the residual PMa after the vector subtraction of the contribution of $\beta$~Pic b is represented in the right panel of Fig.~\ref{bet-Pic-m2r}.
We also display the planetary mass limits summarized by \citetads{2018A&A...612A.108L}, and the combined limit with our PMa analysis.
The permitted properties of an additional planet in the $\beta$~Pic system are represented by the hatched area in Fig.~\ref{bet-Pic-m2r}, and exclude in particular a planet more massive than $m_3 = 5\,M_J$ orbiting beyond 4\,au ($1\sigma$).

%______________ Figure

%__________________________________ Conclusion
\section{Conclusion\label{conclusion}}

The combination of the Hipparcos and Gaia DR2 positions provides extremely accurate long-term PM vectors for a large number of stars.
The improved parameters from the Gaia DR2 (position, parallax, and PM) confirm that Proxima is gravitationally bound to $\alpha$ Cen and the orbital parameters determined by \citetads{2017A&A...598L...7K}.
A divergence of the instantaneous PM vector of a star with respect to this long-term trend can be interpreted in terms of the presence of an orbiting massive companion. Thanks to the generally excellent accuracy of the GDR2 PM vectors, the sensitivity of the PMa indicator to the presence of companions goes well into the planetary mass regime.
As the sensitivity of the PMa is a linear function of the distance, it is most discriminating for the nearest stars, with for example, mass limits below Saturn's mass for the nearest red dwarf Proxima.
We confirm the ubiquity of substellar mass companions, and we find an emerging bimodal distribution in the distribution of the companion masses (Fig.~\ref{GaiaDR2_50pc-m1m2}).
The present work represents a first exploration of the possibilities offered by Gaia on the detection of low mass companions.
The availability of epoch astrometry in the future Gaia data releases will enable a much more detailed survey of the physical properties of the companions of stellar and substellar mass orbiting nearby stars.

%__________________________________ Acknowledgements
\begin{acknowledgements}
We thank Dr Vincent Coud\'e du Foresto and Dr Timothy D. Brandt for fruitful discussions that led to improvements of the present paper.
This work has made use of data from the European Space Agency (ESA) mission {\it Gaia} (\url{http://www.cosmos.esa.int/gaia}), processed by the {\it Gaia} Data Processing and Analysis Consortium (DPAC, \url{http://www.cosmos.esa.int/web/gaia/dpac/consortium}).
Funding for the DPAC has been provided by national institutions, in particular the institutions participating in the {\it Gaia} Multilateral Agreement.
This research made use of Astropy\footnote{Available at \url{http://www.astropy.org/}}, a community-developed core Python package for Astronomy \citepads{2013A&A...558A..33A,2018AJ....156..123A}.
This research has made use of the Washington Double Star Catalog maintained at the U.S. Naval Observatory.
We used the SIMBAD and VIZIER databases and catalog access tool at the CDS, Strasbourg (France), and NASA's Astrophysics Data System Bibliographic Services.
The original description of the VizieR service was published in \citetads{2000A&AS..143...23O}.
The Digitized Sky Surveys were produced at the Space Telescope Science Institute under U.S. Government grant NAG W-2166. The images of these surveys are based on photographic data obtained using the Oschin Schmidt Telescope on Palomar Mountain and the UK Schmidt Telescope. The plates were processed into the present compressed digital form with the permission of these institutions.
\end{acknowledgements}

%__________________________________Bibliography
\bibliographystyle{aa} % style aa.bst
\bibliography{biblioAlfCen}

\begin{appendix}

%__________________________________ Individual
\section{Additional notes on individual targets\label{additional_individual_notes}}

\subsection{Resolved binary stars\label{binary-individuals}}

\subsubsection{HD 173739+HD 173740 (GJ 725 AB)}

\begin{figure}
\centering
\includegraphics[width=\hsize]{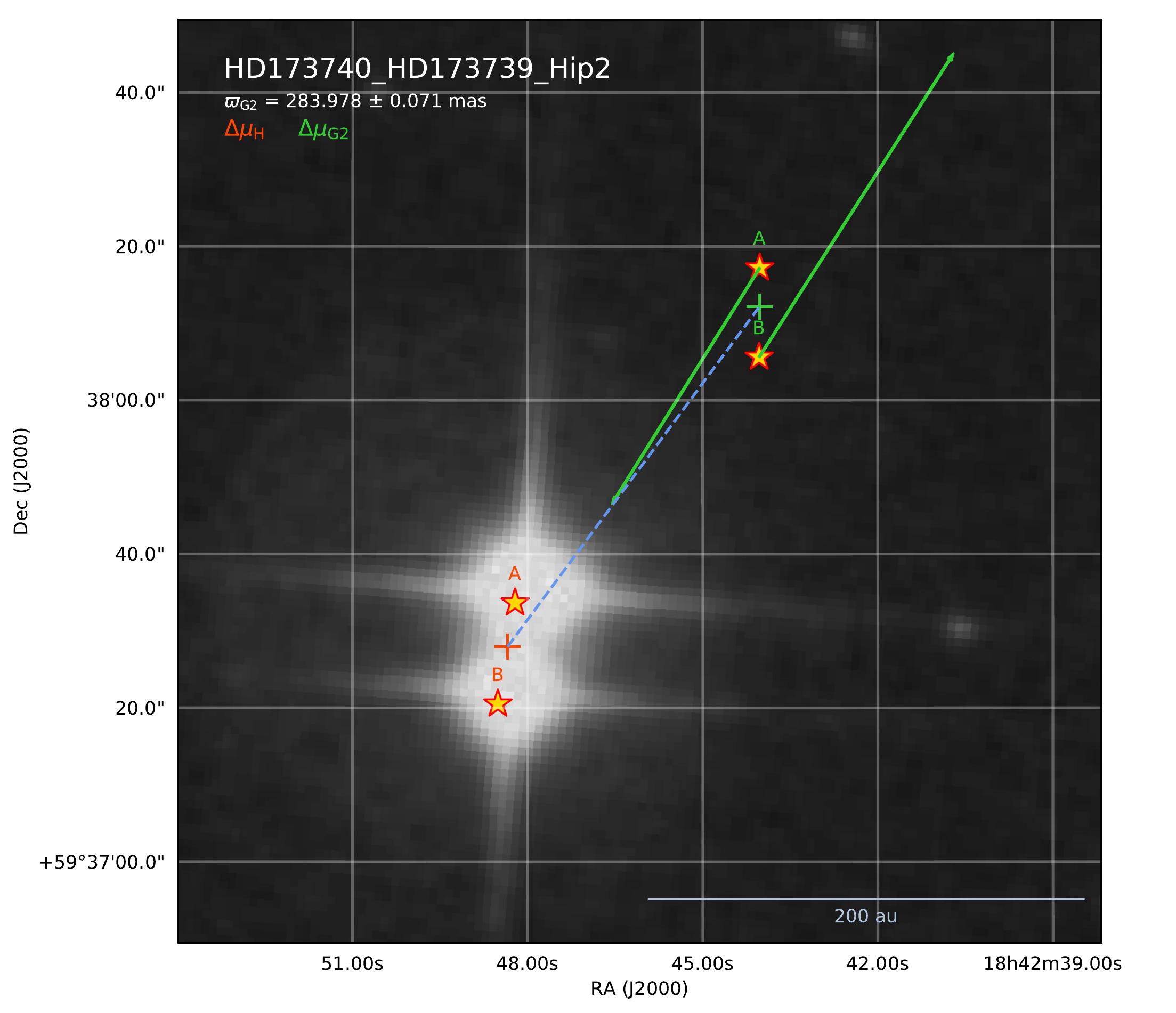}
\caption{Positions of the components of the binary star HD 173740+HD 173739 (M3V+M3.5V) at the Hipparcos and GDR2 epochs.
The proper motion vectors of the two stars are affected by a bias in the Hipparcos catalog and are not represented.
The other symbols are the same as in Fig.~\ref{61CygABfig}. \label{GJ725ABfig}}
\end{figure}

 \object{HD 173739}+\object{HD 173740} (M3V+M3.5V; \object{ADS 11632 AB}, \object{GJ 725} AB, \object{Struve 2398 AB}) is a very low mass binary. For this system (Fig.~\ref{GJ725ABfig}), we derive precisely the same mass ratio $m_B/m_A = 0.767 \pm 0.009$ as the adopted $m_B/m_A = 0.766 \pm 0.022$ from the relations by \citetads{0004-637X-804-1-64}.
We also detect no significant difference in position angle for the PMa from the GDR2 catalog ($\Delta \theta_{AB} = 0.63 \pm 0.47 \deg$). In this case, we do not therefore need to invoke the presence of a third body in the system.
%
%______________ Figure

\subsubsection{HD 79210+HD 79211 (GJ 338 AB)}

\begin{figure}
\centering
\includegraphics[width=\hsize]{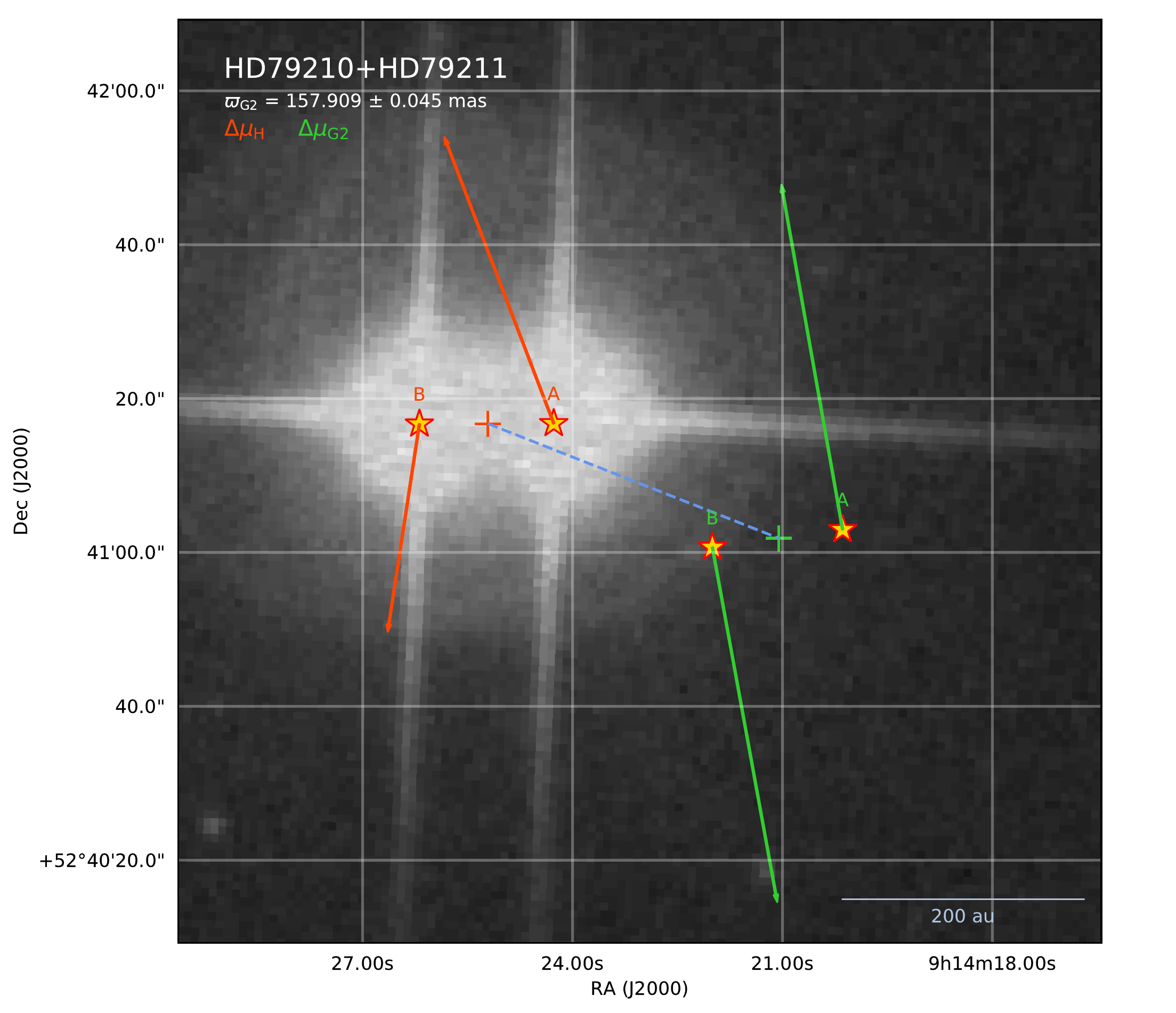}
\caption{Positions of the components of the binary star HD 79210+HD 79211 (M3V+M3.5V) at the Hipparcos and GDR2 epochs.
The symbols are the same as in Fig.~\ref{61CygABfig}. \label{GJ338ABfig}}
\end{figure}

 \object{GJ 338} AB is a binary composed of a K7V primary (\object{HD 79210}) and M0V secondary (\object{HD 79211}). As shown in Fig.~\ref{GJ338ABfig}, the Hip2 catalog has large uncertainties, and the PM vectors of the two components appear discrepant on the figure. However, we retrieve from the GDR2 PMa the same mass ratio $m_B/m_A = 0.970 \pm 0.026$ within $1\sigma$ as the adopted $m_B/m_A =  0.959 \pm 0.027$ from the relations by \citetads{0004-637X-804-1-64}. We also detect no significant offset of the PMa vectors of both components from the GDR2 catalog ($\Delta \theta_{AB} = -0.46 \pm 1.47\,\deg$), and therefore do not detect a third component in the system.
%
%______________ Figure

\subsection{Barnard's star\label{barnard}}

\object{Barnard's star} (\object{GJ 699}, \object{HIP 87937}) is an M4V red dwarf similar in physical properties to Proxima. It is particularly remarkable due to its extremely fast PM of more than 10\,arcsec\,a$^{-1}$.
Several attempts to detect planetary companions to Barnard's star have been conducted either by imaging \citepads{2015MNRAS.452.1677G}, radial velocity \citepads{2013ApJ...764..131C}, or astrometry with the HST-FGS \citepads{1999AJ....118.1086B}.
The most stringent limits are set by the radial velocity technique with a maximum mass of $10\,M_\oplus$ at 1\,au \citepads{2013ApJ...764..131C}.
\citetads{Ribas:2018aa} recently pushed this detection limit even further, as they reported the detection of an exoplanet orbiting \object{Barnard's star} with a period of 233\,days and a minimum mass of $3.2\,M_\oplus$.

\begin{figure}
\centering
\includegraphics[width=9cm]{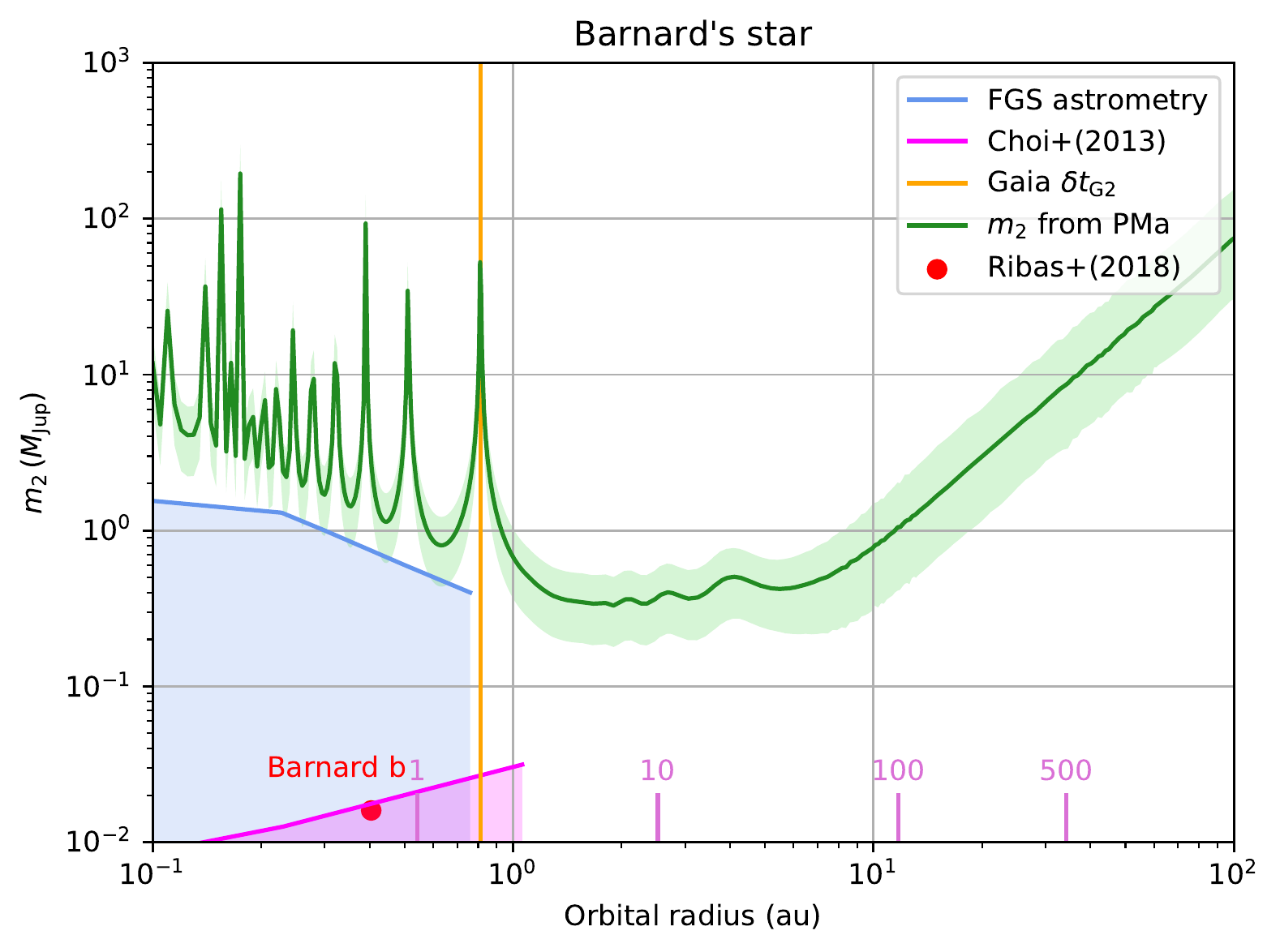}
\caption{Possible companion $(m_2,r)$ combinations for Barnard's star.
The exoplanet Barnard's star b discovered by \citetads{Ribas:2018aa} is shown by a red dot in the left panel.
The permitted range of planetary mass on short-period orbits from radial velocity \citepads{2013ApJ...764..131C} and HST-FGS astrometry \citepads{1999AJ....118.1086B} are represented respectively as shaded pink and blue areas.\label{barnard-m2r}}
\end{figure}

We detect a tangential velocity anomaly of $\Delta v_\mathrm{tan,G2} = 13.5 \pm 6.9$\,m\,s$^{-1}$ at GDR2 epoch (Table~\ref{phys-properties}) significant at a $2.0\sigma$ level. The corresponding $(m_2,r)$  combinations are shown in Fig.~\ref{barnard-m2r}, where the mass limits from radial velocity and FGS astrometry are also represented.
The detected PMa is only marginally significant, and so we do not claim a detection.
This PMa signal could be explained by a $m_2\approx 0.5$ to $1\,M_J$ planet orbiting at $r=1-10$\,au, or a more massive giant planet of $m_2 \approx 3\,M_J$ up to $r=20$\,au.
As shown in Fig.~\ref{barnard-m2r}, the telluric planet detected by \citetads{Ribas:2018aa} is far beyond reach of our PMa search technique by approximately two orders of magnitude.
\citetads{2018arXiv181105920T} studied the feasibility of the astrometric detection of Barnard's star b, and conclude that it will very likely not be detected by Gaia.
However, the future Gaia data releases will make it possible to test the possible presence of a massive giant planet on a long period orbit more stringently.

As for Proxima (Sect.~\ref{proxima}), we also tested the FGS astrometric measurement of Barnard's star obtained by \citetads{1999AJ....118.1086B} for the presence of a PMa.
We obtain a  significant signature at a  signal-to-noise ratio of $\Delta_\mathrm{FGS} = 3.2$, but the reliability of this PM vector is uncertain as the FGS measurement is based on differential astrometry with background stars whose PM vectors were poorly constrained at the time of the original data reduction.
The parallax of $\varpi_\mathrm{FGS}= 545.4 \pm 0.3$\,mas found by \citetads{1999AJ....118.1086B} differs from the GDR2 value ($\varpi_\mathrm{G2} = 547.48 \pm 0.31$\,mas) by $5\sigma$, and also from the Hipparcos parallax ($\varpi_\mathrm{H} = 548.3 \pm 1.5$\,mas). This may indicate a problem in the FGS astrometric solution for  Barnard's star, which could also affect the FGS PM vector.

%______________ Figure

\subsection{Ross 128 \label{ross128}}

\object{Ross 128} (\object{GJ 447}, \object{HIP 57548}) is a nearby red dwarf of spectral type M4V, whose physical properties are very similar to Barnard's star. \citetads{2018A&A...613A..25B} recently reported the discovery of the telluric exoplanet \object{Ross 128 b}, whose mass is estimated to $m\,\sin i = 1.35\,M_\oplus$. It is particularly interesting for exobiology as its orbital period of 9.9\,days places it in the temperate zone of its parent star.
Additionally, Ross 128 is significantly less active than \object{Proxima}, which is generally agreed in the literature to be favorable for the habitability of the planet.

\begin{figure}
\centering
\includegraphics[width=\hsize]{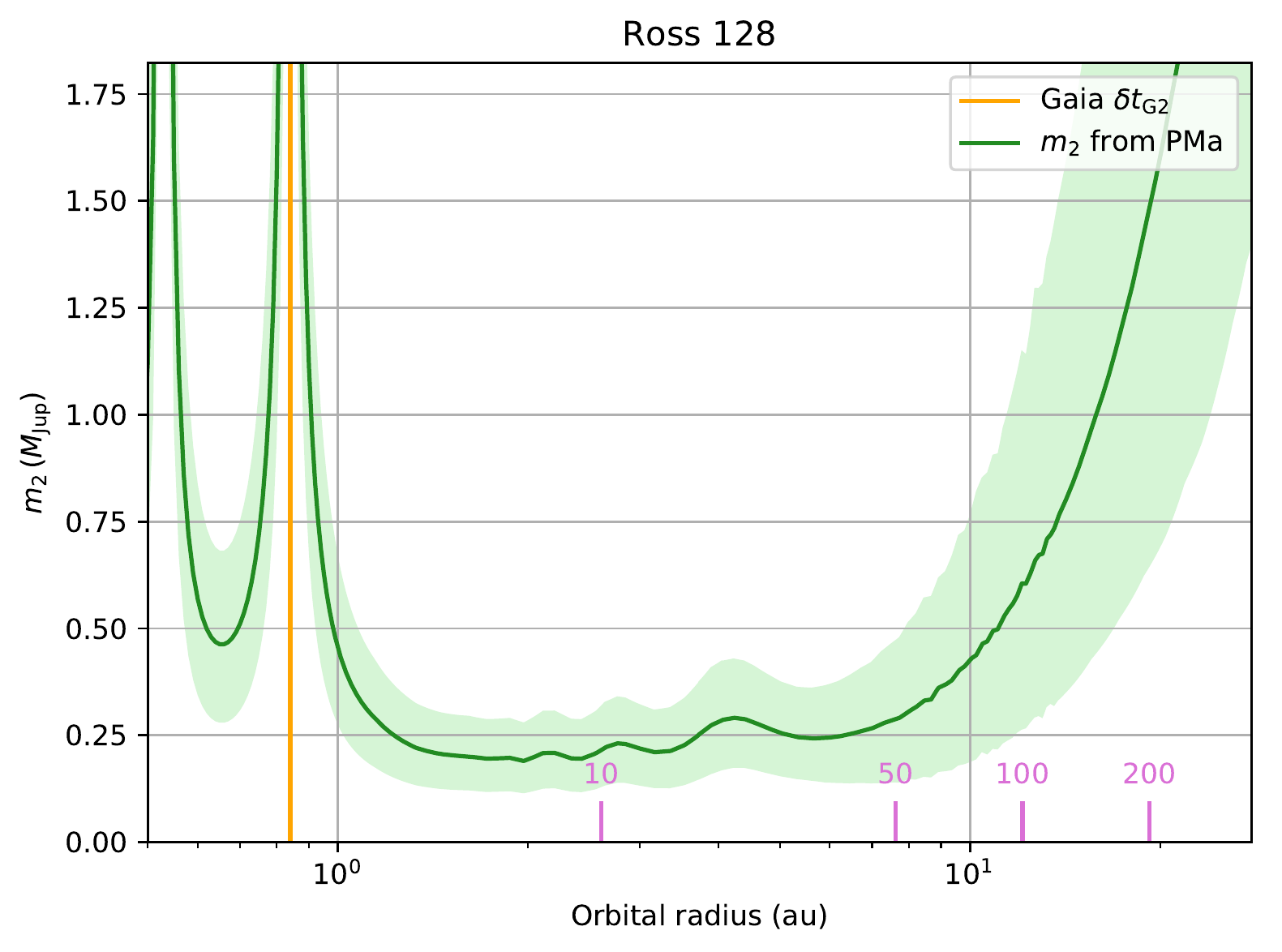}
\caption{Possible $(m_2,r)$ combinations for a companion of the red dwarf Ross 128. \label{Ross-128-m2r}}
\end{figure}

We observe a marginally significant PMa of $\Delta_\mathrm{G2} = 2.2$ on Ross 128 at the GDR2 epoch.
The reflex motion induced by the telluric planet Ross 128 b cannot be the cause of this PMa, due to its very low mass and short orbital period.
The range of possible $(m_2,r)$ combinations for an additional planet is shown in Fig.~\ref{Ross-128-m2r}.
A plausible set of parameters explaining the observed PMa would be a Saturn mass planet orbiting between 1 and 10\,au from Ross 128.
We set an upper limit of $1\,M_J$ to a possible planet orbiting within 10\,au.

%______________ Figure

\subsection{$\epsilon$ Ind}

\begin{figure}
\centering
\includegraphics[width=\hsize]{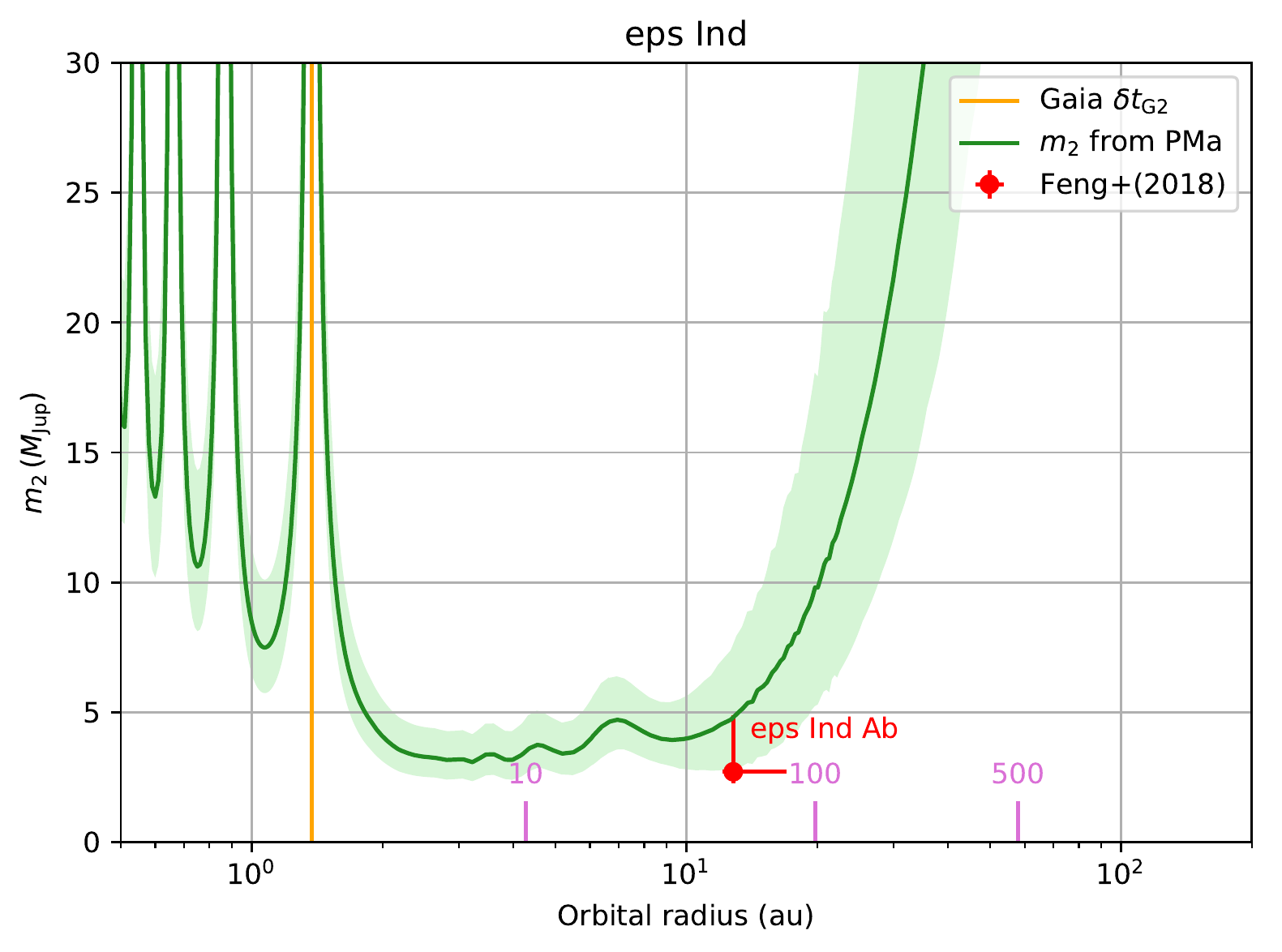}
\caption{Possible companion $(m_2,r)$ combinations for $\epsilon$ Ind A, with the Jovian planet discovered by \citetads{2018arXiv180308163F} marked with a red dot.\label{epsind-m2r}}
\end{figure}

\object{$\epsilon$ Ind} A (\object{GJ 845} A, \object{HD 209100}) is a nearby K5V dwarf forming a triple system with the binary brown dwarf \object{$\epsilon$ Ind B} \citepads{2004A&A...413.1029M, 2010A&A...510A..99K}.
The presence of a dark companion to $\epsilon$ Ind A was suspected based on a radial velocity trend, but an imaging search by \citetads{2009MNRAS.399..377J} did not reveal the unseen companion.
A Jovian mass exoplanet was recently discovered by \citetads{2018arXiv180308163F} from radial velocity measurements, with an estimated minimum mass of $2.7^{+2.2}_{-0.4}\,M_J$ and a semi-major axis of $12.8^{+4.2}_{-0.7}$\,au.
We detect a significant PMa on $\epsilon$ Ind A of $\Delta v_\mathrm{tan,G2} = 44.0 \pm 10.5$\,m\,s$^{-1}$ at the GDR2 epoch. As shown in Fig.~\ref{epsind-m2r}, this tangential velocity anomaly is compatible with the expected contribution from the orbiting planet considering the error bars, although we observe a slightly higher anomaly than predicted from planet Ab alone.

%______________ Figure

The binary brown dwarf $\epsilon$ Ind Ba+Bb is located at a projected separation of 1460\,au \citepads{2010AJ....139..176F}, and has a total mass of $121 \pm 1\,M_J$ \citepads{2010A&A...510A..99K}. Its contribution in terms of tangential velocity on $\epsilon$\,Ind A can be estimated to $\approx 50$\,m\,s$^{-1}$, that is, comparable to the observed $\Delta v_\mathrm{tan,G2}$. $\epsilon$\,Ind B is unfortunately too faint to be detected by Gaia and Hipparcos, so its PMa cannot be determined with a comparable accuracy to that of A.
As the PMa contributions of $\epsilon$ Ind Ab and B sum up vectorially to produce the total observed PMa of $\epsilon$ Ind A, it is difficult to disentangle the origin of the observed signal.

\subsection{Kapteyn's star\label{kapteyn}}

The red dwarf \object{Kapteyn's star} (\object{GJ 191}, \object{HD 33793}) is the nearest star from the halo population.
This property is reflected in its very high tangential PM ($\mu=8644$\,mas\,a$^{-1}$) and space velocity relative to the Sun ($v = 293$\,km\,s$^{-1}$).
Its radial velocity reaches $v_\mathrm{r} = +245$\,km\,s$^{-1}$ \citepads{2002ApJS..141..503N}.
Kapteyn's star is suspected to host two telluric mass planets, that were announced by \citetads{2014MNRAS.443L..89A} from the analysis of radial velocity measurements. Their respective masses are estimated to 5 and $7\,M_\oplus$, at orbital radii of 0.17 and 0.31\,au.
However, \citetads{2015ApJ...805L..22R} noticed a 1:3 commensurability of the inner planet orbital period ($P_b = 48$\,d) and the star's rotation period ($P_\mathrm{rot} = 143$\,d), and showed that the 48\,d period signal is correlated with stellar activity.
This indicates that the observed radial velocity signal, at least for the inner candidate planet, may not be caused by Doppler reflex motion from an orbiting planet (see also \citeads{2016ApJ...821L..19N}).

\begin{figure}
\centering
\includegraphics[width=\hsize]{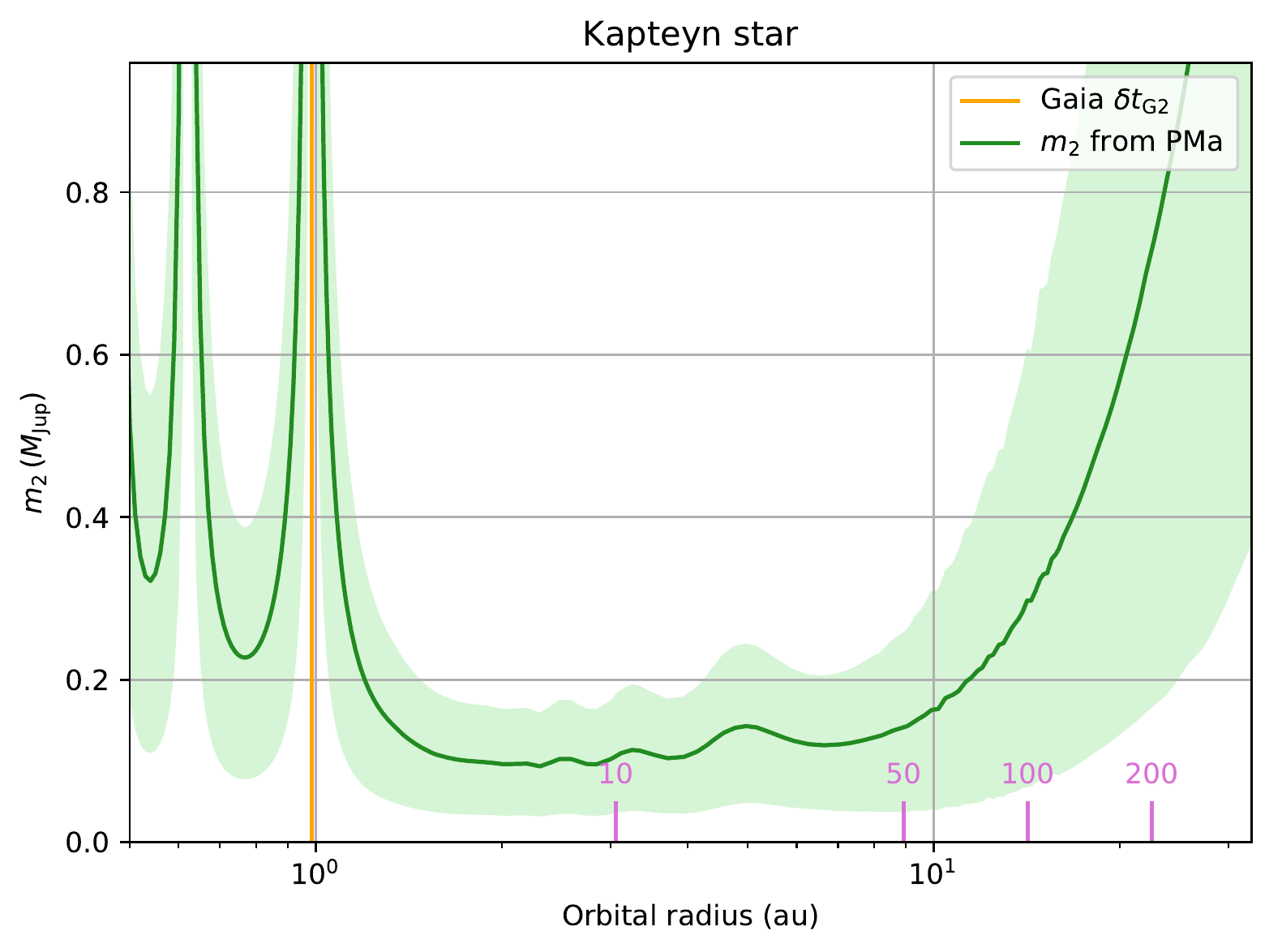}
\caption{Possible companion $(m_2,r)$ combinations for Kapteyn's star (GJ 191).\label{Kapteyn-star-m2r}}
\end{figure}

Kapteyn's star is a stringent test case for the determination of the PMa, due to its extremely fast space velocity.
We determine a tangential velocity anomaly of $\Delta v_\mathrm{tan,G2} = 3.2 \pm 2.0$\,m\,s$^{-1}$ at the GDR2 epoch, corresponding to a signal-to-noise ratio $\Delta_\mathrm{G2}=1.7$.
We present in Fig.~\ref{Kapteyn-star-m2r} the domain of $(m_2,r)$ combinations for a companion that would explain this tangential velocity anomaly.
We determine that no planetary companion more massive than Saturn ($M_S = 0.3\,M_J$) is present around Kapteyn's star between the orbital radii of 1.5 and 10\,au.

%______________ Figure

\subsection{AX Mic}

\object{AX Mic} (\object{Lacaille 8760}, \object{GJ 825}) is a low mass red dwarf of spectral type M1V, with no known planet. \citetads{2018MNRAS.476.5408M} predict that the stellar activity will induce an astrometric position jitter of $\approx 20\,\mu$as for AX Mic, whose effect is negligible on the present PMa analysis.

\begin{figure}
\centering
\includegraphics[width=\hsize]{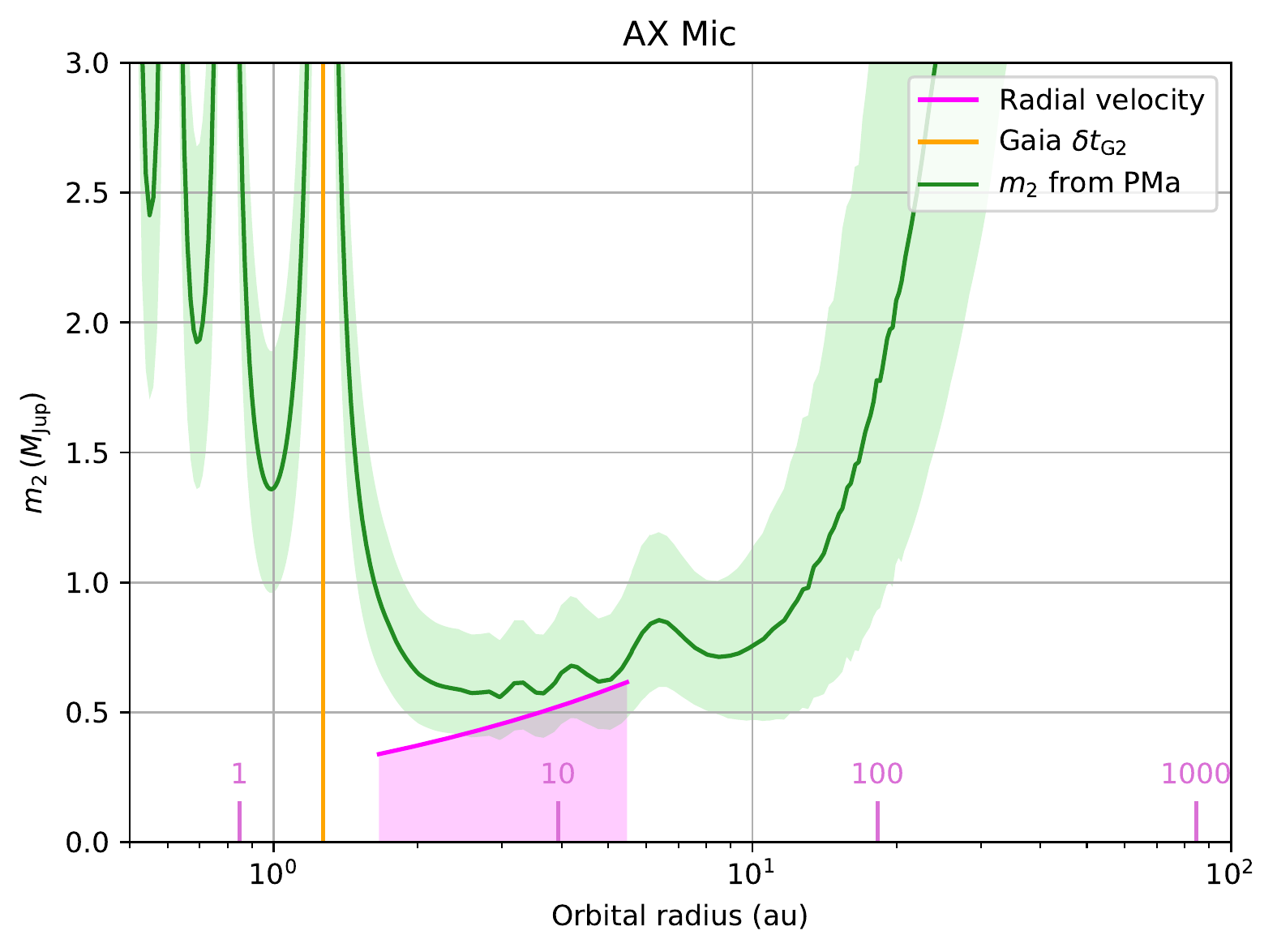}
\caption{Possible companion $(m_2,r)$ combinations for AX-Mic (GJ 825).
The shaded pink area corresponds to the range of possible planets from \citetads{2016ApJ...819...28W} with periods between 1000 and 6000\,days. \label{AX-Mic-m2r}}
\end{figure}

The tangential velocity anomaly of $\Delta v_\mathrm{tan,G2} = 9.8 \pm 2.9$\,m\,s$^{-1}$ (signal-to-noise ratio $\Delta_\mathrm{G2} = 3.4$) that we measure at the GDR2 epoch corresponds to possible companions in the $(m_2,r)$ range shaded in green in Fig.~\ref{AX-Mic-m2r}.
\citetads{2016ApJ...819...28W} established an upper limit to the mean velocity amplitude of $K_\mathrm{max} = 9.2 \pm 0.6$\,m\,s$^{-1}$ for orbital periods between 1000 and 6000\,days (70\% recovery rate).
This limit translates into a permitted $(m_2,r)$ domain shaded in pink in Fig.~\ref{AX-Mic-m2r}.
From the combination of the limits from radial velocity and astrometry, the companion of AX Mic is possibly a giant planet with a mass $m_2 = 0.5$ to $2.5\,M_J$ orbiting between 3 and 10\,au from the red dwarf.

%______________ Figure

\subsection{Ross 614 \label{ross614}}

%______________ Figure
\begin{figure}
\centering
\includegraphics[width=\hsize]{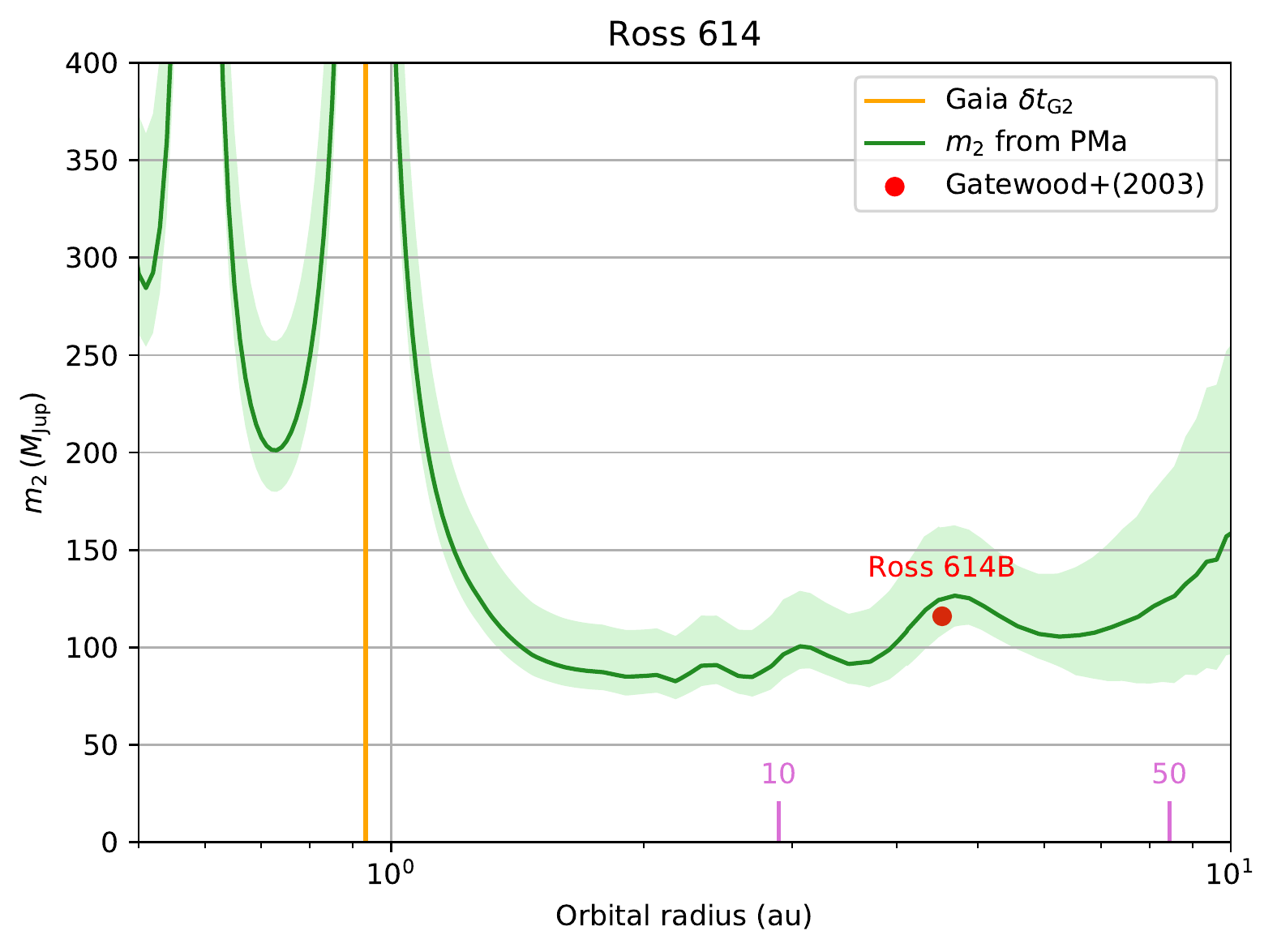}
\caption{Possible companion $(m_2,r)$ combinations for Ross 614.\label{ross614-m2r}}
\end{figure}

\object{Ross 614} (\object{V577 Mon}, \object{GJ 234 AB}) is a pair of very low mass red dwarfs whose fast PM was discovered by \citetads{1927AJ.....37..193R}. Its binarity was the first to be identified by photographic astrometry by \citetads{1936AJ.....45..133R}.
From a combination of historical and modern astrometry, \citetads{2003AJ....125.1530G} determined the period of the system ($P=16.595 \pm  0.0077$\,years), its parallax ($\varpi = 244.07 \pm 0.73$\,mas) and the masses of the two stars ($m_1 = 0.2228 \pm 0.0055\,M_\odot$ and $m_2 = 0.1107 \pm 0.0028 M_\odot$).
We note that the GDR2 parallax measurement ($\varpi_\mathrm{G2} = 243.00 \pm 0.88$\,mas), although perturbed by the orbital motion and the unresolved contribution of Ross 614 B, confirms their result.
The semi-major axis is $a = 1101.2 \pm 8.2$\,mas, corresponding to $4.53 \pm 0.03$\,au.
Fig.~\ref{ross614-m2r} presents the range of possible $(m_2,r)$ combinations determined from the PMa of Ross 614. For an orbital radius of 4.53\,au, we determine a mass $m_2 = 0.109^{+0.030}_{-0.012}\,M_\odot$ for the secondary, in remarkable agreement with the actual mass determined by \citetads{2003AJ....125.1530G}.
This very good level of correspondence is due, in particular, to the inclination of the orbital plane ($i=54^\circ$), which is close to the most statistically
probable value of $60^\circ$.
It is also favored by the orbital period of the system, which is near the optimum for the Hip2-GDR2 PMa analysis (Sect.~\ref{inclination}).

\subsection{Wolf 28 (van Maanen's star)\label{wolf28}}

\object{Wolf 28} (\object{van Maanen 2}, \object{GJ 35}, \object{WD 0046+051}) is the third nearest WD after Sirius B and Procyon B, and the nearest single WD. It was discovered in 1917 by A.~van Maanen \citepads{1917PASP...29..258V}, and has a relatively low effective temperature of $T_\mathrm{eff} \approx 6000$\,K \citepads{2008MNRAS.386L...5B}.

The presence of a $m_2 = 0.08\,M_\odot$ companion orbiting around Wolf 28 on a $P\approx 1.5$\,year (maximum apparent separation $\approx 0.3\arcsec$ corresponding to $\approx 1.3$\,au), equator-on orbit ($i=89^\circ$) was announced by \citetads{2004ApJ...600L..71M}, from the analysis of the Hipparcos data.
This mass estimate was computed using a WD mass of $m_1 = 0.83\,M_\odot$ that is likely too high by $\approx 25\%$. The presence of such a companion is however disproved by \citetads{2004ApJ...608L.109F} from adaptive optics imaging.
\citetads{2008MNRAS.386L...5B} confirm this non-detection from deep imaging in the near-infrared $J$ band and established an upper limit of  $m_2 < 7 \pm 1\,M_J$ between 3 and 50\,au from the WD. They also exclude the presence of companions more massive than $10\,M_J$ based on the absence of significant infrared excess in the Spitzer IRAC photometry at $4.5\,\mu$m.

%______________ Figure
\begin{figure}[htbp]
\centering
\includegraphics[width=\hsize]{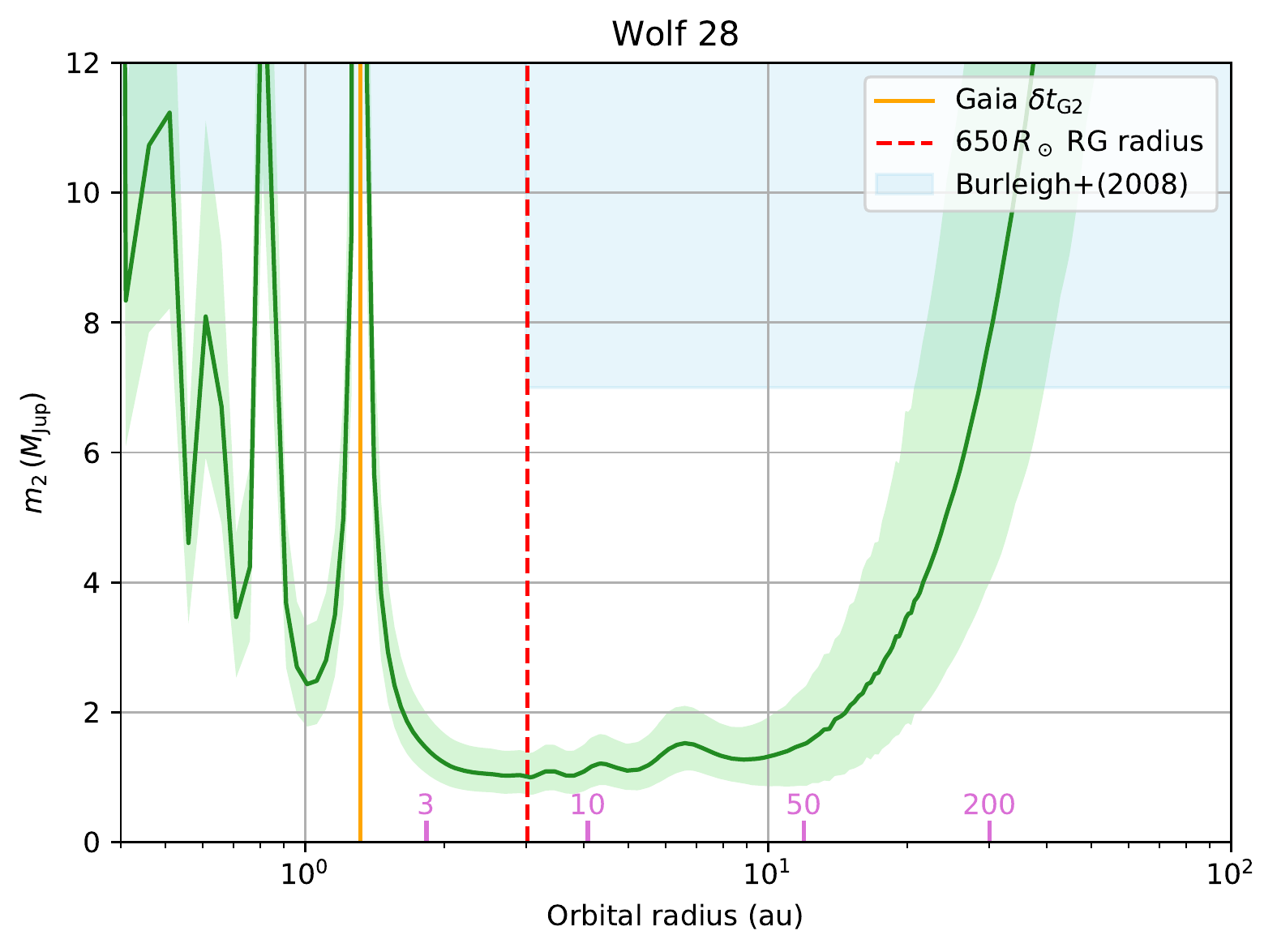}
\caption{Possible companion $(m_2,r)$ combinations for the white dwarf Wolf 28 (van Maanen's star).
Excluded $(m_2,r)$ combinations from \citetads{2008MNRAS.386L...5B} are represented with the shaded blue area.
\label{Wolf-28-m2r}}
\end{figure}

We do not identify any resolved common PM candidate of Wolf 28 in the GDR2 catalog, between a minimum separation of $\approx 0.5\arcsec$ and $10^\circ$, down to the limiting magnitude of the catalog ($G = 21$), corresponding to an absolute magnitude $M_G = 22.8$ at the distance of Wolf 28.
From the grid by \citetads{2013ApJS..208....9P}\footnote{\url{http://www.pas.rochester.edu/~emamajek/EEM_dwarf_UBVIJHK_colors_Teff.txt}} (see also \citeads{2012ApJ...746..154P}), this corresponds to a hottest possible spectral type around L4.5V, and a maximum mass of $\approx 75\,M_J$.
We also searched the 2MASS catalog \citepads{2006AJ....131.1163S} for very red sources close to Wolf 28, but did not find any down to the limiting magnitude of $m_K \approx 15.5$, that corresponds to $m_2< 40\,M_J$.
The conversion of the $K$ band infrared magnitude into mass for planetary mass objects was taken from the AMES-Cond isochrone by \citetads{2012RSPTA.370.2765A}\footnote{\url{http://perso.ens-lyon.fr/france.allard/}} for an age of 4\,Ga.

We detect a marginal PMa in Wolf 28 with a signal-to-noise ratio $\Delta_\mathrm{G2} = 2.0$. The quality of the GDR2 astrometric solution is good with a renormalized unit weight error of $\varrho = 1.2$ and a null excess noise $\epsilon_i$.
Fig.~\ref{Wolf-28-m2r} shows the $(m_2,r)$ domain corresponding to the observed PMa, together with the pre-existing constraints from \citetads{2008MNRAS.386L...5B}  on the possible mass and orbital radius of the companion.
It is unlikely that the orbital radius of the companion is smaller than $\approx 650\,R_\odot$ ($\approx 3$\,au) due to the past expansion of Wolf 28 during its red giant phase up to this radius.
The determined possible range of orbital radii however includes separations larger than the red giant radius.
We exclude the presence of an orbiting companion more massive than $2\,M_J$ between 3 and 10\,au.

\subsection{LAWD 37 (GJ 440) \label{lawd37}}

\object{LAWD 37} (\object{GJ 440}, \object{HIP 57367}, \object{LTT 4364}) is a type C2 white dwarf with a DQ6 spectral type, that is routinely used as a spectrophotometric standard \citepads{1992PASP..104..533H}. It has an estimated mass of $m_1 = 0.61\,M_\odot$.
\citeads{2000AJ....119..906S} observed LAWD 37 using the HST-WFPC2 to search for companions, but the achieved detection limits were relatively high ($30-40\,M_J$ down to $1\arcsec$).
We detect a significant PMa at a signal-to-noise ratio $\Delta_\mathrm{G2}=4.9$, and a corresponding tangential velocity of $\Delta v_\mathrm{tan,G2}=14.1 \pm 2.9$\,m\,s$^{-1}$ at a position angle $\theta = 306 \pm 9\,\deg$. We also examined the Gaia DR1 record of LAWD 37, and observe a comparable tangential velocity anomaly $\Delta v_\mathrm{tan,G1}=11.5 \pm 5.7$\,m\,s$^{-1}$ giving a signal-to-noise ratio $\Delta_\mathrm{G2}=2.0$.

It should be noted, however, that no RV measurement is available in the literature for LAWD 37, and we therefore assumed a zero RV in this computation.
The tangential velocity of LAWD 37 is $v_\mathrm{tan,G2} = 59$\,km\,s$^{-1}$. Assuming a RV range of $v_\mathrm{r} = \pm 60$\,km\,s$^{-1}$, we obtain a tangential velocity anomaly of $\Delta v_\mathrm{tan,G2}=9$ to 31\,m\,s$^{-1}$ at the GDR2 epoch, still significant at a $\Delta_\mathrm{G2} = 3.2$ to 10.4 level.
A spectroscopic measurement of the RV would allow a more accurate computation, but the near absence of spectral features in the spectrum of LAWD 37 makes it a particularly difficult enterprise. \citetads{1999A&A...348.1040D} derived an astrometric RV of $v_\mathrm{r} = +43 \pm 106$\,km\,s$^{-1}$ that corresponds to a tangential velocity anomaly signal-to-noise ratio of $\Delta_\mathrm{G2} = 2.4$.

\begin{figure}
\centering
\includegraphics[width=\hsize]{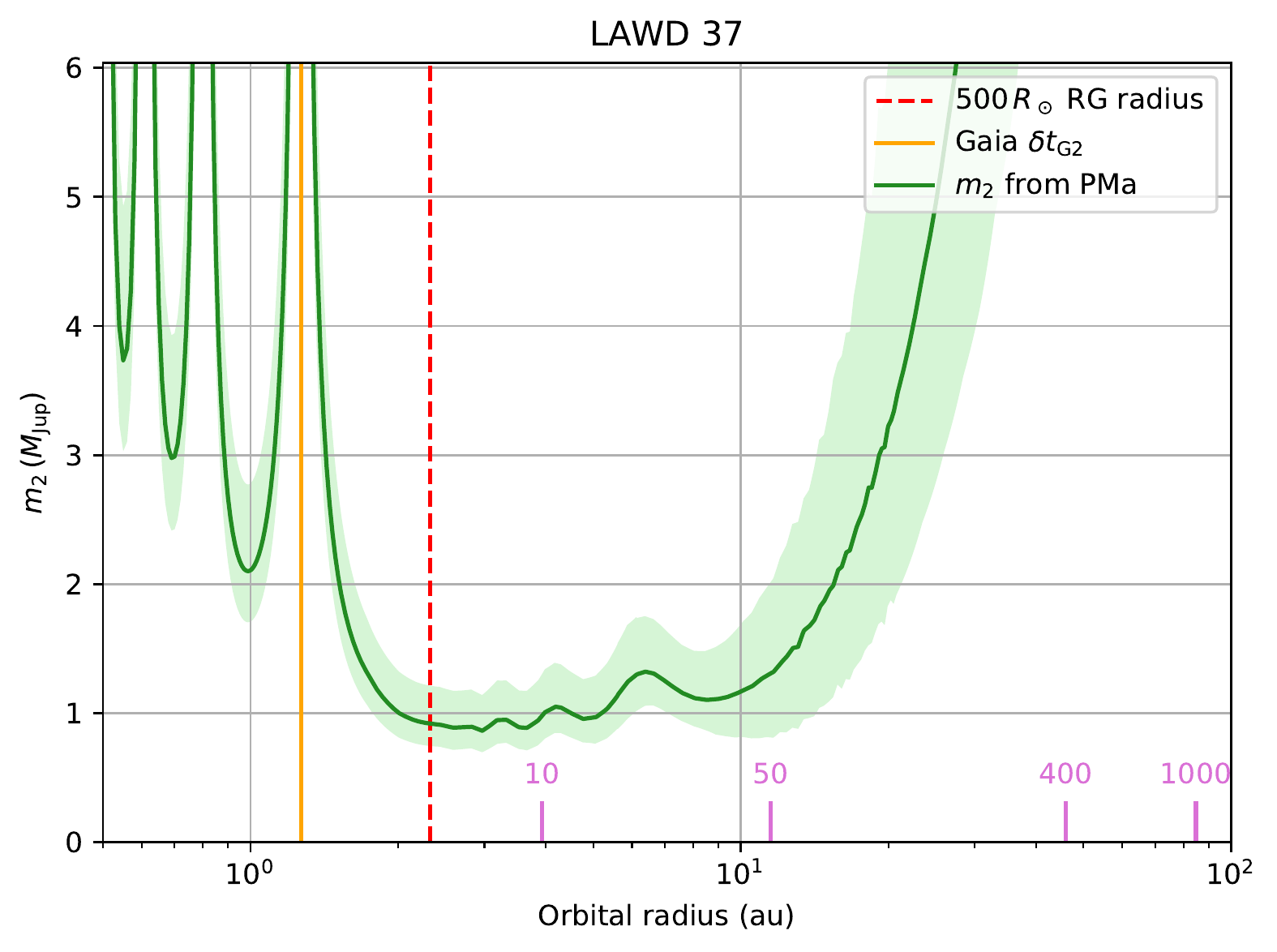}
\caption{Possible companion $(m_2,r)$ combinations for the white dwarf LAWD 37.
\label{LAWD-37-m2r}}
\end{figure}

By analogy with Wolf 28 (Sect.~\ref{wolf28}), the radius of its progenitor while on the red giant branch was on the order of $500\,R_\odot$, therefore setting a minimum orbital radius of $\approx 2.5$\,au. Fig.~\ref{LAWD-37-m2r} shows the range of possible $(m_2,r)$ pairs.
The presence of a companion with a mass of $1-3\,M_J$ orbiting at a radius of $1-20$\,au (that is, an angular separation of $0.2\arcsec-4\arcsec$) would explain the observed signal.
Interestingly, \citetads{2018MNRAS.478L..29M} recently predicted a microlensing event that will occur in November 2019 and involve LAWD 37 as the lens. Based on the GDR2 catalog, \citetads{2018A&A...618A..44B} further predicted that LAWD 37 will create nine microlensing events in the time to 2026.
The observation of these events may confirm the presence of an orbiting planet.

%______________ Figure

\subsection{HD 42581 (GJ 229)}

\begin{figure}
\centering
\includegraphics[width=\hsize]{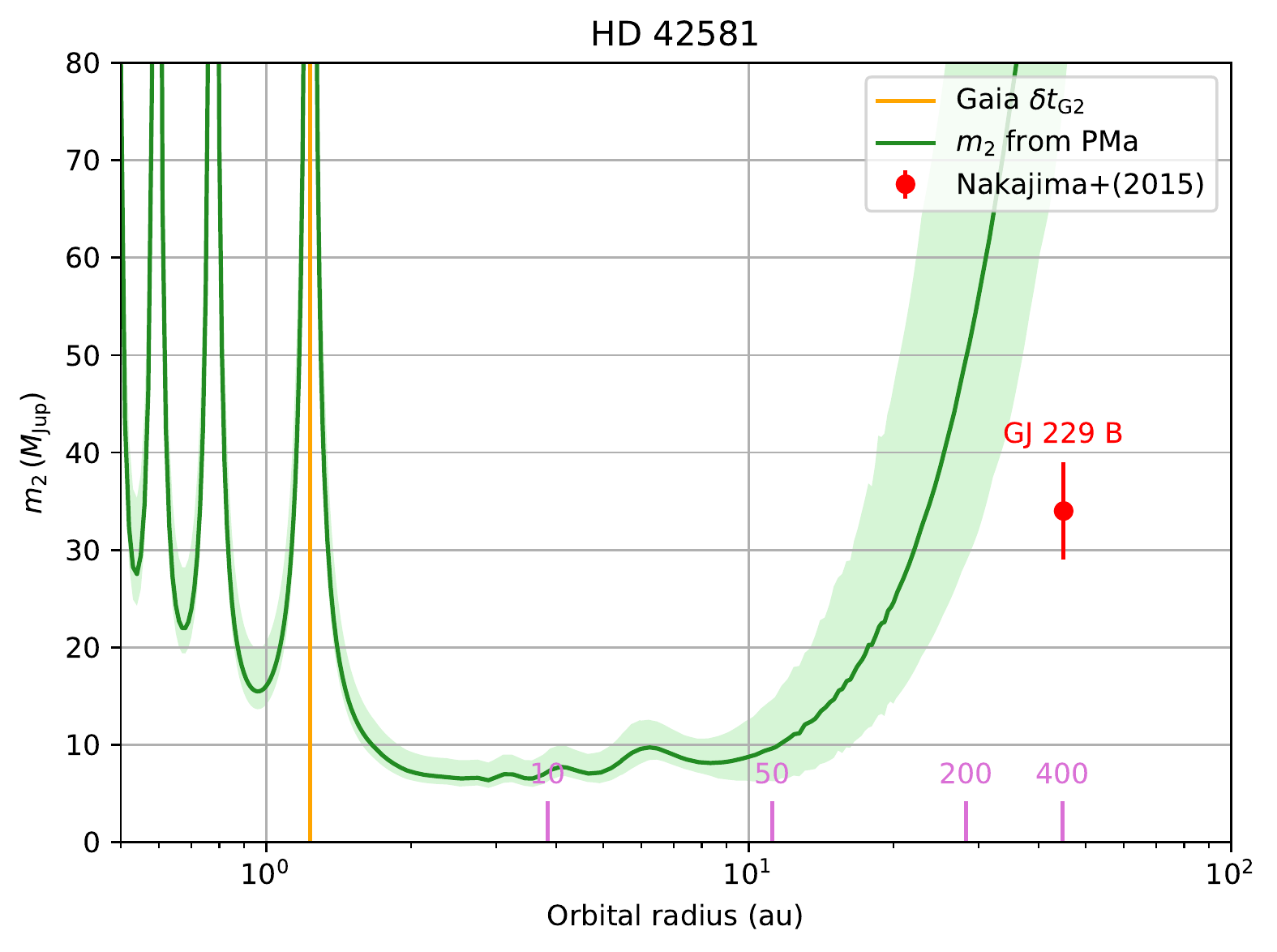}
\caption{Diagram of the possible companion $(m_2,r)$ combinations for the brown dwarf host star HD 42581 (GJ 229 A). The position of the brown dwarf GJ 229 B is shown, adopting its projected separation at discovery of 45\,au.
\label{HD-42581-m2r}}
\end{figure}

\object{HD 42581} (\object{GJ 229}) hosts the brown dwarf \object{GJ 229 B} \citepads{1995Natur.378..463N,1995Sci...270.1478O}. The brown mass is estimated to $29-39\,M_J$ from the two best-fit models of \citetads{2015AJ....150...53N}. Its discovery separation from its parent star is $7.8\arcsec$ from the main star, that is, $r=45$\,au at the distance of HD 42581 ($\varpi_\mathrm{G2} = 173.725 \pm 0.054$\,mas).
Fig.~\ref{HD-42581-m2r} shows the range of $(m_2,r)$ pairs that explain the observed PMa, together with the position of the brown dwarf in this diagram. One unknown of the determination of its parameters is the true value of its orbital semi-major axis. The value we adopt is the measured separation of $r=45$\,au, but it may be larger or smaller depending on the orbital phase and eccentricity.

%______________ Figure

\subsection{e Eri (HD 20794)}

The solar analog \object{e Eri} (\object{GJ 139}, \object{HD 20794}, spectral type G6V) was found by \citetads{2017A&A...605A.103F} to show evidence in radial velocity for the presence of at least three telluric planets with masses of a few times the Earth and orbital radii of 0.1 to 0.5\,au. Indications were also found for the existence of three additional planets, the most distant orbiting at 0.9\,au with a minimum mass around $10\,M_\oplus$.

\begin{figure}
\centering
\includegraphics[width=\hsize]{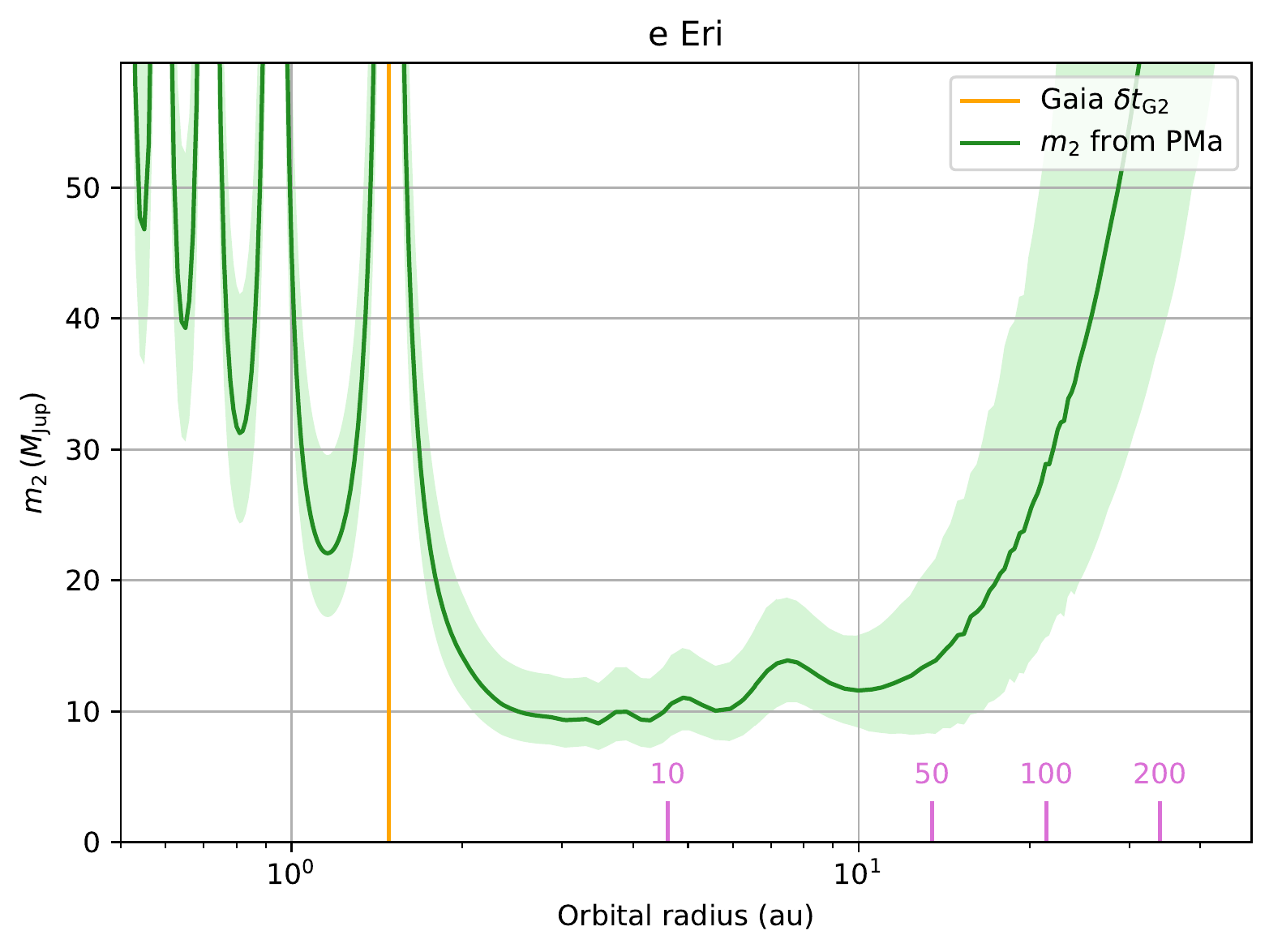}
\caption{Possible companion $(m_2,r)$ combinations for the solar analog e Eri (HD 20794).
\label{e-Eri-m2r}}
\end{figure}

We detect a significant PMa on e Eri, with a residual tangential velocity of $\Delta v_\mathrm{tan,H} = 25.0 \pm 8.3$\,m\,s$^{-1}$ at Hipparcos epoch, and $\Delta v_\mathrm{tan,G2} = 111.0 \pm 24.7$\,m\,s$^{-1}$ at GDR2 epoch. We note however the presence of a significant excess noise in the GDR2 astrometric solution ($\epsilon_i = 1.3$\,mas), although the RUWE is low at $\varrho = 1.0$.
As for $\beta$~Pic (Sect.~\ref{betapic}), the Hip2 data is more precise than the GDR2 PM vector, and we therefore examine the PMa from the Hip2 values. Figure~\ref{e-Eri-m2r}, the residual observed at GDR2 epoch corresponds to a plausible mass range of $10\,M_J$ for orbital radii of 2 to 10\,au. The presence of a massive planet on a wide orbit, in addition to inner telluric planets would make e Eri a promising analog of the solar system.

%______________ Figure

\subsection{TW PsA (Fomalhaut B)}

\citetads{2012ApJ...754L..20M} demonstrate that the K4Ve dwarf \object{TW PsA} (\object{Fomalhaut B}, \object {HD 216803}, \object{GJ 879}) is likely a bound stellar companion of the bright planet host star \object{Fomalhaut A} (\object{$\alpha$ PsA}, \object{GJ 881}). It forms a triple system with the M4V low mass dwarf \object{Fomalhaut C} (\object{LP 876-10}) \citepads{2013AJ....146..154M}.

\begin{figure}
\centering
\includegraphics[width=\hsize]{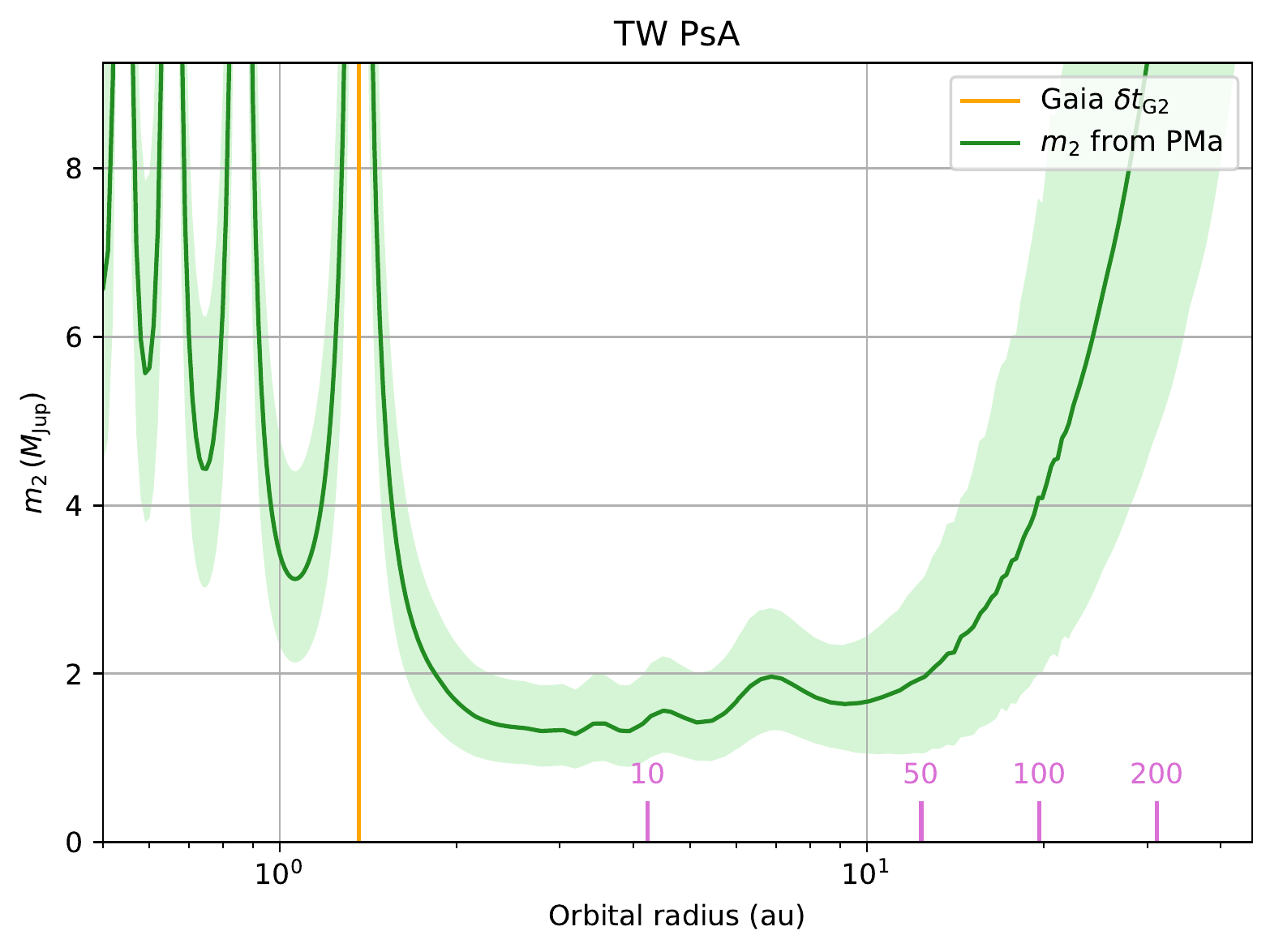}
\caption{Possible companion $(m_2,r)$ combinations for TW PsA (Fomalhaut B).
We note that the observed tangential velocity anomaly may be due to gravitational interaction with Fomalhaut A, located at a physical separation of only 0.28\,pc.
\label{TW-PsA-m2r}}
\end{figure}

We detect a significant tangential velocity anomaly of $\Delta v_\mathrm{tan,G2} = 18.7 \pm 6.4$\,m\,s$^{-1}$ on TW PsA in the GDR2. This relatively low velocity could indicate the presence of a moderately massive companion in orbit (Fig.~\ref{TW-PsA-m2r}). But the gravitational interaction with its massive stellar neighbor Fomalhaut A, located at a distance of only 0.28\,pc ($\approx 58$\,kau), is expected to induce an orbital velocity anomaly on TW\,PsA at a level comparable to the detected value.
This object is therefore an interesting example of a star in a relatively complex gravitational environment, for which the astrometric detection of planetary mass companions will require modeling of the multiple stellar system.

%______________ Figure

\subsection{HD 103095 (Gmb 1830)}

\object{HD 103095} (\object{Gmb 1830}, \object{GJ 451}, \object{HIP 57939}) is a very metal-poor dwarf ([Fe/H]=-1.29; \citeads{2017MNRAS.469.4378M}, see also \citeads{2014A&A...564A.133J}) with a spectral type K1V \citepads{2012A&A...545A..17C,2018MNRAS.475L..81K} and a very high space velocity of $v_\mathrm{tot} =322$\,km\,s$^{-1}$. HD 103095 is one of the 34 Gaia benchmark stars that were selected to serve as references for the calibration of the Gaia catalog \citepads{2015A&A...582A..49H}.

We do not detect any significant tangential velocity anomaly on this star ($\Delta v_\mathrm{tan,G2} = 1.0 \pm 5.0$\,m\,s$^{-1}$).
This allows us to set a limit to the mass of possible planetary companions of $0.5\,M_J$ between 2 and 10\,au, and $1\,M_J$ up to 20\,au.
We formally exclude the existence of the M dwarf companion on a $r\approx 13$\,au orbit discussed by \citetads{1974ApJ...194..637B}.

Although spectroscopic evidence for planetary contamination of some metal-poor stars has been recently presented by \citetads{2018MNRAS.475.3502R}, the formation of massive planets around metal poor stars is still a largely open question.
Due to the low density and depth of absorption lines in their spectra, these stars are unfavorable targets for high precision radial velocity surveys.
Astrometric measurements are however unaffected, and future Gaia releases will provide a statistical view of the presence of massive planets around old population, metal-poor stars.

%______________ Figure
\begin{figure}
\centering
\includegraphics[width=\hsize]{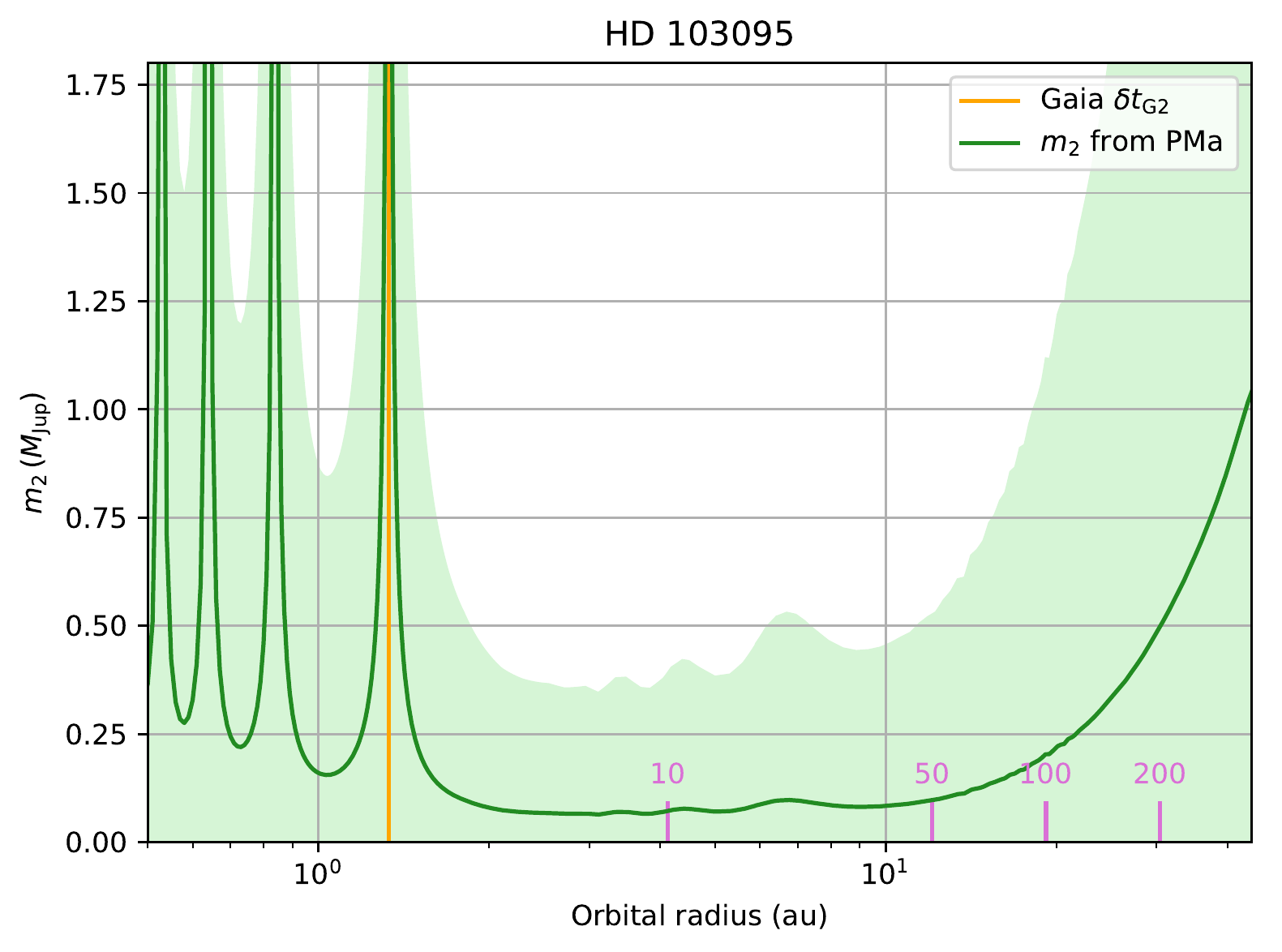}
\caption{Possible companion $(m_2,r)$ combinations for the very metal-poor dwarf star HD 103095 (Gmb 1830).
\label{HD-103095-m2r}}
\end{figure}

\subsection{51 Peg}

\begin{figure}
\centering
\includegraphics[width=\hsize]{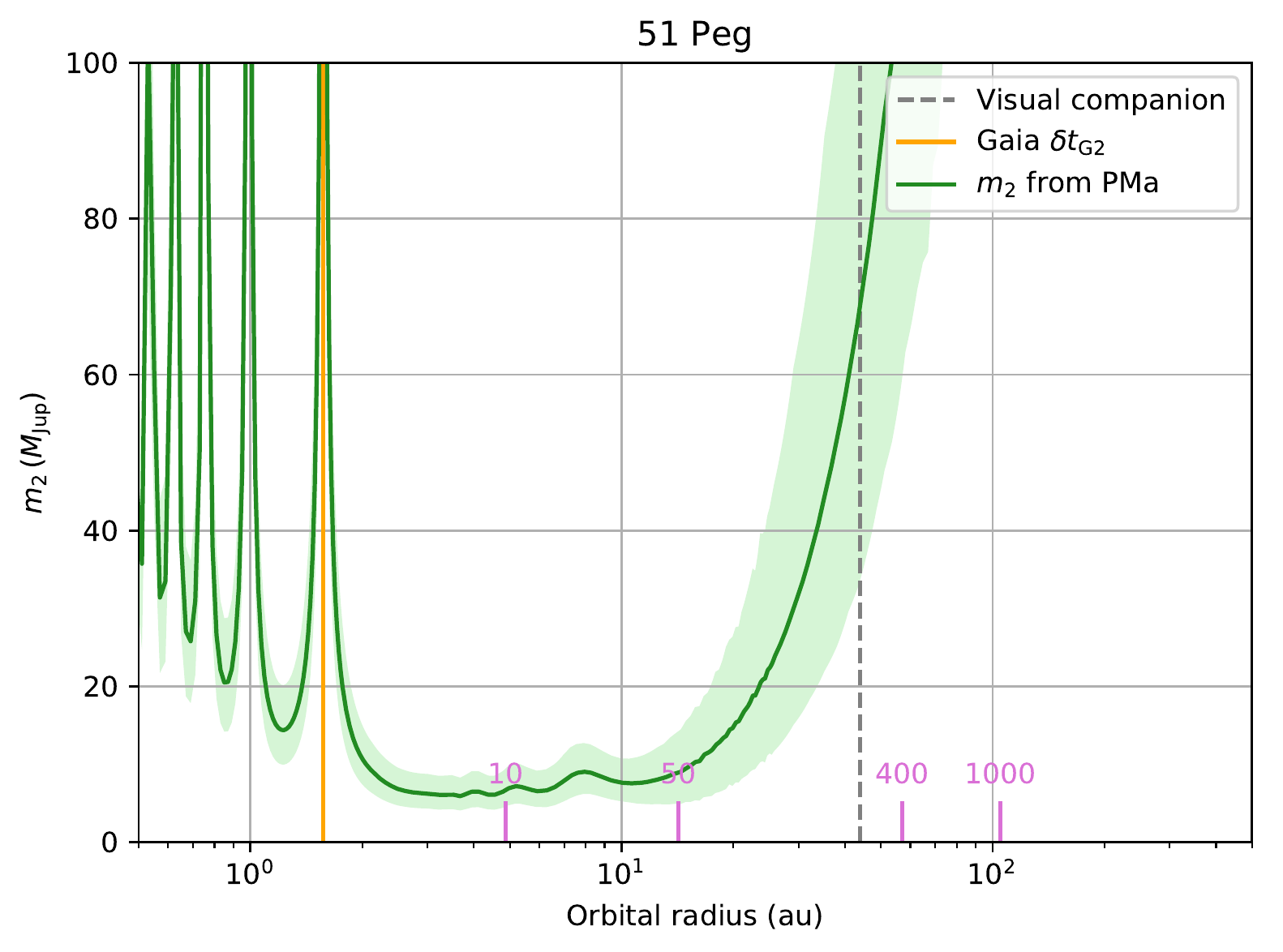}
\caption{Possible companion $(m_2,r)$ combinations for the exoplanet host star 51 Peg.
\label{51-Peg-m2r}}
\end{figure}

The discovery of the sub-Jovian mass companion of the otherwise unremarkable G2IV subgiant \object{51 Peg} by \citetads{1995Natur.378..355M} triggered the spectacular development of the field of exoplanet research.
We detect a relatively strong PMa on 51 Peg at a signal-to-noise ratio $\Delta_\mathrm{G2}=3.0$, indicative of the presence of an orbiting companion.
Fig.~\ref{51-Peg-m2r} shows the range of $(m_2,r)$ combinations that would explain the observed PMa.
\object{51 Peg b}, with an estimated mass of $m\,\sin i = 0.47\,M_J$ and an orbital semi-major axis of $a=0.052$\,au (period $P_\mathrm{orb} = 4.2$\,d) has a negligible influence on the PMa.

A visual companion to 51 Peg has been identified by \citetads{2011AJ....142..175R} at a projected separation of $2.87\arcsec$, that is, at a linear separation of $r=44$\,au (dashed gray line in Fig.~\ref{51-Peg-m2r}).
At this orbital radius, the PMa that we measure from the GDR2 would correspond to a mass of $m_2 \approx 70^{+70}_{-40}\,M_J$ for this companion, that could be compatible with the observed magnitude difference $\Delta I = 10 \pm 0.7$ listed by \citetads{2011AJ....142..175R}. The visual companion may therefore be gravitationally bound to 51\,Peg, although further characterization is needed to conclude on this matter.

%______________ Figure

\subsection{$\tau$ Boo}

%______________ Figure
\begin{figure}
\centering
\includegraphics[width=\hsize]{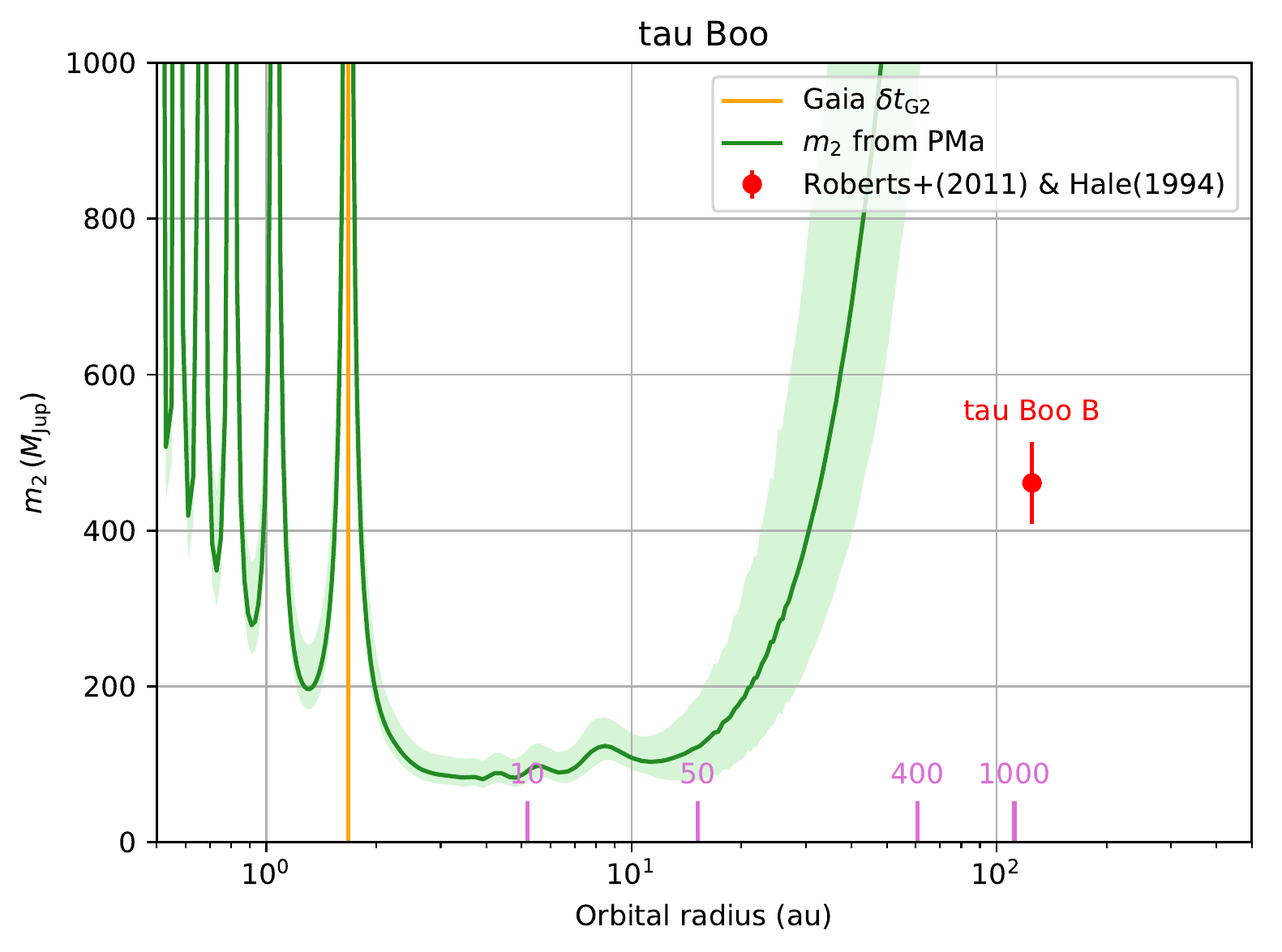}
\caption{Companion $(m_2,r)$ combinations for the exoplanet host star $\tau$ Boo. The known stellar companion of $\tau$\,Boo is shown in red. Its mass is estimated from the spectral type of M2V determined by \citetads{1994AJ....107..306H} and its semi-major axis of $r=125$\,au is from \citetads{2011AJ....142..175R}.
\label{tau-Boo-m2r}}
\end{figure}

$\tau$ Boo (\object{HD 120136}, \object{GJ 527}) is a known binary stellar system, with a stellar companion orbiting at a semi-major axis of $8.01\arcsec$ \citepads{2011AJ....142..175R} corresponding to 125\,au.
The spectral type of the companion determined by \citetads{1994AJ....107..306H} corresponds to an M2V star, of a mass of $m_2 \approx 0.44 \pm 0.05\,M_\odot$ \citepads{2013ApJS..208....9P}.
The orbital motion induced by the stellar companion $\tau$\,Boo B appears too slow to explain the observed PMa (Fig.~\ref{tau-Boo-m2r}), which may indicate the presence of another massive body in the system.
The exoplanet \object{$\tau$ Boo b} discovered by \citetads{1997ApJ...474L.115B}, with a mass of $4.1\,M_J$, an orbital radius $r=0.046$\,au, and an orbital period $P_\mathrm{orb} = 3.3$\,d, has a negligible  signature on the PMa. 

\end{appendix}
\end{document}

%% file: Tables/PM-table.tex
\begin{sidewaystable*}
 \caption{Proper motion of nearby stars from the Hip2 and GDR2 catalogs.
The PM vectors at the Hipparcos ($\mu_\mathrm{H}$) and GDR2 ($\mu_\mathrm{G2}$) epochs are compared to the mean PM from the Hip2 and GDR2 positions ($\mu_\mathrm{HG}$).
 The differences are expressed in S/N $\Delta_\mathrm{H}$ and $\Delta_\mathrm{G2}$.
$\Delta$ is set to $\star$ for a strong detection ($\Delta_\mathrm{G2}$>5), $\bullet$ for 3<$\Delta_\mathrm{G2}$<5 and $\circ$ for 2<$\Delta_\mathrm{G2}$<3.
$\epsilon_i$ is the astrometric excess noise and $\varrho$ the renormalized unit weight error (Sect.~\ref{excess-noise}).
The stars are listed in order of decreasing parallax, and the names of the stars discussed in Sect.~\ref{individual_notes} are emphasized using bold characters.
The full table is available from the CDS.}
 \label{pm_binaries1}
 \small
 \centering
 \renewcommand{\arraystretch}{1.0}
\setlength\tabcolsep{3pt}
\begin{tabular}{lcrrrrrrrrrrcccr}
 \hline
 \hline
Name  & $\varpi_\mathrm{G2}$  & $v_\mathrm{r}$ & \multicolumn{2}{c}{$\mu_\mathrm{HG}[\mathrm{H}]$ (mas\,a$^{-1}$)} & \multicolumn{2}{c}{$\mu_\mathrm{H} - \mu_\mathrm{HG}[\mathrm{H}]$ (mas\,a$^{-1}$)} & $\Delta_\mathrm{H}$ & \multicolumn{2}{c}{$\mu_\mathrm{HG}[\mathrm{G2}]$ (mas\,a$^{-1}$)} & \multicolumn{2}{c}{$\mu_\mathrm{G2} - \mu_\mathrm{HG}[\mathrm{G2}]$ (mas\,a$^{-1}$)} & $\Delta_\mathrm{G2}$ & $\Delta$ & $\epsilon_i$ & $\varrho$ \\
 & (mas) & (km\,s$^{-1}$) & \multicolumn{1}{c}{$\mu_\alpha$} & \multicolumn{1}{c}{$\mu_\delta$} & \multicolumn{1}{c}{$\mu_\alpha$} & \multicolumn{1}{c}{$\mu_\delta$} & & \multicolumn{1}{c}{$\mu_\alpha$} & \multicolumn{1}{c}{$\mu_\delta$} & \multicolumn{1}{c}{$\mu_\alpha$} & \multicolumn{1}{c}{$\mu_\delta$} & & & (mas) & \\
  \hline \noalign{\smallskip}
\textbf{\object{Proxima}} & $768.529_{\,0.220}$ & $ -22.20_{\, 0.03}^{\,K}$ & $-3779.090_{\,0.046}$ & $+765.519_{\,0.065}$ & $+3.34_{\,1.63}$ & $+0.02_{\,2.01}$ &  1.3 & $-3781.629_{\,0.048}$ & $+769.421_{\,0.052}$ & $+0.22_{\,0.11}$ & $+0.38_{\,0.21}$ &  1.8 &  & 0.34 & 1.0 \\
\textbf{\object{Barnard star}} & $547.480_{\,0.314}$ & $-110.40_{\, 0.30}^{\,a}$ & $-798.937_{\,0.035}$ & $+10330.695_{\,0.037}$ & $+0.36_{\,1.72}$ & $-2.58_{\,1.22}$ &  1.2 & $-801.413_{\,0.040}$ & $+10361.664_{\,0.040}$ & $-1.34_{\,0.69}$ & $+0.79_{\,0.39}$ &  2.0 &  & 0.14 & 1.1 \\
\object{Ross 154} & $336.152_{\,0.072}$ & $ -10.41_{\, 0.30}^{\,a}$ & $+639.381_{\,0.050}$ & $-193.866_{\,0.048}$ & $-2.36_{\,2.80}$ & $+2.23_{\,1.70}$ &  1.0 & $+639.499_{\,0.051}$ & $-193.878_{\,0.045}$ & $-0.15_{\,0.15}$ & $+0.22_{\,0.13}$ &  1.4 &  & 0.14 & 1.0 \\
\textbf{\object{eps Eri}} & $312.248_{\,0.505}$ & $  16.20_{\, 0.30}^{\,a}$ & $-975.327_{\,0.013}$ & $+19.964_{\,0.013}$ & $+0.16_{\,0.21}$ & $-0.47_{\,0.20}$ &  1.7 & $-975.082_{\,0.013}$ & $+19.978_{\,0.013}$ & $-0.23_{\,0.61}$ & $+0.29_{\,0.60}$ &  0.4 &  & 1.36 & 0.9 \\
\object{HD 217987} & $304.248_{\,0.053}$ & $   8.79_{\, 0.30}^{\,a}$ & $+6767.535_{\,0.017}$ & $+1326.511_{\,0.021}$ & $+0.67_{\,0.59}$ & $+1.01_{\,0.68}$ &  1.3 & $+6765.878_{\,0.018}$ & $+1330.225_{\,0.021}$ & $+0.18_{\,0.09}$ & $+0.03_{\,0.09}$ &  1.4 &  & 0.00 & 1.0 \\
\textbf{\object{Ross 128}} & $296.336_{\,0.078}$ & $ -30.99_{\, 0.30}^{\,a}$ & $+607.045_{\,0.078}$ & $-1222.424_{\,0.070}$ & $-1.78_{\,2.32}$ & $+3.14_{\,1.97}$ &  1.2 & $+607.321_{\,0.076}$ & $-1222.981_{\,0.047}$ & $+0.40_{\,0.17}$ & $-0.22_{\,0.10}$ &  2.3 & $\circ$ & 0.19 & 1.0 \\
\textbf{\object{61 Cyg B}} & $286.175_{\,0.067}$ & $ -64.04_{\, 0.30}^{\,a}$ & $+4103.596_{\,0.014}$ & $+3149.652_{\,0.013}$ & $+3.30_{\,0.32}$ & $-4.97_{\,0.44}$ & 11.0 & $+4108.549_{\,0.013}$ & $+3150.925_{\,0.015}$ & $-2.73_{\,0.11}$ & $+4.69_{\,0.12}$ & 34.0 & $\star$ & 0.30 & 1.0 \\
\textbf{\object{61 Cyg A}} & $285.975_{\,0.111}$ & $ -65.73_{\, 0.30}^{\,a}$ & $+4156.698_{\,0.236}$ & $+3252.805_{\,0.218}$ & $+11.61_{\,6.57}$ & $+16.39_{\,12.08}$ &  1.5 & $+4161.853_{\,0.225}$ & $+3254.206_{\,0.230}$ & $+2.36_{\,0.31}$ & $-4.36_{\,0.36}$ & 10.5 & $\star$ & 0.46 & 0.8 \\
\textbf{\object{HD 173739}} & $283.978_{\,0.071}$ & $  -0.82_{\, 0.30}^{\,a}$ & $-1310.882_{\,0.070}$ & $+1795.416_{\,0.077}$ & $-21.15_{\,2.23}$ & $+12.06_{\,2.75}$ &  6.9 & $-1311.369_{\,0.065}$ & $+1795.092_{\,0.089}$ & $-0.44_{\,0.21}$ & $-3.01_{\,0.25}$ &  9.2 & $\star$ & 0.00 & 1.1 \\
\textbf{\object{HD 173740}} & $283.891_{\,0.117}$ & $   1.23_{\, 0.30}^{\,a}$ & $-1400.171_{\,0.102}$ & $+1858.954_{\,0.124}$ & $+61.49_{\,3.29}$ & $-49.80_{\,4.06}$ & 15.1 & $-1400.669_{\,0.089}$ & $+1858.528_{\,0.136}$ & $+0.75_{\,0.36}$ & $+3.77_{\,0.44}$ &  6.8 & $\star$ & 0.15 & 1.2 \\
\object{GJ 15 A} & $280.719_{\,0.051}$ & $  11.81_{\, 0.30}^{\,a}$ & $+2890.548_{\,0.020}$ & $+412.262_{\,0.013}$ & $-1.63_{\,0.60}$ & $-2.16_{\,0.48}$ &  3.5 & $+2890.209_{\,0.020}$ & $+411.245_{\,0.015}$ & $+1.28_{\,0.08}$ & $+0.55_{\,0.06}$ & 14.0 & $\star$ & 0.00 & 0.9 \\
\textbf{\object{tau Cet}} & $277.545_{\,0.559}$ & $ -16.75_{\, 0.30}^{\,a}$ & $-1721.610_{\,0.017}$ & $+854.509_{\,0.017}$ & $+0.56_{\,0.18}$ & $-0.35_{\,0.15}$ &  2.8 & $-1721.958_{\,0.017}$ & $+854.806_{\,0.017}$ & $-7.70_{\,1.42}$ & $+0.62_{\,0.85}$ &  4.7 & $\bullet$ & 1.68 & 1.1 \\
\textbf{\object{eps Ind}} & $274.834_{\,0.270}$ & $ -40.10_{\, 0.30}^{\,b}$ & $+3960.985_{\,0.008}$ & $-2538.753_{\,0.010}$ & $-0.06_{\,0.24}$ & $-0.48_{\,0.17}$ &  1.6 & $+3964.958_{\,0.009}$ & $-2537.320_{\,0.011}$ & $+2.12_{\,0.41}$ & $+1.42_{\,0.45}$ &  4.2 & $\bullet$ & 1.18 & 1.2 \\
\object{YZ Cet} & $269.392_{\,0.087}$ & $  28.21_{\, 0.30}^{\,a}$ & $+1205.210_{\,0.153}$ & $+637.846_{\,0.187}$ & $+3.32_{\,5.57}$ & $+2.88_{\,3.71}$ &  0.7 & $+1204.728_{\,0.196}$ & $+637.658_{\,0.181}$ & $+0.52_{\,0.27}$ & $+0.02_{\,0.23}$ &  1.5 &  & 0.26 & 1.2 \\
\object{Kapteyn star} & $254.255_{\,0.035}$ & $ 245.22_{\, 0.30}^{\,a}$ & $+6507.242_{\,0.024}$ & $-5731.700_{\,0.027}$ & $-2.16_{\,0.98}$ & $+0.86_{\,0.96}$ &  1.7 & $+6491.492_{\,0.024}$ & $-5709.016_{\,0.027}$ & $+0.10_{\,0.07}$ & $-0.14_{\,0.07}$ &  1.7 &  & 0.00 & 1.1 \\
\textbf{\object{AX Mic}} & $251.858_{\,0.064}$ & $  20.08_{\, 0.10}^{\,c}$ & $-3259.437_{\,0.020}$ & $-1147.107_{\,0.021}$ & $-0.85_{\,0.92}$ & $+0.89_{\,0.40}$ &  1.2 & $-3258.974_{\,0.016}$ & $-1145.813_{\,0.021}$ & $+0.44_{\,0.12}$ & $+0.27_{\,0.10}$ &  3.4 & $\bullet$ & 0.00 & 0.9 \\
\textbf{\object{Ross 614}} & $242.995_{\,0.954}$ & $  16.75_{\, 0.20}^{\,e}$ & $+724.650_{\,0.097}$ & $-673.608_{\,0.079}$ & $-19.37_{\,2.66}$ & $+61.69_{\,2.40}$ & 18.0 & $+724.506_{\,0.090}$ & $-673.469_{\,0.070}$ & $+25.68_{\,1.78}$ & $-129.38_{\,1.59}$ & 55.1 & $\star$ & 3.70 & 11.9 \\
\object{BD-12 4523} & $232.238_{\,0.071}$ & $ -21.20_{\, 0.30}^{\,a}$ & $-94.195_{\,0.048}$ & $-1183.523_{\,0.050}$ & $-0.61_{\,2.30}$ & $+0.09_{\,1.74}$ &  0.2 & $-94.221_{\,0.045}$ & $-1183.812_{\,0.043}$ & $+0.20_{\,0.15}$ & $-0.01_{\,0.10}$ &  1.1 &  & 0.12 & 1.1 \\
\textbf{\object{Wolf 28}} & $231.766_{\,0.046}$ & $  15.00_{\,20.00}^{\,A}$ & $+1231.412_{\,0.160}$ & $-2711.868_{\,0.108}$ & $+5.49_{\,5.26}$ & $+2.68_{\,3.18}$ &  1.0 & $+1231.163_{\,0.157}$ & $-2711.417_{\,0.132}$ & $+0.19_{\,0.18}$ & $-0.51_{\,0.15}$ &  2.3 & $\circ$ & 0.00 & 1.2 \\
\object{HD 225213} & $230.162_{\,0.068}$ & $  25.53_{\, 0.30}^{\,b}$ & $+5633.905_{\,0.019}$ & $-2338.235_{\,0.024}$ & $+0.78_{\,0.86}$ & $+0.53_{\,0.71}$ &  0.8 & $+5633.447_{\,0.024}$ & $-2334.707_{\,0.021}$ & $+0.01_{\,0.19}$ & $-0.20_{\,0.09}$ &  0.9 &  & 0.00 & 1.2 \\
\object{CD-46 11540} & $219.830_{\,0.057}$ & $  -2.90_{\, 0.30}^{\,b}$ & $+572.459_{\,0.040}$ & $-880.484_{\,0.031}$ & $-1.20_{\,1.77}$ & $-0.35_{\,0.74}$ &  0.7 & $+572.540_{\,0.041}$ & $-880.471_{\,0.036}$ & $-0.02_{\,0.12}$ & $+0.15_{\,0.09}$ &  1.0 &  & 0.00 & 1.0 \\
\object{BD+68 946} & $219.810_{\,0.041}$ & $ -28.79_{\, 0.30}^{\,a}$ & $-320.766_{\,0.027}$ & $-1269.513_{\,0.042}$ & $+0.20_{\,1.14}$ & $+0.18_{\,1.39}$ &  0.1 & $-320.746_{\,0.029}$ & $-1269.942_{\,0.042}$ & $+0.24_{\,0.10}$ & $+0.34_{\,0.10}$ &  3.1 & $\bullet$ & 0.08 & 1.2 \\
\textbf{\object{LAWD 37}} & $215.766_{\,0.041}$ & $   0.00_{\,50.00}^{\,}$ & $+2661.823_{\,0.057}$ & $-346.882_{\,0.073}$ & $+2.09_{\,2.21}$ & $-0.50_{\,2.07}$ &  0.7 & $+2662.054_{\,0.062}$ & $-345.108_{\,0.065}$ & $-0.52_{\,0.09}$ & $+0.38_{\,0.09}$ &  5.0 & $\bullet$ & 0.00 & 1.0 \\
\object{BD-15 6290} & $213.896_{\,0.085}$ & $  -1.59_{\, 0.30}^{\,b}$ & $+957.891_{\,0.052}$ & $-673.629_{\,0.045}$ & $+1.95_{\,3.36}$ & $-1.70_{\,1.68}$ &  0.7 & $+957.926_{\,0.048}$ & $-673.613_{\,0.038}$ & $+0.08_{\,0.14}$ & $-0.16_{\,0.12}$ &  0.9 &  & 0.20 & 1.1 \\
\object{HD 88230} & $205.421_{\,0.042}$ & $ -25.74_{\, 0.30}^{\,a}$ & $-1363.047_{\,0.008}$ & $-505.336_{\,0.011}$ & $+0.73_{\,0.43}$ & $+0.82_{\,0.32}$ &  2.0 & $-1363.309_{\,0.008}$ & $-505.724_{\,0.013}$ & $+0.02_{\,0.07}$ & $+0.20_{\,0.07}$ &  2.0 & $\circ$ & 0.00 & 0.9 \\
\object{HD 204961} & $201.436_{\,0.051}$ & $  13.15_{\, 0.30}^{\,b}$ & $-46.048_{\,0.022}$ & $-816.381_{\,0.026}$ & $-0.00_{\,0.95}$ & $-1.25_{\,0.59}$ &  1.1 & $-46.047_{\,0.023}$ & $-816.273_{\,0.029}$ & $+0.24_{\,0.09}$ & $-0.47_{\,0.09}$ &  4.2 & $\bullet$ & 0.00 & 1.0 \\
\object{CD-44 11909} & $199.732_{\,0.092}$ & $ -34.84_{\, 0.30}^{\,b}$ & $-705.519_{\,0.082}$ & $-937.569_{\,0.078}$ & $-3.46_{\,2.55}$ & $+0.17_{\,1.88}$ &  1.1 & $-705.839_{\,0.081}$ & $-937.836_{\,0.077}$ & $-0.34_{\,0.22}$ & $-0.15_{\,0.16}$ &  1.4 &  & 0.21 & 1.1 \\
\object{omi02 Eri} & $198.595_{\,0.551}$ & $ -42.42_{\, 0.30}^{\,a}$ & $-2239.659_{\,0.014}$ & $-3420.303_{\,0.014}$ & $-0.46_{\,0.23}$ & $+0.03_{\,0.20}$ &  1.5 & $-2240.716_{\,0.011}$ & $-3421.654_{\,0.011}$ & $+0.25_{\,0.49}$ & $+0.25_{\,0.39}$ &  0.6 &  & 1.40 & 1.3 \\
\object{EV Lac} & $198.040_{\,0.046}$ & $   0.30_{\, 0.30}^{\,b}$ & $-706.130_{\,0.037}$ & $-458.808_{\,0.045}$ & $+0.77_{\,1.12}$ & $-1.91_{\,1.33}$ &  1.2 & $-706.091_{\,0.042}$ & $-458.864_{\,0.048}$ & $-0.07_{\,0.10}$ & $-0.05_{\,0.09}$ &  0.6 &  & 0.11 & 1.0 \\
\object{GJ 445} & $190.291_{\,0.055}$ & $-111.61_{\, 0.30}^{\,a}$ & $+747.288_{\,0.062}$ & $+480.542_{\,0.040}$ & $-3.68_{\,1.87}$ & $+0.86_{\,1.51}$ &  1.6 & $+748.287_{\,0.051}$ & $+480.720_{\,0.046}$ & $-0.09_{\,0.12}$ & $+0.00_{\,0.11}$ &  0.6 &  & 0.22 & 1.4 \\
%----------
\noalign{\smallskip}  \hline  \noalign{\smallskip}
\textbf{\object{HD 42581}} & $173.724_{\,0.054}$ & $   4.70_{\, 0.30}^{\,a}$ & $-136.428_{\,0.014}$ & $-714.929_{\,0.021}$ & $-0.66_{\,0.50}$ & $+1.27_{\,0.81}$ &  1.5 & $-136.427_{\,0.017}$ & $-714.899_{\,0.020}$ & $+0.52_{\,0.12}$ & $-4.10_{\,0.17}$ & 19.7 & $\star$ & 0.00 & 1.0 \\
\textbf{\object{e Eri}} & $166.590_{\,0.254}$ & $  87.77_{\, 0.30}^{\,b}$ & $+3037.463_{\,0.008}$ & $+726.558_{\,0.008}$ & $+0.88_{\,0.20}$ & $+0.02_{\,0.21}$ &  3.0 & $+3035.017_{\,0.006}$ & $+727.044_{\,0.010}$ & $-1.17_{\,0.55}$ & $+3.72_{\,0.67}$ &  4.5 & $\bullet$ & 1.31 & 1.0 \\
\textbf{\object{HD 79211}} & $157.914_{\,0.049}$ & $  12.47_{\, 0.30}^{\,a}$ & $-1570.602_{\,0.190}$ & $-660.666_{\,0.208}$ & $+18.19_{\,9.46}$ & $+19.79_{\,6.94}$ &  2.3 & $-1570.289_{\,0.192}$ & $-660.982_{\,0.228}$ & $-2.82_{\,0.21}$ & $+1.00_{\,0.24}$ &  9.4 & $\star$ & 0.00 & 1.0 \\
\textbf{\object{HD 79210}} & $157.909_{\,0.045}$ & $  11.12_{\, 0.30}^{\,a}$ & $-1548.565_{\,0.132}$ & $-568.636_{\,0.170}$ & $+12.85_{\,6.94}$ & $-7.95_{\,4.53}$ &  1.8 & $-1548.294_{\,0.131}$ & $-568.956_{\,0.175}$ & $+2.21_{\,0.15}$ & $-0.03_{\,0.19}$ &  9.1 & $\star$ & 0.00 & 1.0 \\
\textbf{\object{TW PsA}} & $131.467_{\,0.095}$ & $   7.14_{\, 0.30}^{\,b}$ & $+330.153_{\,0.018}$ & $-158.551_{\,0.017}$ & $+0.96_{\,0.65}$ & $-0.43_{\,0.48}$ &  1.3 & $+330.142_{\,0.015}$ & $-158.535_{\,0.018}$ & $-0.51_{\,0.13}$ & $+0.11_{\,0.12}$ &  2.9 & $\circ$ & 0.00 & 1.0 \\
\textbf{\object{HD 103095}} & $108.984_{\,0.057}$ & $ -98.08_{\, 0.30}^{\,a}$ & $+4002.617_{\,0.012}$ & $-5813.190_{\,0.009}$ & $+1.36_{\,0.37}$ & $-0.43_{\,0.23}$ &  3.3 & $+4002.625_{\,0.013}$ & $-5817.733_{\,0.009}$ & $+0.02_{\,0.09}$ & $-0.01_{\,0.07}$ &  0.2 & $\bullet$ & 0.00 & 1.0 \\
\textbf{\object{51 Peg}} & $ 64.678_{\,0.135}$ & $ -33.22_{\, 0.30}^{\,a}$ & $+207.532_{\,0.009}$ & $+61.091_{\,0.009}$ & $-0.28_{\,0.31}$ & $-0.75_{\,0.30}$ &  1.9 & $+207.554_{\,0.011}$ & $+61.096_{\,0.009}$ & $-0.17_{\,0.23}$ & $+0.86_{\,0.18}$ &  3.0 & $\bullet$ & 0.40 & 1.1 \\
\textbf{\object{tau Boo}} & $ 63.893_{\,0.369}$ & $ -16.25_{\, 0.10}^{\,c}$ & $-475.815_{\,0.011}$ & $+57.972_{\,0.009}$ & $-3.71_{\,0.16}$ & $-4.48_{\,0.13}$ & 28.2 & $-475.840_{\,0.011}$ & $+57.967_{\,0.011}$ & $+8.01_{\,0.77}$ & $+6.79_{\,0.59}$ & 10.9 & $\star$ & 1.79 & 1.3 \\
\textbf{\object{bet Pic}} & $ 50.652_{\,0.361}$ & $  19.89_{\, 0.70}^{\,e}$ & $+4.939_{\,0.011}$ & $+83.935_{\,0.018}$ & $-0.29_{\,0.11}$ & $-0.83_{\,0.15}$ &  4.7 & $+4.939_{\,0.011}$ & $+83.930_{\,0.018}$ & $-2.33_{\,0.74}$ & $-1.27_{\,0.74}$ &  2.5 & $\bullet$ & 2.14 & 1.2 \\
\noalign{\smallskip} \hline
 \end{tabular}
 \tablebib{The reference for the radial velocities $v_\mathrm{r}$ are:
  $K$: \citetads{2017A&A...598L...7K},
 $A$: \citetads{1993AJ....105.1033A},
 $a$: \citetads{2002ApJS..141..503N},
 $b$: \citetads{2018A&A...616A...7S},
 $c$: \citetads{2007A&A...475..519H},
 $d$: \citetads{2018A&A...616A...1G},
 $e$: \citetads{2012AstL...38..331A},
 $f$: radial velocity unavailable.}
\end{sidewaystable*}

%% file: Tables/Phys-table.tex
\begin{table*}
\caption{Physical properties of the stars tested for the presence of a PMa.
The stars are listed in order of decreasing parallax, and the names of the stars discussed in Sect.~\ref{individual_notes} are emphasized using bold characters.
$\Delta v_\mathrm{T,G2}$ is the norm of the tangential velocity anomaly, $\theta_\mathrm{G2}$ is the position angle of the observed PMa vector, and $m^\diamond_{2}$ is the mass of a companion explaining the PMa, normalized at an orbiting radius of $r=1$\,au.
$\Delta_\mathrm{G2}$ is the S/N of the PMa, $\Delta$ represents the detection level ($\circ$ for $2<\Delta_\mathrm{G2}<3$, $\bullet$ for $3<\Delta_\mathrm{G2}<5$ and $\star$ for $\Delta_\mathrm{G2}>5$).
$r_\mathrm{G2}$ is the radius of a circular orbit whose period is equal to $\delta t_\mathrm{G2}=668$\,days.
The stars present in the Washington Double Star catalog \citepads{2001AJ....122.3466M} are marked with a $\diamond$ symbol in the `Bin.' column, and those present in the Double and Multiple Star Annex (DMSA) of the original Hipparcos catalog \citepads{1997ESASP1200.....E} are marked with a $\ddag$ symbol.
The full table is available from the CDS.
}
 \label{phys-properties}
 \small
 \centering
 \renewcommand{\arraystretch}{1.2}
 \begin{tabular}{llccrrrcccc}
 \hline
 \hline
  Name  & Spectral & $m_1$ & $R$ & $\Delta v_\mathrm{T,G2}$  & $\theta_\mathrm{G2}$ & $\Delta_\mathrm{G2}$   & $\Delta$  & $r_\mathrm{G2}$  & $m^\diamond_{2}$ & Bin. \\
      & Type & ($M_\odot$) & ($R_\odot$) & (m\,s$^{-1}$)  & ($\deg$) &  &  & (au) & ($M_J\,\mathrm{au}^{-1/2}$) &  \\
  \hline  \noalign{\smallskip}
\textbf{\object{Proxima}} & M5.5Ve & $0.122_{\,0.003}$ & $0.154_{\,0.005}$ & $   2.7_{\, 1.5}$ & $30_{24}$ & $1.8$  &  & 0.74 & $  0.040^{\,+0.020}_{\,-0.020}$ &  $\diamond$ \\
\textbf{\object{Barnard star}} & M4V & $0.160_{\,0.003}$ & $0.194_{\,0.006}$ & $  13.5_{\, 6.9}$ & $301_{20}$ & $2.0$  &  & 0.81 & $  0.220^{\,+0.110}_{\,-0.100}$ &   \\
\object{Ross 154} & M3.5Ve & $0.179_{\,0.004}$ & $0.210_{\,0.006}$ & $   3.8_{\, 2.8}$ & $325_{27}$ & $1.4$  &  & 0.84 & $  0.060^{\,+0.050}_{\,-0.040}$ &  $\diamond$ \\
\textbf{\object{eps Eri}} & K2V & $0.847_{\,0.042}$ & $0.702_{\,0.035}$ & $   5.7_{\,13.1}$ & $321_{48}$ & $0.4$  &  & 1.42 & $  0.210^{\,+0.430}_{\,-0.420}$ &  $\diamond$ \\
\object{HD 217987} & M2V & $0.473_{\,0.036}$ & $0.462_{\,0.034}$ & $   2.9_{\, 2.0}$ & $80_{32}$ & $1.4$  &  & 1.17 & $  0.080^{\,+0.050}_{\,-0.050}$ &   \\
\textbf{\object{Ross 128}} & M4V & $0.178_{\,0.004}$ & $0.210_{\,0.006}$ & $   7.3_{\, 3.2}$ & $119_{17}$ & $2.3$  & $\circ$ & 0.84 & $  0.120^{\,+0.060}_{\,-0.050}$ &   \\
\textbf{\object{61 Cyg B}} & K7V & $0.657_{\,0.057}$ & $0.644_{\,0.067}$ & $  89.9_{\, 2.6}$ & $330_{1}$ & $34.0$  & $\star$ & 1.30 & $  2.950^{\,+0.830}_{\,-0.340}$ &  $\diamond$ \\
\textbf{\object{61 Cyg A}} & K5V & $0.708_{\,0.053}$ & $0.702_{\,0.068}$ & $  82.1_{\, 7.8}$ & $152_{4}$ & $10.5$  & $\star$ & 1.33 & $  2.790^{\,+0.820}_{\,-0.380}$ &  $\diamond$ \\
\textbf{\object{HD 173739}} & M3V & $0.340_{\,0.007}$ & $0.347_{\,0.010}$ & $  50.8_{\, 5.5}$ & $188_{4}$ & $9.2$  & $\star$ & 1.04 & $  1.200^{\,+0.350}_{\,-0.170}$ &  $\ddag$$\diamond$ \\
\textbf{\object{HD 173740}} & M3.5V & $0.261_{\,0.006}$ & $0.281_{\,0.008}$ & $  64.2_{\, 9.4}$ & $11_{5}$ & $6.8$  & $\star$ & 0.95 & $  1.330^{\,+0.410}_{\,-0.220}$ &  $\ddag$$\diamond$ \\
\object{GJ 15 A} & M2V & $0.409_{\,0.008}$ & $0.406_{\,0.012}$ & $  23.5_{\, 1.7}$ & $67_{3}$ & $14.0$  & $\star$ & 1.11 & $  0.610^{\,+0.170}_{\,-0.070}$ &  $\diamond$ \\
\textbf{\object{tau Cet}} & G8V & $0.900_{\,0.045}$ & $0.751_{\,0.038}$ & $ 131.9_{\,28.2}$ & $275_{7}$ & $4.7$  & $\bullet$ & 1.44 & $  5.060^{\,+1.700}_{\,-1.090}$ &  $\diamond$ \\
\textbf{\object{eps Ind}} & K5V & $0.778_{\,0.039}$ & $0.707_{\,0.035}$ & $  44.0_{\,10.5}$ & $56_{10}$ & $4.2$  & $\bullet$ & 1.38 & $  1.570^{\,+0.550}_{\,-0.370}$ &  $\diamond$ \\
\object{YZ Cet} & M4.0Ve & $0.135_{\,0.003}$ & $0.169_{\,0.005}$ & $   9.1_{\, 6.2}$ & $88_{28}$ & $1.5$  &  & 0.77 & $  0.140^{\,+0.090}_{\,-0.080}$ &   \\
\object{HIP 24186} & M1VIp & $0.286_{\,0.006}$ & $0.302_{\,0.009}$ & $   3.2_{\, 1.9}$ & $145_{22}$ & $1.7$  &  & 0.98 & $  0.070^{\,+0.040}_{\,-0.040}$ &   \\
\textbf{\object{AX Mic}} & M1V & $0.609_{\,0.042}$ & $0.592_{\,0.046}$ & $   9.8_{\, 2.9}$ & $58_{12}$ & $3.4$  & $\bullet$ & 1.27 & $  0.310^{\,+0.120}_{\,-0.090}$ &   \\
\textbf{\object{Ross 614}} & M4.5V & $0.242_{\,0.005}$ & $0.265_{\,0.008}$ & $2573.2_{\,46.7}$ & $169_{1}$ & $55.1$  & $\star$ & 0.93 & $ 51.170^{\,+14.270}_{\,-5.420}$ &  $\diamond$ \\
\object{BD-12 4523} & M3V & $0.310_{\,0.007}$ & $0.322_{\,0.010}$ & $   4.1_{\, 3.8}$ & $92_{34}$ & $1.1$  &  & 1.01 & $  0.090^{\,+0.080}_{\,-0.070}$ &  $\diamond$ \\
\textbf{\object{Wolf 28}} & DZ7.5 & $0.680_{\,0.020}$ & $0.011_{\,0.000}$ & $  11.0_{\, 4.8}$ & $160_{15}$ & $2.3$  & $\circ$ & 1.31 & $  0.370^{\,+0.170}_{\,-0.150}$ &   \\
\object{HD 225213} & M2V & $0.397_{\,0.008}$ & $0.396_{\,0.012}$ & $   4.1_{\, 4.4}$ & $177_{22}$ & $0.9$  &  & 1.10 & $  0.100^{\,+0.100}_{\,-0.100}$ &   \\
\object{CD-46 11540} & M3V & $0.360_{\,0.007}$ & $0.365_{\,0.011}$ & $   3.3_{\, 3.3}$ & $351_{28}$ & $1.0$  &  & 1.06 & $  0.080^{\,+0.070}_{\,-0.070}$ &   \\
\object{BD+68 946} & M3.0V & $0.410_{\,0.008}$ & $0.407_{\,0.012}$ & $   8.9_{\, 2.9}$ & $35_{13}$ & $3.1$  & $\bullet$ & 1.11 & $  0.230^{\,+0.090}_{\,-0.070}$ &  $\diamond$ \\
\textbf{\object{LAWD 37}} & DQ & $0.610_{\,0.010}$ & $0.015_{\,0.000}$ & $  14.1_{\, 2.8}$ & $306_{8}$ & $5.0$  & $\bullet$ & 1.27 & $  0.440^{\,+0.150}_{\,-0.090}$ &   \\
\object{BD-15 6290} & M3.5V & $0.346_{\,0.007}$ & $0.352_{\,0.011}$ & $   4.0_{\, 4.2}$ & $152_{34}$ & $0.9$  &  & 1.05 & $  0.090^{\,+0.090}_{\,-0.090}$ &   \\
\object{HD 88230} & K6VeFe-1 & $0.709_{\,0.049}$ & $0.703_{\,0.062}$ & $   4.6_{\, 2.3}$ & $5_{15}$ & $2.0$  & $\circ$ & 1.33 & $  0.160^{\,+0.080}_{\,-0.070}$ &  $\diamond$ \\
\object{HD 204961} & M2/3V & $0.449_{\,0.009}$ & $0.441_{\,0.013}$ & $  12.5_{\, 2.9}$ & $153_{10}$ & $4.2$  & $\bullet$ & 1.15 & $  0.340^{\,+0.120}_{\,-0.080}$ &   \\
\object{CD-44 11909} & M3.5 & $0.281_{\,0.006}$ & $0.298_{\,0.009}$ & $   8.8_{\, 6.5}$ & $246_{29}$ & $1.4$  &  & 0.98 & $  0.190^{\,+0.130}_{\,-0.120}$ &  $\diamond$ \\
\object{omi02 Eri} & K0V & $0.900_{\,0.045}$ & $0.788_{\,0.039}$ & $   8.5_{\,14.9}$ & $45_{43}$ & $0.6$  &  & 1.44 & $  0.330^{\,+0.510}_{\,-0.500}$ &  $\diamond$ \\
\object{EV Lac} & M4.0V & $0.328_{\,0.007}$ & $0.337_{\,0.010}$ & $   1.9_{\, 3.2}$ & $235_{42}$ & $0.6$  &  & 1.03 & $  0.050^{\,+0.070}_{\,-0.060}$ &  $\diamond$ \\
\object{G 254-29} & M4.0Ve & $0.250_{\,0.006}$ & $0.272_{\,0.008}$ & $   2.3_{\, 4.1}$ & $272_{47}$ & $0.6$  &  & 0.94 & $  0.050^{\,+0.070}_{\,-0.070}$ &   \\
%------
\noalign{\smallskip}  \hline  \noalign{\smallskip}
\textbf{\object{HD 42581}} & M1V & $0.563_{\,0.042}$ & $0.547_{\,0.044}$ & $ 112.7_{\, 5.7}$ & $173_{2}$ & $19.7$  & $\star$ & 1.24 & $  3.420^{\,+0.970}_{\,-0.400}$ &  $\diamond$ \\
\textbf{\object{e Eri}} & G6V & $0.980_{\,0.049}$ & $0.838_{\,0.042}$ & $ 111.0_{\,24.7}$ & $343_{8}$ & $4.5$  & $\bullet$ & 1.49 & $  4.440^{\,+1.510}_{\,-0.980}$ &   \\
\textbf{\object{TW PsA}} & K4Ve & $0.757_{\,0.038}$ & $0.743_{\,0.037}$ & $  18.7_{\, 6.4}$ & $282_{13}$ & $2.9$  & $\circ$ & 1.36 & $  0.660^{\,+0.270}_{\,-0.210}$ &   \\
\textbf{\object{HD 103095}} & K1V-Fe-1.5 & $0.703_{\,0.013}$ & $0.696_{\,0.021}$ & $   1.0_{\, 5.0}$ & $107_{48}$ & $0.2$  & $\bullet$ & 1.33 & $  0.030^{\,+0.150}_{\,-0.150}$ &   \\
\textbf{\object{51 Peg}} & G2IV & $1.158_{\,0.058}$ & $1.183_{\,0.059}$ & $  64.7_{\,21.5}$ & $349_{11}$ & $3.0$  & $\bullet$ & 1.57 & $  2.820^{\,+1.130}_{\,-0.870}$ &  $\diamond$ \\
\textbf{\object{tau Boo}} & F6IV+M2 & $1.400_{\,0.070}$ & $1.307_{\,0.065}$ & $ 778.9_{\,71.6}$ & $50_{4}$ & $10.9$  & $\star$ & 1.67 & $ 37.260^{\,+10.820}_{\,-4.970}$ &  $\diamond$ \\
\textbf{\object{bet Pic}} & A6V & $1.700_{\,0.085}$ & $1.459_{\,0.073}$ & $ 248.5_{\,97.9}$ & $241_{17}$ & $2.5$  & $\bullet$ & 1.78 & $ 13.100^{\,+5.790}_{\,-4.700}$ &   \\
\noalign{\smallskip} \hline
\end{tabular}
\end{table*}